\documentclass[aps,prd,floats,preprint,nofootinbib]{revtex4}
\usepackage{graphicx}
\usepackage{amsmath}
\usepackage{amsfonts}
\usepackage{amssymb}
\usepackage{multirow}
\begin{document}
\def\beq{\begin{equation}}
\def\eeq{\end{equation}}
\def\bea{\begin{eqnarray}}
\def\eea{\end{eqnarray}}
\def\ben{\begin{enumerate}}
\def\een{\end{enumerate}}
\def\ie{{\it i.e.}}
\def\etc{{\it etc.}}
\def\eg{{\it e.g.}}
\def\lsim{\mathrel{\raise.3ex\hbox{$<$\kern-.75em\lower1ex\hbox{$\sim$}}}}
\def\gsim{\mathrel{\raise.3ex\hbox{$>$\kern-.75em\lower1ex\hbox{$\sim$}}}}
\def\ifmath#1{\relax\ifmmode #1\else $#1$\fi}
\def\half{\ifmath{{\textstyle{1 \over 2}}}}
\def\threehalf{\ifmath{{\textstyle{3 \over 2}}}}
\def\quarter{\ifmath{{\textstyle{1 \over 4}}}}
\def\eigth{\ifmath{{\textstyle{1\over 8}}}}
\def\sixth{\ifmath{{\textstyle{1 \over 6}}}}
\def\third{\ifmath{{\textstyle{1 \over 3}}}}
\def\twothirds{{\textstyle{2 \over 3}}}
\def\fivethirds{{\textstyle{5 \over 3}}}
\def\fourth{\ifmath{{\textstyle{1\over 4}}}}
\def\chitil{\wt\chi}
\def\fbi{~{\mbox{fb}^{-1}}}
\def\fb{~{\mbox{fb}}}
\def\br{BR}
\def\gev{~{\mbox{GeV}}}
\def\calm{\mathcal{M}}
\def\mll{m_{\ell^+\ell^-}}
\def\tanb{\tan\beta}

\def\wtil{\widetilde}
\def\cnone{\wt\chi^0_1}
\def\cnonestar{\wt\chi_1^{0\star}}
\def\cntwo{\wt\chi^0_2}
\def\cnthree{\wt\chi^0_3}
\def\cnfour{\wt\chi^0_4}
\def\snu{\wt\nu}
\def\snul{\wt\nu_L}
\def\msnul{m_{\snul}}
\def\se{\wt e}
\def\smu{\wt\mu}
\def\snu{\wt\nu}
\def\snul{\wt\nu_L}
\def\msnul{m_{\snul}}

\def\snue{\wt\nu_e}
\def\snuel{\wt\nu_{e\,L}}
\def\msnuel{m_{\snul}}

\def\snubar{\ov{\snu}}
\def\msnu{m_{\snu}}

\def\snue{\wt\nu_e}
\def\snuel{\wt\nu_{e\,L}}
\def\msnuel{m_{\snul}}

\def\snubar{\ov{\snu}}
\def\msnu{m_{\snu}}
\def\mcnone{m_{\cnone}}
\def\mcntwo{m_{\cntwo}}
\def\mcnthree{m_{\cnthree}}
\def\mcnfour{m_{\cnfour}}
\def\wt{\widetilde}
\def\anti{\overline}
\def\wh{\widehat}
\def\cpone{\wt \chi^+_1}
\def\cmone{\wt \chi^-_1}
\def\cpmone{\wt \chi^{\pm}_1}
\def\mcpone{m_{\cpone}}
\def\mcpmone{m_{\cpmone}}

\def\staur{\wt \tau_R}
\def\staul{\wt \tau_L}
\def\stau{\wt \tau}
\def\mstaur{m_{\staur}}
\def\stauone{\wt \tau_1}
\def\mstauone{m_{\stauone}}

\def\gl{\wt g}
\def\mgl{m_{\gl}}
\def\stl{{\wt t_L}}
\def\str{{\wt t_R}}
\def\mstl{m_{\stl}}
\def\mstr{m_{\str}}
\def\sbl{{\wt b_L}}
\def\sbr{{\wt b_R}}
\def\msbl{m_{\sbl}}
\def\msbr{m_{\sbr}}
\def\sbot{\wt b}
\def\msbot{m_{\sbot}}
\def\sq{\wt q}
\def\sqbar{\ov{\sq}}
\def\msq{m_{\sq}}
\def\slep{\wt \ell}
\def\slepbar{\ov{\slep}}
\def\mslep{m_{\slep}}
\def\slepl{\wt \ell_L}
\def\mslepl{m_{\slepl}}
\def\slepr{\wt \ell_R}
\def\mslepr{m_{\slepr}}
\def\jet{{\rm jet}}
\def\filt{{\rm filt}}
\def\cut{{\rm cut}}
\def\sub{{\rm sub}}
\def\sig{\mbox{sig}}

\def\nsub{ n_{\text{sub}}}
\def\CC{{C\nolinebreak[4]\hspace{-.05em}\raisebox{.4ex}{\tiny\bf ++}}}
\newcommand{ \slashchar }[1]{\setbox0=\hbox{$#1$}   
   \dimen0=\wd0                                     
   \setbox1=\hbox{/} \dimen1=\wd1                   
   \ifdim\dimen0>\dimen1                            
      \rlap{\hbox to \dimen0{\hfil/\hfil}}          
      #1                                            
   \else                                            
      \rlap{\hbox to \dimen1{\hfil$#1$\hfil}}       
      /                                             
   \fi}
\widowpenalty=1000
\clubpenalty=1000
\vspace*{3cm}
\title{$W$-jet Tagging: Optimizing the Identification of Boosted Hadronically-Decaying $W$ Bosons }

\author{Yanou Cui,  Zhenyu Han, and Matthew D. Schwartz}

\affiliation{ \small \sl Center for the Fundamental Laws of Nature\\ Harvard University,\\ Cambridge, MA 02138, USA}

\def\thesection{\arabic{section}}
\def\thetable{\arabic{table}}

\begin{abstract}
A method is proposed for distinguishing highly boosted hadronically decaying $W$'s ($W$-jets) from QCD-jets using jet substructure.
Previous methods, such as the filtering/mass-drop method, can give a factor of $\sim2$ improvement in $S/\sqrt{B}$ for jet $p_T \gsim 200$ GeV.
In contrast, a multivariate approach including new discriminants such as $R$-cores,
which characterize the shape of the $W$-jet, subjet planar flow, and grooming-sensitivities
is shown to provide a much larger factor of $\sim 5$ improvement in $S/\sqrt{B}$.
For longitudinally polarized $W$'s, such as those coming from many new physics models, the discrimination is even better.
Comparing different Monte Carlo simulations, we observe a sensitivity of some variables to the underlying event;
however, even with a conservative estimates, the multivariate approach is very powerful.
Applications to semileptonic $WW$ resonance searches
and all-hadronic $W$+jet searches at the LHC are also discussed.
Code implementing our $W$-jet tagging algorithm is publicly available at http://jets.physics.harvard.edu/wtag.
\end{abstract}

\maketitle
\thispagestyle{empty}

\pagenumbering{arabic}
\section{Introduction}
\label{sec:introduction}
Highly energetic  $W$ and $Z$ bosons
 appear in many interesting physics processes at the TeV scale to be explored at the Large Hadron Collider (LHC).
For example, $WW$ scattering at high energy is a direct probe of the electroweak breaking mechanism \cite{Chanowitz:1985hj, Butterworth:2002tt}.
Heavy resonances, such as a $Z'$, a $W'$, a heavy Higgs or fourth generation quarks, often decay
to electroweak gauge bosons. Since the energy scales of these processes are much higher than the electroweak scale,
the $W$ and $Z$ bosons are often highly boosted. When decaying hadronically, a highly boosted $W$ or $Z$ boson
then appears as a single jet, called a $W$-jet or $Z$-jet. Since high energy QCD-jets (jets initiated by a quark or gluon)
 will be copiously produced at the LHC, $W$ or $Z$-jets may be overwhelmed by the QCD background,
 making it difficult to explore the nature of TeV scale physics. Therefore, being able to distinguish efficiently
 $W$ and $Z$-jets from QCD-jets could significantly improve our ability to understand the nature of TeV scale physics.

A number of recent studies have explored the hadronic decays of boosted objects, including not only $W$'s and $Z$'s
\cite{Butterworth:2002tt, Almeida:2008yp, pruning,  Thaler:2010tr,  Hackstein:2010wk, Katz:2010mr} but also
boosted light Higgses \cite{filtering, Plehn:2009rk, Kribs:2009yh, Soper:2010xk, Chen:2010wk, Falkowski:2010hi, Kribs:2010hp,  Almeida:2010pa, Katz:2010iq, Kim:2010uj} and
top quarks \cite{Thaler:2008ju, Kaplan:2008ie, Almeida:2008tp, pruning, Krohn:2009wm, Almeida:2010pa, Plehn:2010st, Bhattacherjee:2010za, Rehermann:2010vq}.
These studies have led to a general understanding of some of the  essential differences between a QCD-jet
and a jet initiated from a boosted massive particle decay. For example, a massive particle decay
often contains more than one hard subjet, {\it i.e.}
regions within the jet where energy is concentrated.
On the contrary, the energy distribution of a QCD-jet is more often dominated by one and only one such region.
Due to collinear singularities, QCD-jets tend to comprise particles with hierarchical energies,
while the energies of particles in a massive particle jet are usually more balanced.
These ideas were used in one of the first jet-substructure studies, Ref.~\cite{Butterworth:2002tt},
which attempted to identify $W$-jets in $WW$ scattering.
Some of the most poignant applications of substructure techniques include reviving the light Higgs to $b\bar{b}$ search \cite{filtering},
which has been implemented by ATLAS \cite{atlas}, and reducing the backgrounds to boosted hadronic tops by a factor of 10,000~\cite{Kaplan:2008ie}
which was implemented by CMS \cite{cms}.

\looseness -1 Boosted jets are often highly collimated, with characteristic sizes of order $R=0.4$ or smaller.
The basic trick to using jet substructure is, rather than starting with $R=0.4$ jets,
one starts with much larger jets, say $R=1.2$, and
then parses the jet using its clustering history. The goal is to
keep decay products from the boosted object, throwing out
contamination from initial state radiation and the underlying event. Some general algorithms
for doing this include filtering~\cite{filtering}, trimming~\cite{trimming}, and pruning~\cite{pruning}.
While these grooming techniques seem to help, it is not clear they are in any way optimal.
It was shown in~\cite{Soper:2010xk} that the different methods extract
overlapping but also at least partially complimentary information. In~\cite{Black:2010dq}, it was shown that even one
algorithm, trimming, is at least partially complimentary to itself if different sets of parameters are used.
Moreover, an interesting but underappreciated point about grooming that we demonstrate here (see Figure~\ref{fig:moneyplot})
is that grooming, by itself, does not produce significance improvements much better than simply using narrower jets.
For example, while filtering with a mass-drop criteria can produce up to a factor of 2.3 improvement in $S/\sqrt{B}$ in a $p_T\sim500\gev$  boosted-$W$ sample, simply
using a narrow jet size ($R=0.4$) can itself already do nearly as well, with a $S/\sqrt{B}$ improvement of order $2$.

\looseness -1 It is the goal of this paper to explore the optimization of boosted $W$-tagging by using
much more of the jets' substructure than what comes out of grooming.
For example, the decay products of a highly boosted $W$ are confined to
a small region around the $W$ momentum, while the radiation of a QCD-jet with the same $p_T$ is much more scattered. This effect is not
taken into account if we only consider the leading subjets after jet grooming.
To optimize the discriminating power, we attempt a comprehensive examination of the properties of a decaying color singlet particle
and its QCD-jet background.
 We define a set of variables which characterize jet radiation
patterns. These include what we call mass- and $p_T$ $R$-cores, which measure how the mass and $p_T$ of a jet change
when it is reclustered with different $R$'s.
 We also consider variables describing jet shapes including
planar flow \cite{Almeida:2008yp, Almeida:2010pa} and pull~\cite{Gallicchio:2010sw}.
In addition, we do use the jet grooming algorithms to extract some useful information, such as the masses and $p_T$'s of the groomed jets, the number of subjets, and the subjet $p_T$'s and masses.

To quantify and compare variables, we use the Significance Improvement Characteristic (SIC)~\cite{Black:2010dq}, defined as
the ratio of the signal efficiency to the square root of the background efficiency, $\varepsilon_S/\sqrt{\varepsilon_B}$.
As discussed in~\cite{Black:2010dq} SIC curves facilitate a visual comparison of various potential discriminants.
We find that filtering gives a SIC around 2.0. Starting from the samples after filtering, the additional shape and substructure variables each add at most an
additional 20\% when individually used.  However, we find that when the variables are combined in a multivariate analysis (MVA) using
 Boosted Decision Trees (BDT), the significance improvement can be as high as $3.4\sim 6.7$,
for jets with $p_T$ from  $200\sim1000$ GeV. In other words, for a signal efficiency of 40\%, we can reject around 4 times
as much of the background as filtering alone. This allows for substantial improvement in the reach for diboson resonances,
as well as the possibility of seeing the hadronic $W$-decay mode in the $W$+jets sample. Figure~\ref{fig:moneyplot} shows a summary of
our method's efficiency.

\begin{figure}
\begin{center}
\includegraphics[width=0.6\textwidth]{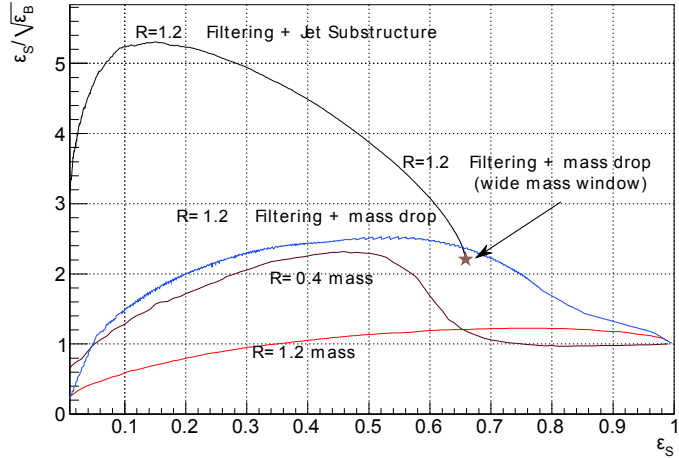}
\caption{
Significance Improvement Characteristics ($\varepsilon_S/\sqrt{\varepsilon_B}$) for leptonic-$W$+$W$-jet events (signal)
versus their leptonic-$W$+QCD-jet background, for $p_T^{\text{jet}}\in(500, 550)\gev$. The bottom two curves
show the effect of an optimized simple mass window for $R=1.2$ and $R=0.4$ Cambridge/Aachen jets. The falloff of the $R=0.4$ efficiencies
is due to events in which the $W$-subjets are well separated.
The next curve up shows
the efficiency of the filtering-with-mass-drop method of~\cite{filtering}, optimized over the filtering
parameters. The top curve is the result of
our multivariate analysis, including many variables on top of the filtered result. The starting point
for the multivariate analysis is a filtered sample with a window slightly wider than what is optimal for filtering,
as indicated by the star.
\label{fig:moneyplot}
}
\end{center}
\end{figure}

\looseness -1 This article is organized as follows. In Section \ref{sec:events}, the sample we use to optimize $W$-jet tagging is described.
Section~\ref{sec:grooming} reviews the jet-grooming algorithms and describes to what extent they are useful for $W$-jet tagging.
Section~\ref{sec:variables} describes the jet-substructure and jet-shape variables we use on top of grooming. In
Section~\ref{sec:MVA},  we describe how to combine the variables in a multivariate analysis to optimize $W$-jet tagging.
 In Section \ref{sec:wpol}, we discuss the difference in performance for different $W$ polarizations,
which has implications for applications to new physics searches. Then in Section~\ref{sec:mc}
we explore the robustness of our method using different Monte Carlo tools.
Section~\ref{sec:applications} contains applications to two interesting processes: $Z'$ boson discovery and $W$-jet identification in dijet events. We conclude in Section \ref{sec:conclusion}.

 \section{Event Samples}
 \label{sec:events}
Although we are more interested in boosted $W$'s from new physics, we use the standard model (SM) processes, $WW$ and $W$+jet to illustrate our method. As we will show, the properties of the $W$-jet and therefore the distinguishing power is fairly insensitive to the particular process. The results (cuts, parameters, {\it etc.}) of our analysis can be applied directly to processes with boosted $W$-jets. It is also straightforward to apply the same procedure for other boosted hadronically-decaying particles, such as a $Z$ or Higgs, although the optimal cuts will differ. For simplicity, we stick to $W$'s in this work.

For the optimization procedure we take as the signal process $WW$ production in the standard model, with one of the
$W$'s decaying hadronically and the other one leptonically. The background is $W$+jet production with the $W$ decaying leptonically.
At large $p_T$, each signal event contains a $W$-jet while each background event contains a high $p_T$ QCD-jet. We simulate the hard $WW$ process in $pp$ collisions at 14 TeV center of mass energy with both $W$'s decayed using Madgraph/Madevent~{\sc v4.4.32}~\cite{madgraph}, which includes the full $2\rightarrow 4$ matrix elements.
Thus, spin correlations and polarization effects are included.
 The Madgraph events  are then
fed into Pythia {\sc v8.142} \cite{pythia8}, where showering, hadronization and the underlying event are added. The $W$+jet events are generated with Pythia 8 alone.

In order to simulate the detector response, we divide the $(\eta, \phi)$ plane to $0.1\times 0.1$ calorimeter cells and restrict $\eta$ to be within $[-5, 5]$, roughly corresponding to the hadronic calorimeter resolution of the LHC detectors. We sum over the energy of particles entering each calorimeter cell and replace it with a massless particle of the same energy, pointing to the center of the cell. We have excluded neutrinos and charged leptons from leptonic $W$ decays
when summing over the energy.

The calorimeter cells are clustered first with a relatively large radius $R=1.2$ using Cambridge/Aachen algorithm as
implemented in FastJet {\sc v2.4.2}~\cite{Cacciari:2005hq}
 to identify the high $p_T$ jets. Only the leading jet in each event is kept in our analysis.
We then separate the sample by $p_T$ in 50 GeV bins from 200 GeV to
1050 GeV. We have also included a single bin for $p_T>1050\gev$,
 to account for higher $p_T$ jets appearing occasionally in the applications considered in Section \ref{sec:applications}.\footnote{Due to PDF suppression, this bin is dominated by jets with $p_T$ just above $1050\gev$ and gives similar results as the $(1000, 1050)\gev$ bin.
Special care is needed to optimize extremely high $p_T$ $W$-jets ($\gtrsim 1200\gev$) because all or most of the decay products can enter the same calorimeter cell, making it very difficult to extract the mass.
This regime is beyond the scope of this article.}

To characterize the effectiveness of different methods, we first calculate the signal and background efficiencies.
Let $n_S^i$ and $n_B^i$ denote respectively the initial number of signal and background jets within a particular $p_T$ bin. At the end of our analysis, after various cuts we are left with $n_s$ signal jets and $n_B$ background jets. Then the signal and background efficiencies
are defined as
\begin{equation}
  \varepsilon_S \equiv \frac{n_S}{n_S^i}, \qquad
  \varepsilon_B \equiv \frac{n_B}{n_B^i}.
\end{equation}
By comparing the efficiences, the conclusions are luminosity-independent.
Having a lower $\varepsilon_B$ at the same value of $\varepsilon_S$ is the indication of a superior discriminant. To
visualize the effectiveness of discriminants, we will look at the Significance Improvement Characteristic
\begin{equation}
  \text{SIC} \equiv \frac{\varepsilon_S}{\sqrt{\varepsilon_B}},
\end{equation}
which is a rough proxy for the improvement in significance. One advantage of using this characteristic, as explained
in~\cite{Black:2010dq} is that it gives a well-defined quantitative measure of how good a variable does.
For a given analysis, one will often choose cuts on a variable or multivariable discriminant away from the optimal SIC. In that case,
for any $\varepsilon_S$, the SIC curves let you easily read off the corresponding $\varepsilon_B$.

We choose to analyze for each $p_T$ bin separately
because we eventually want to use our method to identify boosted $W$'s from new physics processes, which may have a very different $p_T$ distribution from the SM $WW$. As we will show, the optimal cuts are $p_T$-dependent,
and we can obtain the best distinguishing power by treating the $p_T$ bins separately.

\section{Grooming: filtering, pruning and trimming}
\label{sec:grooming}
The first step in our optimization procedure is to identify subjets and reduce the number of background events using existing jet grooming algorithms. These algorithms include filtering (we always use the mass drop method together with filtering), pruning and trimming. These algorithms are qualitatively similar but differ in details, which we briefly review in Appendix \ref{app:pars}. More details can be found in Refs.~\cite{filtering, trimming, pruning}.

\begin{figure}
\begin{center}
\begin{tabular}{cc}
\includegraphics[width=0.35\textwidth]{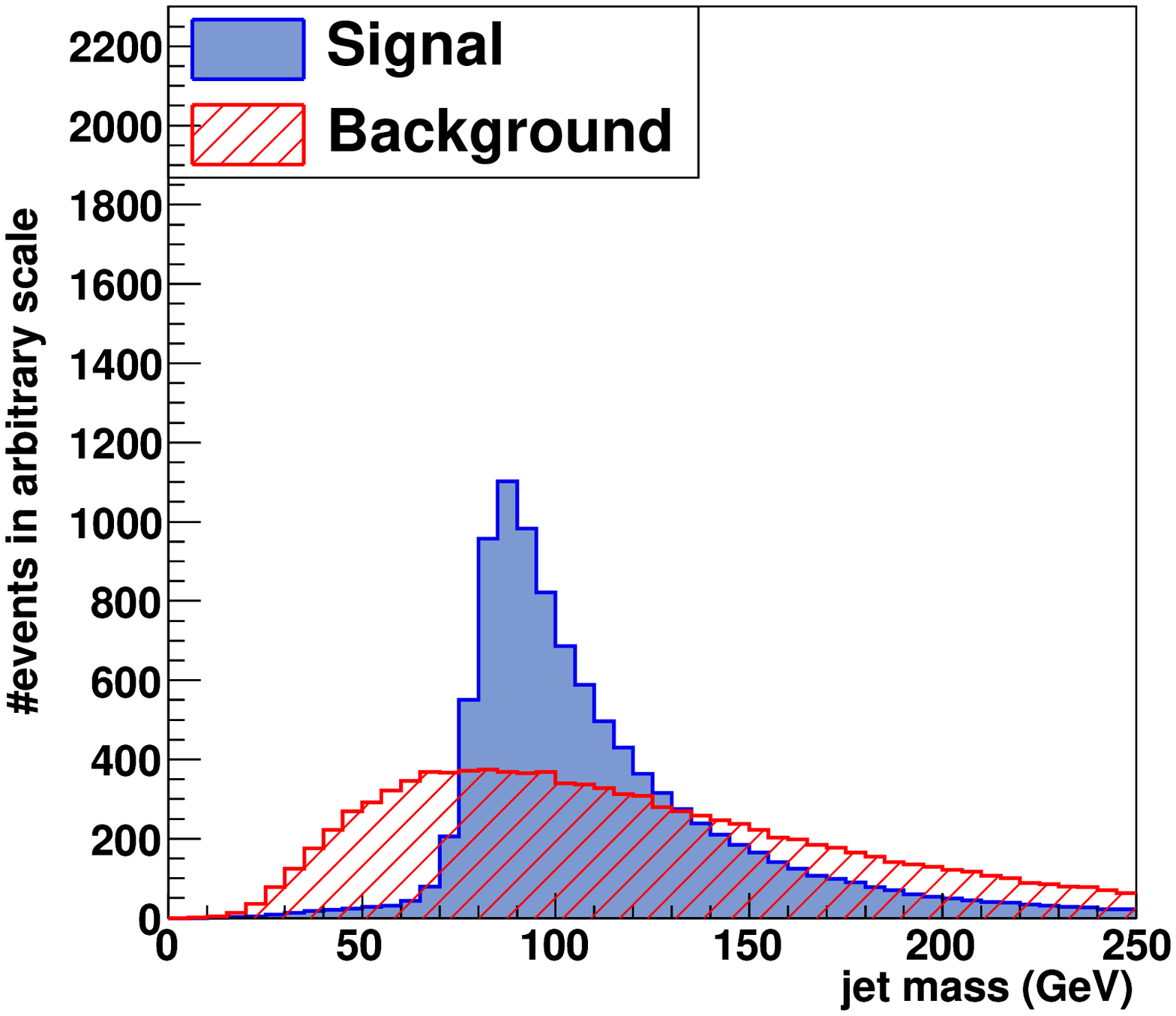}
&\includegraphics[width=0.35\textwidth]{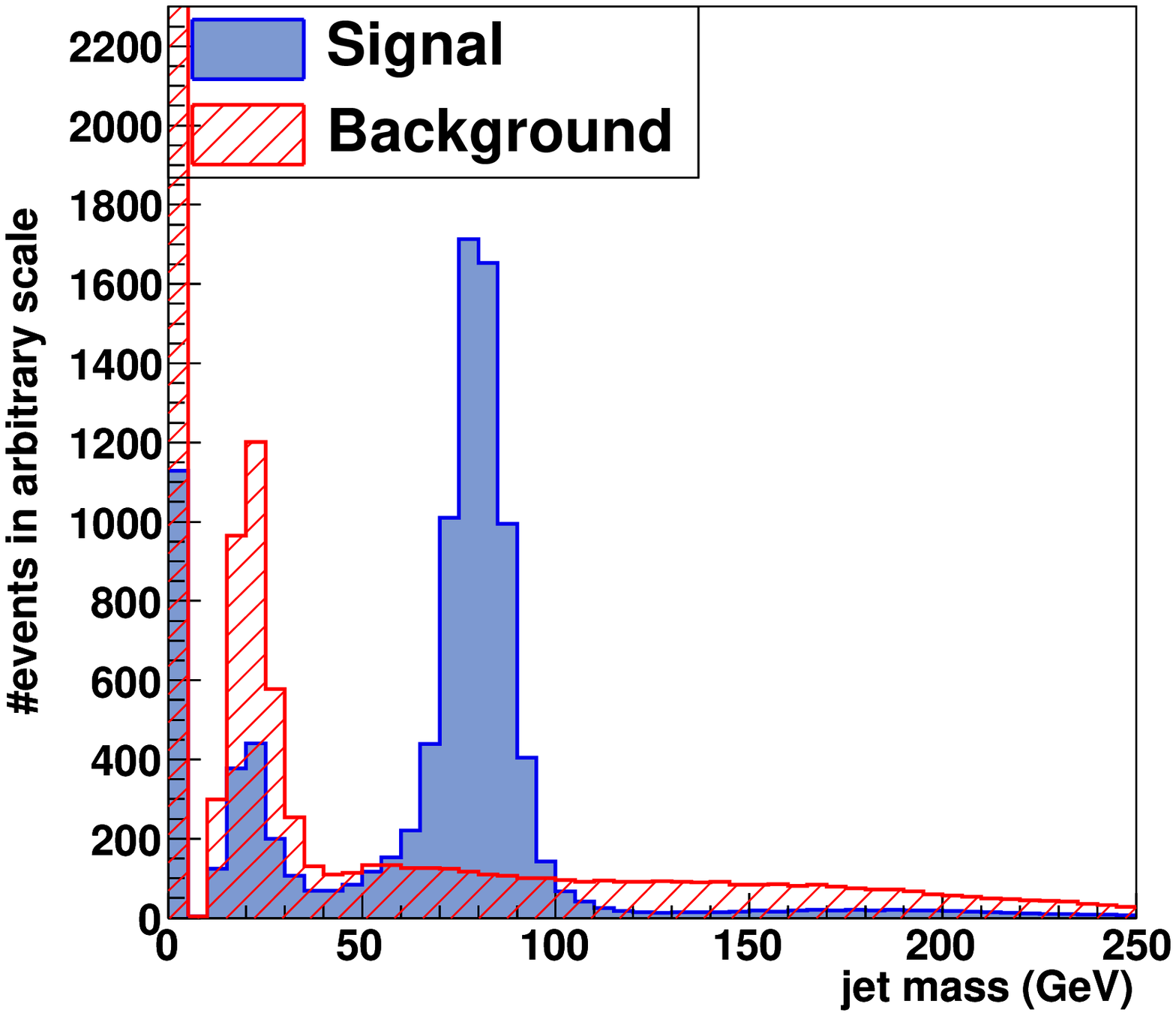}
\\(a) &(b)
\end{tabular}
\caption{Jet masses before and after filtering/mass-drop for $p_T^\jet\in(500, 550)\gev$. The numbers of events are normalized to be the same for the signal and the background. (a) Before filtering;
(b) after filtering with $\mu =0.71 $ and $y_\cut=0.09$. When a mass-drop is not found, we add an entry in the zero mass bin such that the total number of jets is unchanged.\label{fig:filtered_mass}}
\end{center}
\end{figure}

Besides the jet size  $R$ one uses to cluster the original jets, each of the three jet grooming algorithms involves two tunable parameters. We will scan the parameters to maximize $n_S/\sqrt{n_B}$, where the numbers of signal and background events after jet grooming are defined as follows. After jet grooming, the jet mass is always shifted lower, with signal jets concentrated around the $W$ mass and background jets concentrated around much lower values. See Figure~\ref{fig:filtered_mass} for an example. Therefore, we can apply a mass window cut to efficiently reduce the number of background events. Then $n_S$ and $n_B$ are defined as the number of signal and background events in the mass window.

Obviously, the significance also depends on the mass window we choose, so we scan over the mass window too. The filtering result presented in Figure~\ref{fig:moneyplot} is from such scans. For example, the optimal mass window for $p_T^\jet\in(500, 550)\gev$ is  $m_\filt \in (70, 90)\gev$ with filtering parameters $\mu = 0.71$ and $y_\cut = 0.09$, where $m_\filt$ is the jet mass after filtering. However, as we will further improve the distinguishing power by conducting a multivariate analysis using jet-substructure variables in the following sections, it is desirable to keep more events at this stage. Therefore, we choose a relatively large mass window, $m_\filt \in (60, 100)\gev$ and scan the grooming parameters to maximize $n_S/\sqrt{n_B}$ in this window for all $p_T$'s. It turns out by doing so we obtain an equal or larger significance improvement
after the multivariate analysis than what we would have gotten with the window which is optimal for filtering alone.

We have scanned the parameters for all three algorithms and all $p_T$ bins. The optimal parameters are given in Table \ref{tab:jet_pars} in Appendix \ref{app:pars}. In Figure~\ref{fig:filtering_contour}, we show the contour plot for the significance improvement characteristics as a function of the filtering parameters $\mu$ and $y_\cut$,
for $p_T^\jet\in(500, 550)\gev$ and $m_{\text{filt}}  \in (60, 100)\gev$. Note that the contours do not close on the right where the significance is insensitive to the $\mu$ parameter. This is because the $y_\cut$, which constrains how ``imbalanced'' the two subjets can be, effectively yields a lower bound on the mass drop ratio, making larger $\mu$ parameters ineffective. The filtering parameters that maximize the significance for all $p_T$ bins are shown in Figure~\ref{fig:filter_parameters} (a), and the corresponding signal and background efficiencies, as well as the SICs are shown in Figure~\ref{fig:filter_parameters} (b). We see that we typically gain a factor of $\sim 2$
in significance from filtering using the best parameters. This is also true for trimming and pruning. See Appendix \ref{app:pars} for more details. It turns out that filtering yields slightly better significance. Therefore, in the following, we will apply the mass window cut $m_{\text{filt}} \in (60, 100)\gev$ on the filtered jet mass, and examine further the events passing the cut.

\begin{figure}
\begin{center}
 \includegraphics[width=0.6\textwidth]{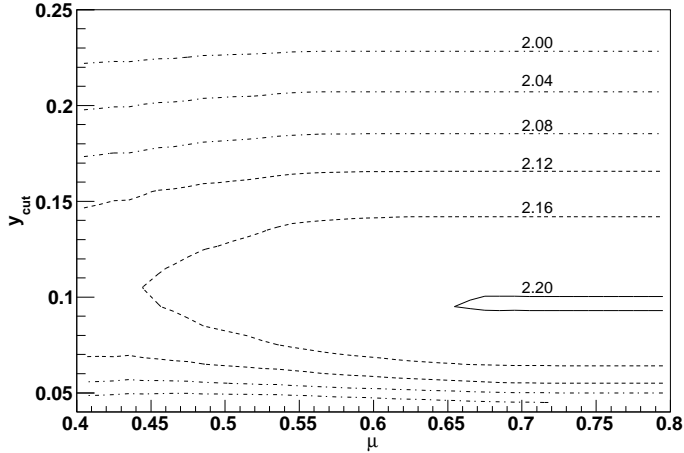}
\caption{\label{fig:filtering_contour} The significance improvement characteristic (SIC$\equiv\varepsilon_S/\sqrt{\varepsilon_B}$)
as a function of the filtering parameters, $\mu$ and $y_\cut$, for $p_T^\jet\in(500, 550)\gev$.}
\end{center}
\end{figure}

\begin{figure}
\begin{center}
\begin{tabular}{ccc}
\includegraphics[width=0.43\textwidth]{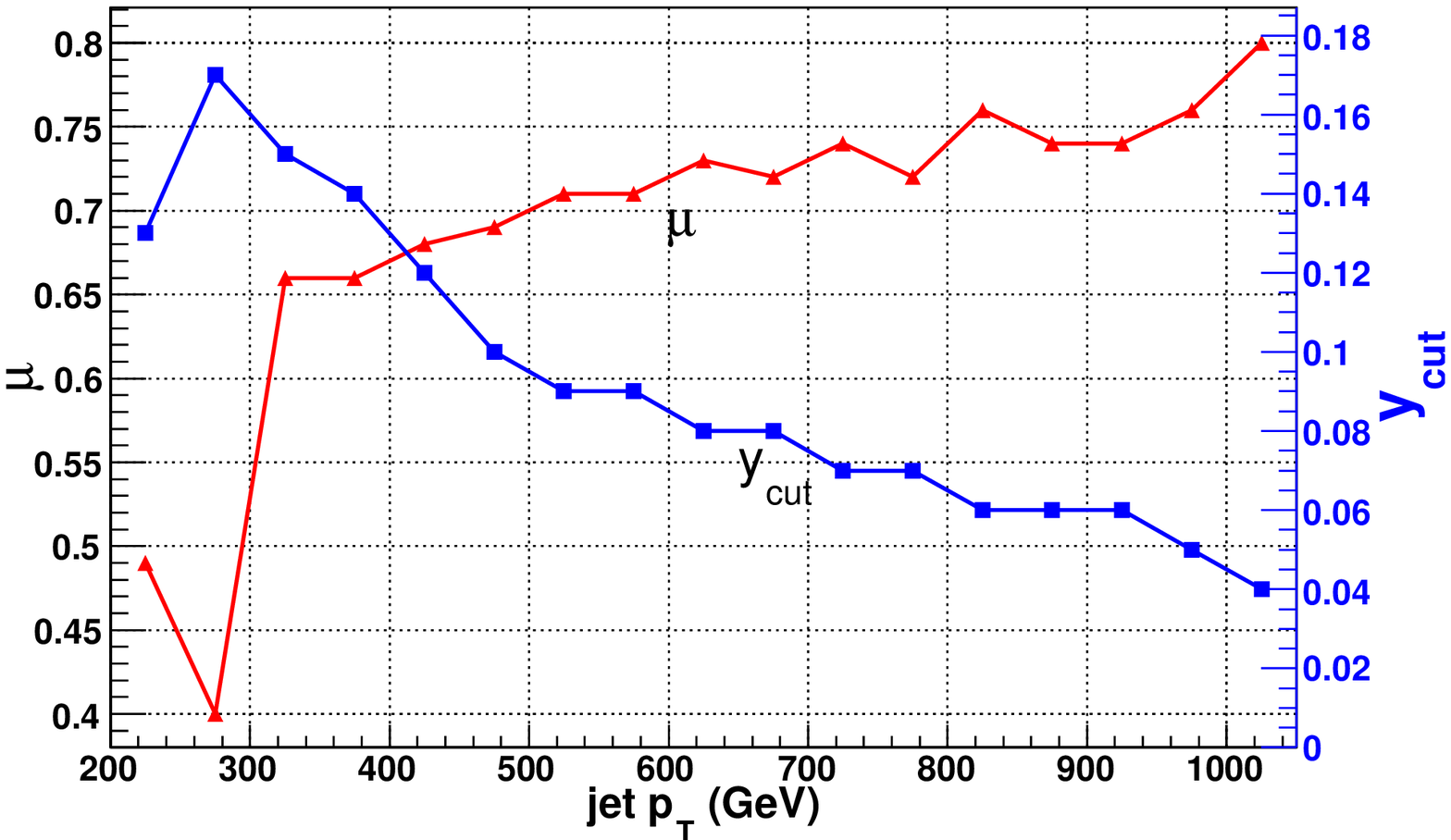}&
&\advance\rightskip1cm\includegraphics[width=0.43\textwidth]{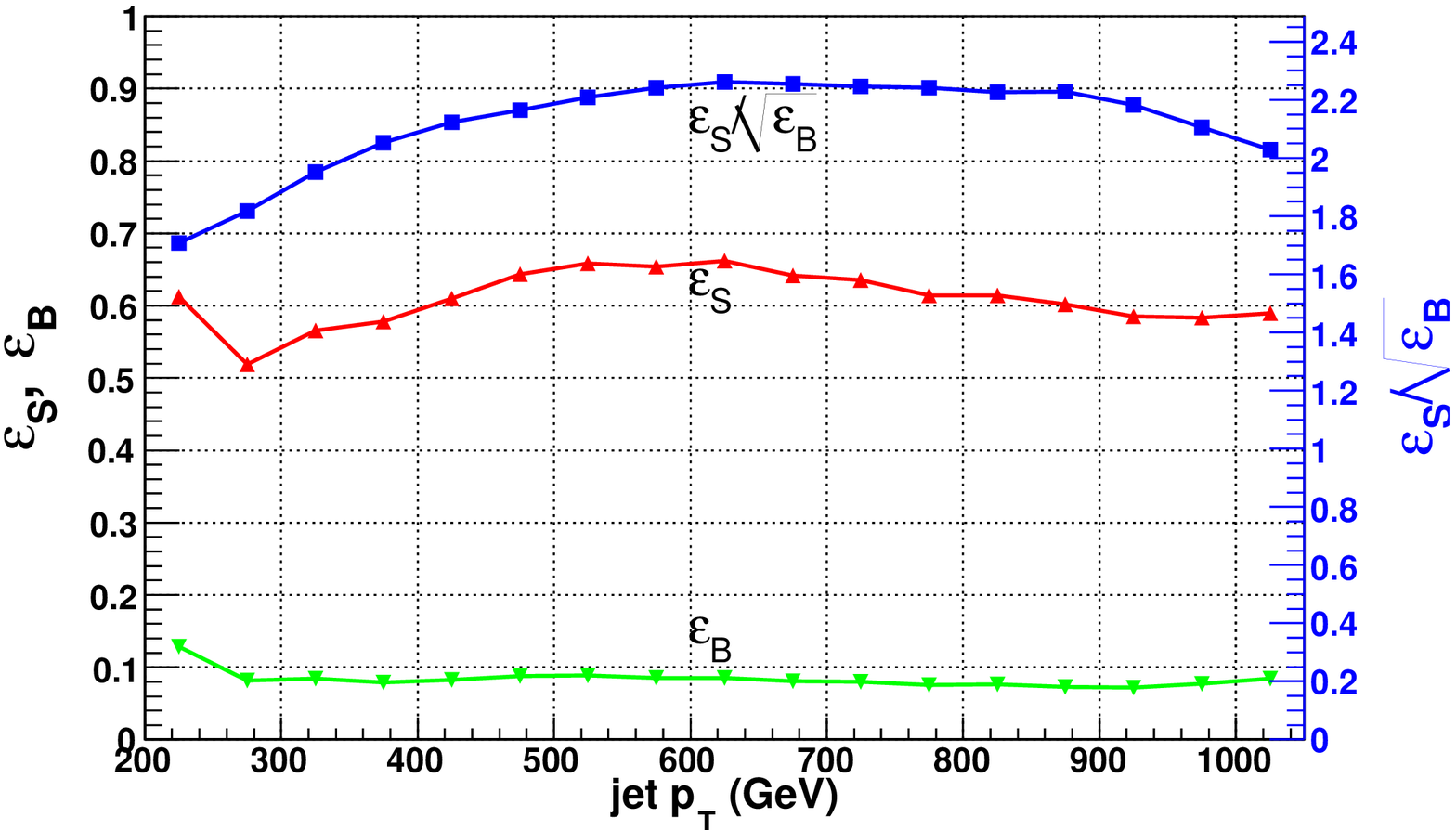}
\\(a) Optimized filtering parameters &&
(b) Optimized filtering efficiences and SICs.
\end{tabular}
\caption{Tuning of filtering parameters for $W$-jets versus QCD-jets in the standard model. \label{fig:filter_parameters}}
\end{center}
\end{figure}

\section{Jet substructure and Jet shape variables}
\label{sec:variables}
As discussed in the previous section, the first step in our analysis is to require that the candidate $W$-jet,
after filtering, has a mass $m_{\text{filt}} \in (60,100)$ GeV. Even after this cut, $W$-jets and QCD-jets
still differ in many aspects. In this section, we define a set of observables
which help further boost the significance. Some of these variables have been proposed in recent works on jet substructure, as will be briefly reviewed. There are also other variables which we find very useful yet have not been mentioned or emphasized in existing references. We first classify relevant variables according to the physics they represent, then present results based on a set of principle variables which gives major significance gain. As mentioned before, the discrimination power depends on the jet $p_T$, so we always work on data samples in separate 50 GeV $p_T$ bins.

Keep in mind, the jets studied in this section are the original unfiltered $R=1.2$ ``fat'' jets,
but we have thrown out jets not passing the filtered mass window. The efficiency for the filtering mass cut
is indicated by the point marked $\star$ in Figure~\ref{fig:moneyplot}.

\subsection{Jet and subjet mass}
       \begin{figure}
      \begin{tabular}{cc}
      \includegraphics[width=0.35\textwidth]{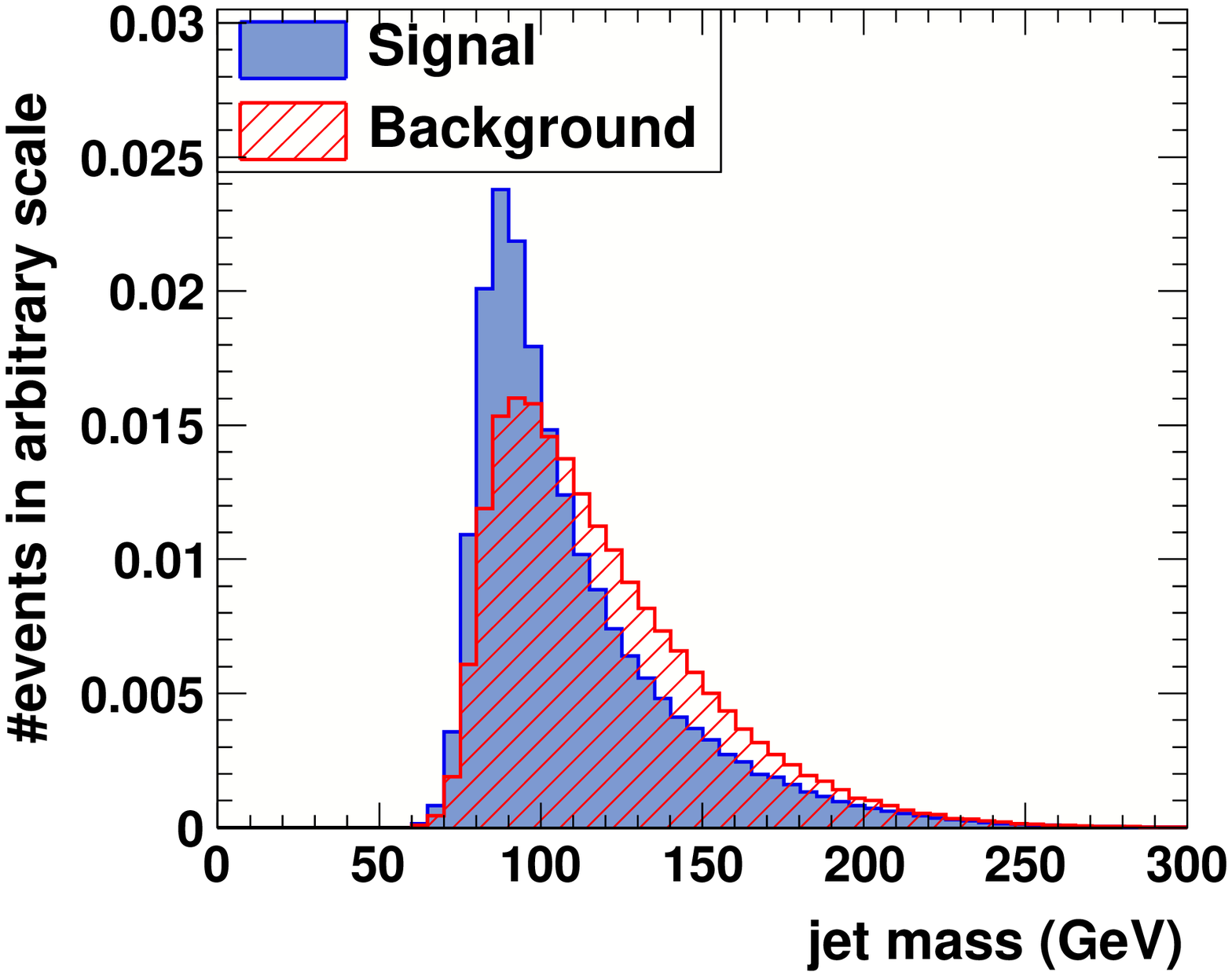} & \includegraphics[width=0.35\textwidth]{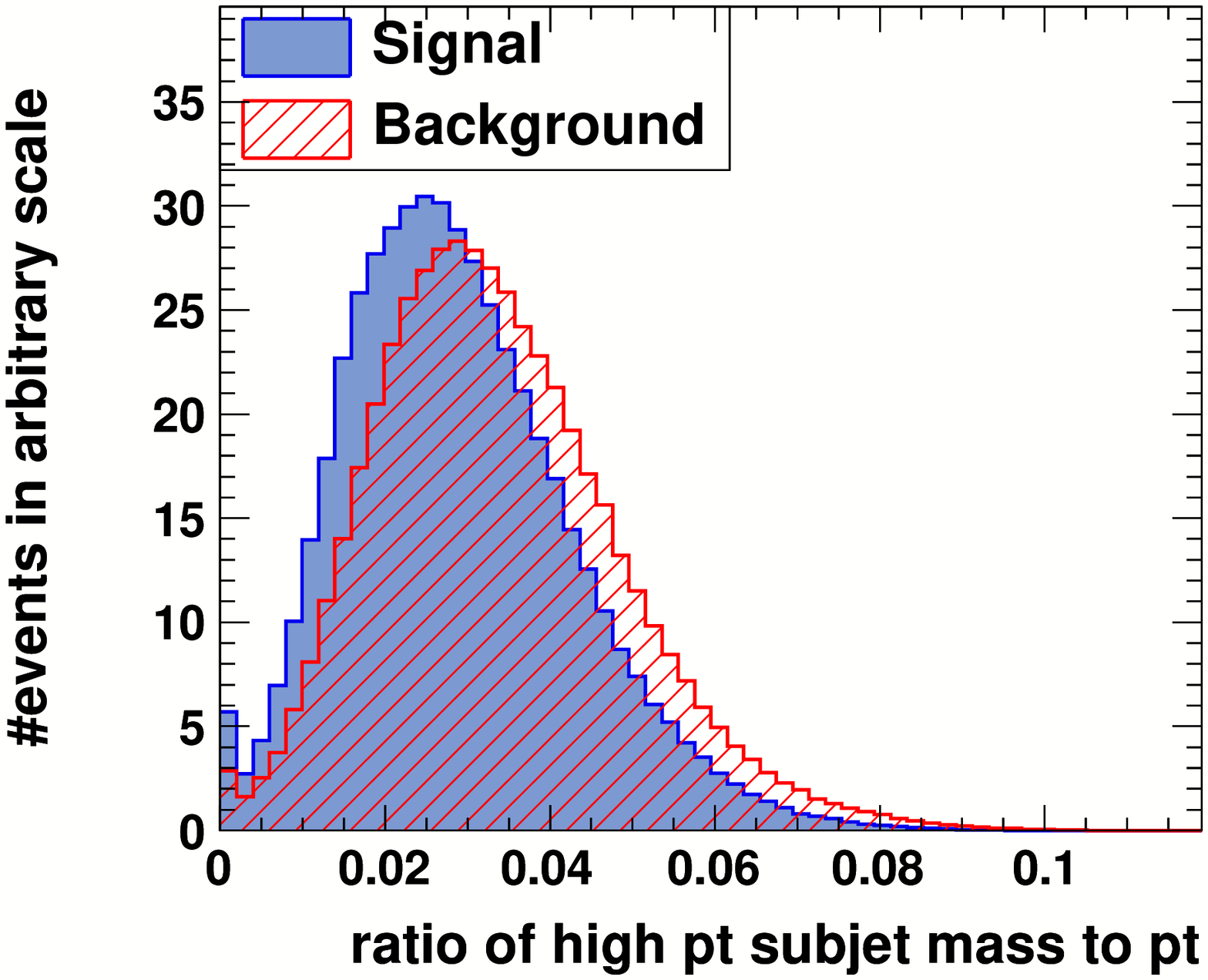}
      \end{tabular}
      \caption{Distributions for the fat-jet mass and hardest subjet mass for signal ($W$-jets) and background (QCD-jets) with
 $p_T^\jet\in(500,550)$ GeV. The edge at 60 GeV in the jet mass plot follows from a preselection cut on the filtered mass,
$m_{\text{filt}} \in (60,100)\gev$.}
      \label{fig:masses}
      \end{figure}
For samples with the same $p_T$, a QCD-jet originates from a highly off-shell quark or gluon, with no definite mass scale, while a hard jet from resonance decay such as a $W$-jet is associated with a definite mass scale $m_W$. As a result, a QCD-jet's mass ($m_{\jet}$) is expected to be roughly proportional to its $p_T$, while the mass of a boosted $W$ is mostly set by $m_W$ with milder dependence on its $p_T$. In the same way, if a jet can be decomposed into two hard subjets, for example via filtering, the masses of these subjets ($m_\sub$) are roughly set by $p_T^\jet$ in the case of QCD while by $m_W$ in the case of $W$-jets.
In our samples, both the QCD-jets and the $W$-jets have already passed the filtered mass window cut.
Nevertheless, there is still distinguishing power in both
 $m_\jet$ and $m_\sub$. For illustration, see Figure~\ref{fig:masses}. It is natural to also ask about the relationship between the fat-jet mass and the mass
after grooming. We call observables describing this relationship {\it grooming sensitivities}, and they will be described below.

\subsection{Color connections and $R$-cores}
Another difference between a QCD-jet and a $W$-jet is that the $W$-jet originates from a color singlet, while the QCD-jet does not.
By looking at the leading order matrix element of related processes,
one can see in QCD (for example $q\bar{q}\rightarrow g\rightarrow q\bar{q}$)
final state partons are color-connected to initial state partons. On the other hand, the two partons from a $W$
 decay are color-connected to each other.  This picture is exact at large $N_C$, and gets  $O(1/N_C^2)\sim 10\%$ corrections in practice.
The difference in color-flow was exploited in~\cite{Gallicchio:2010sw}, which observed that the subsequent radiation pattern had a characteristic
first moment vector which was called {\bf pull}. Projections of the various pull vectors,
such as pull-angles and pull-size~\cite{Black:2010dq} were shown to have discrimination power. Recently, pull has been
measured by D0 in $Z+$jet events with $Z\to \nu\nu$~\cite{D0pull}.

       \begin{figure}
   \begin{tabular}{ccc}
   \includegraphics[width=0.3\textwidth]{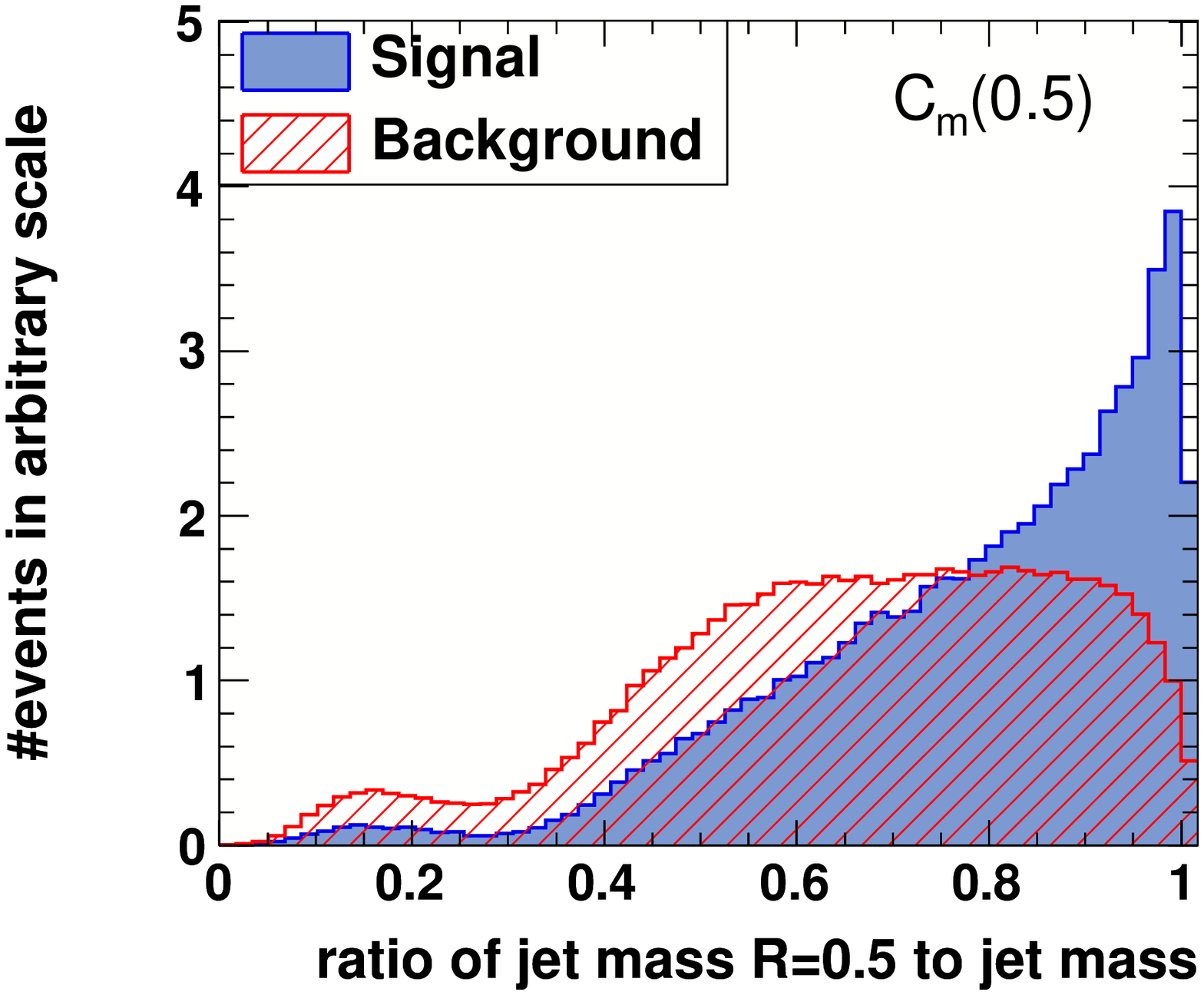} &
    \includegraphics[width=0.3\textwidth]{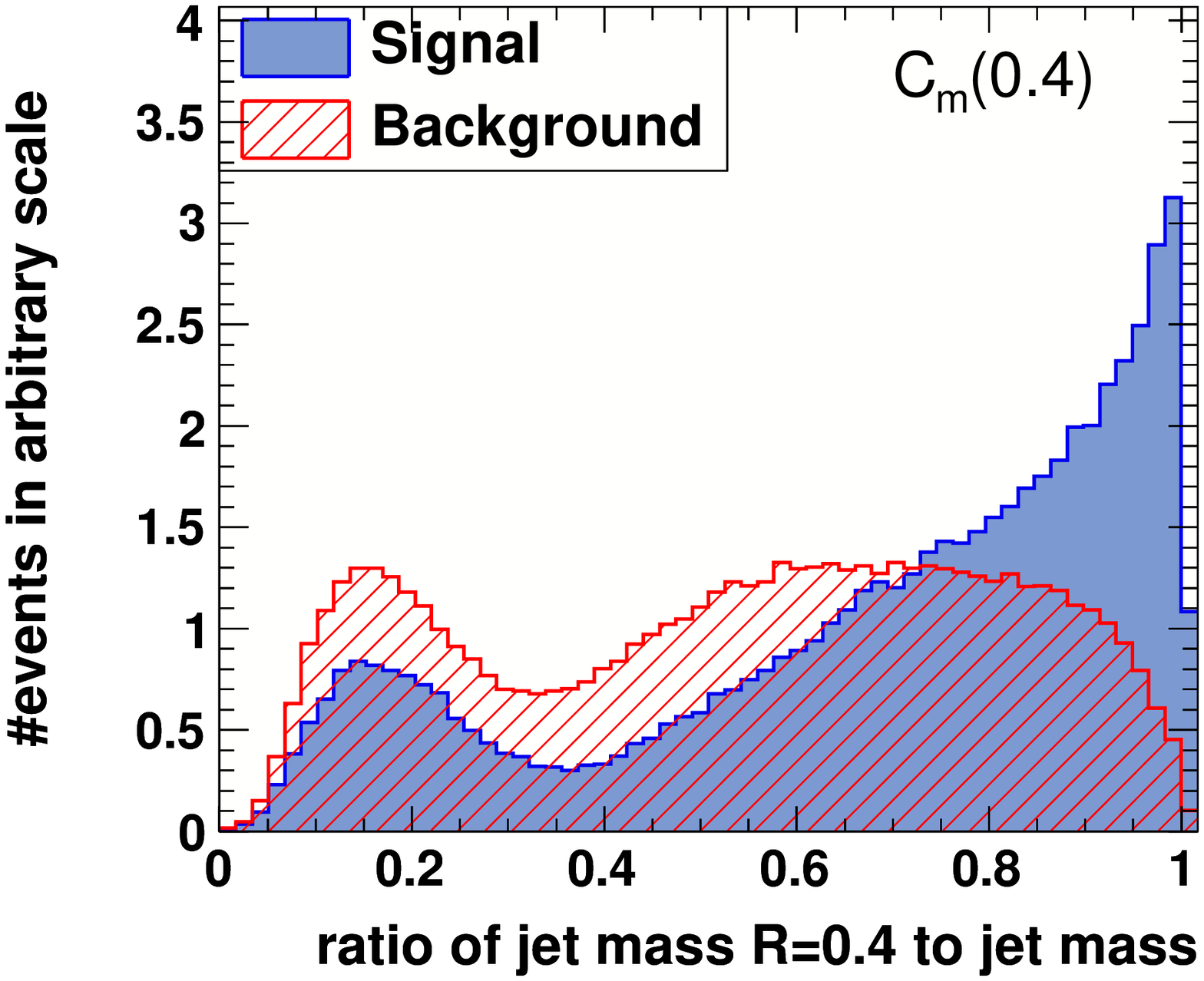}&
     \includegraphics[width=0.3\textwidth]{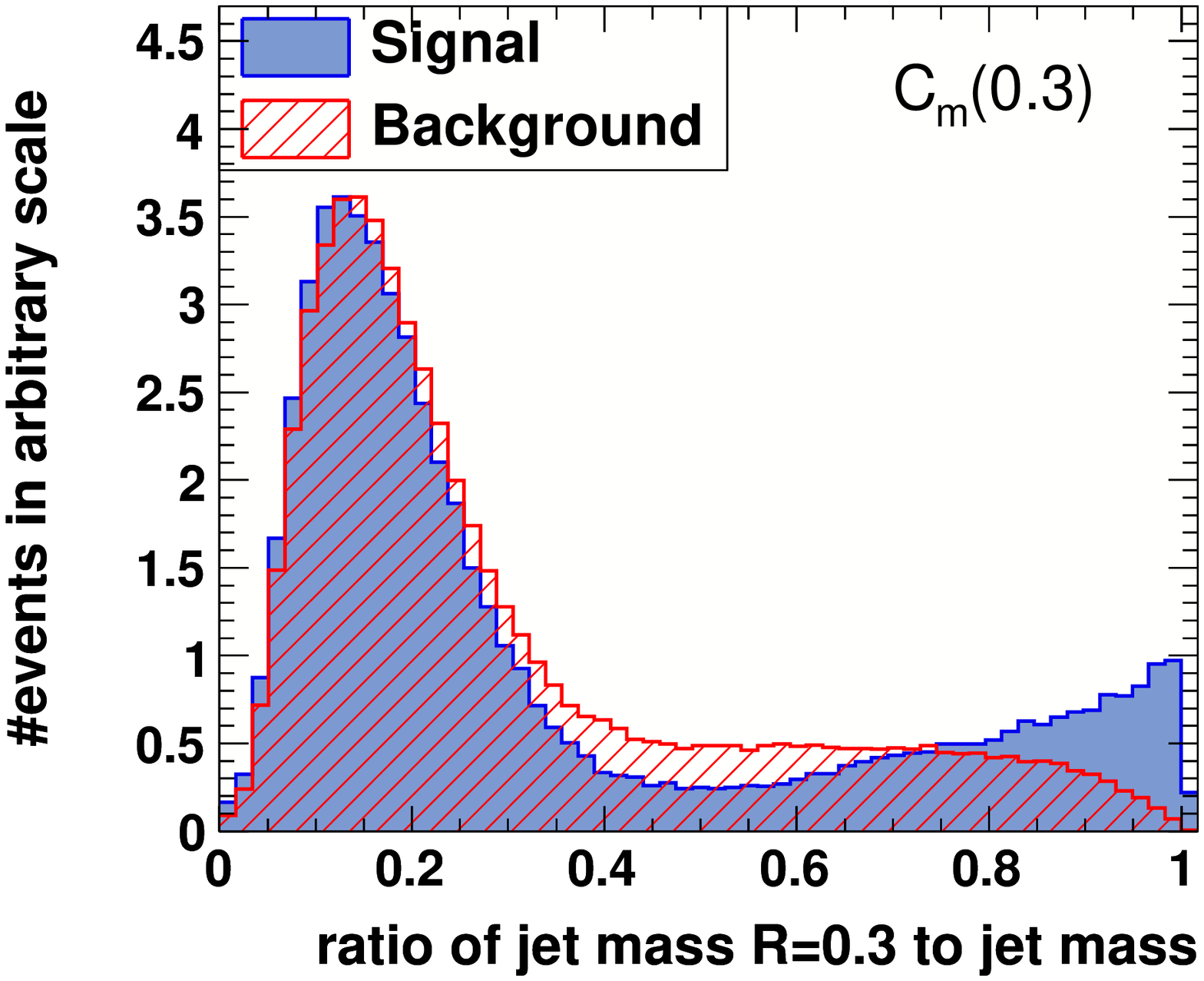}\\
\includegraphics[width=0.3\textwidth]{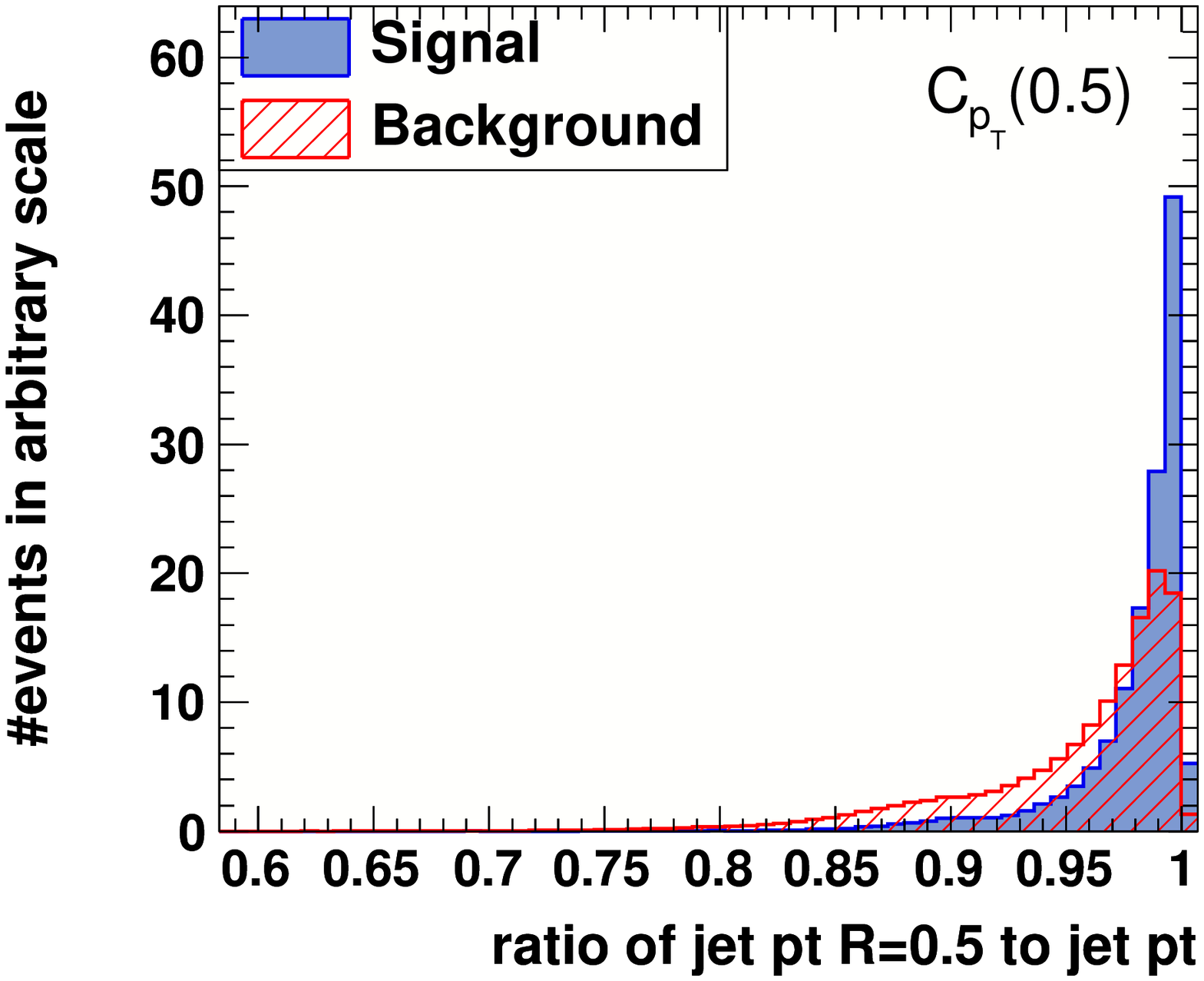}&
\includegraphics[width=0.3\textwidth]{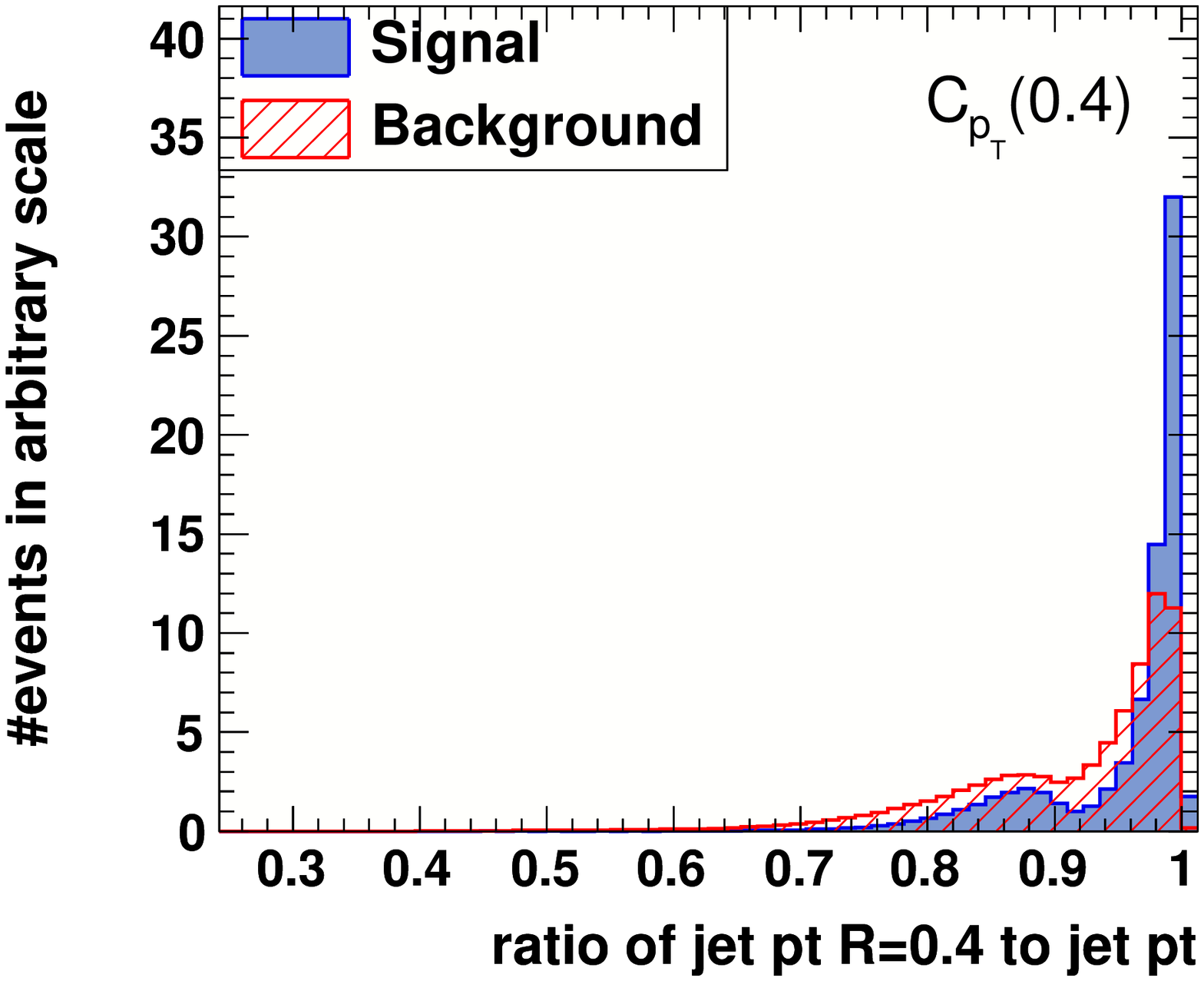}&
\includegraphics[width=0.3\textwidth]{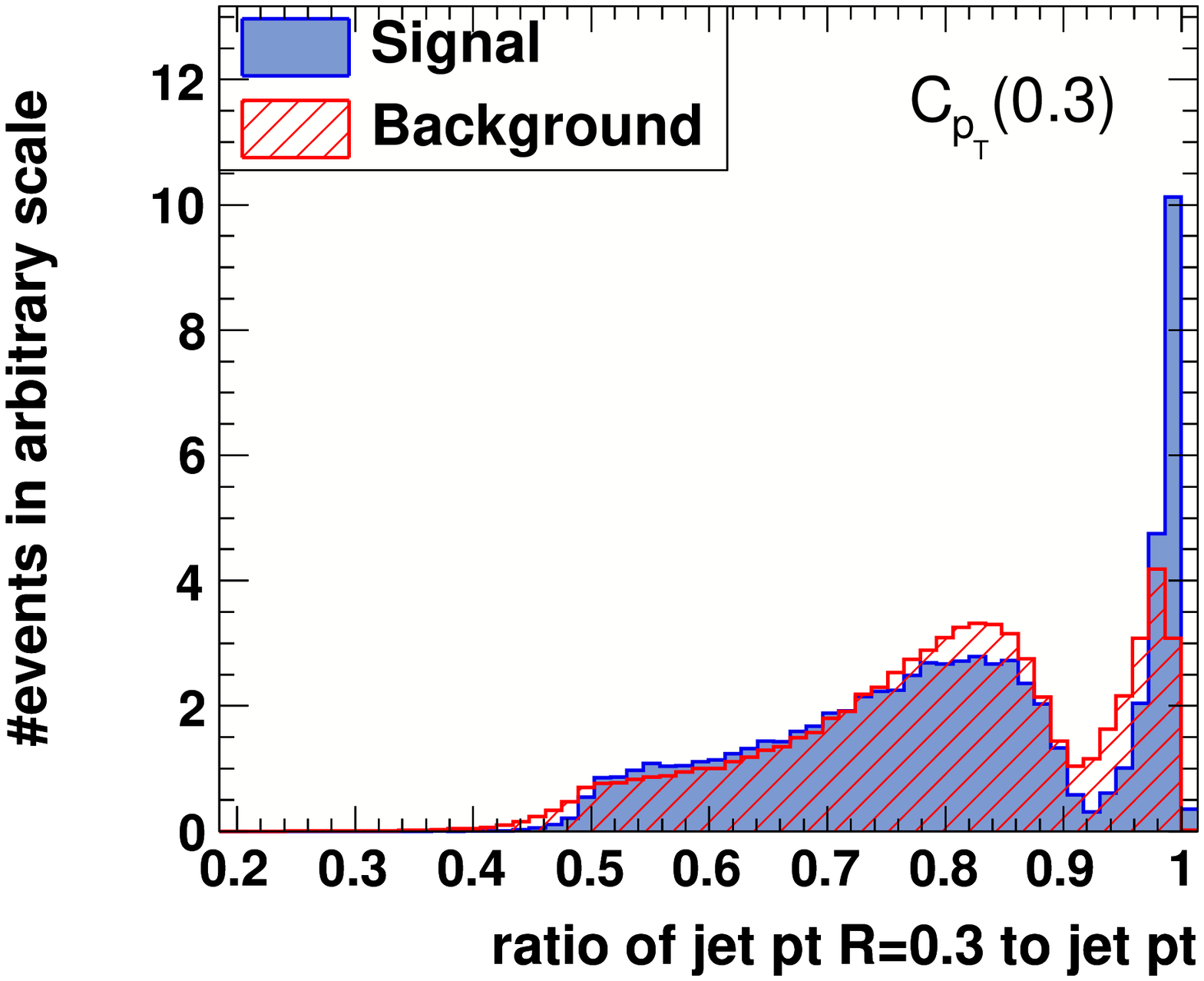}
   \end{tabular}
   \caption{Representative $R$-core distributions for $R=1.2$ fat jets with $p_T^\jet\in(500,550)\gev$ and $m_\filt\in(60, 100)\gev$. A dissection of the physics producing these shapes is discussed in the text.
}
   \label{fig:ptRcores}
   \end{figure}

While pull is a useful, general purpose measure of color flow, there may be better ways to capitalize on the color singlet nature
of the $W$ boson in the boosted case.
Here we propose a new set of variables {\bf {\textit R}-cores} inspired by color connection considerations, but which
are sensitive to aspects of the energy balance in $W$-jets and QCD-jets as well.
For a jet of given $p_T$ to have  mass $m_{\text{jet}}$, it must have at least two subjets. The characteristic separation between the
subjets is then $\Delta R_{\text{sub}} \sim 2 m_{\text{jet}}/p_T$. In the case that the jet originates from a color singlet, one expects
the additional radiation to be within this radius, while for a QCD-jet, which is color-connected to the beam, one expects the
additional radiation to be outside this radius. To characterize this radiation pattern in an infrared safe way, we define
$R$-cores as follows.
\begin{itemize}
\item Recluster the fat-jet with a smaller $R < R_{\text{fat}}$.
\item Take the highest $p_T$ subjet after reclustering, call its mass $m(R)$ and its transverse momentum $p_T(R)$.
\item The {\bf mass {\textit R}-cores} are defined as  $c_m(R) \equiv m(R)/m(R_{\text{fat}})$.
\item The  ${\mathbf p}_{\mathbf T}$ {\bf {\textit R}-cores} are defined as $c_{p_T}(R) \equiv  p_T(R)/p_T(R_{\text{fat}})$.
\end{itemize}
For the application to boosted $W$'s, we have $R_\text{fat}=1.2$ and we consider $R$-cores with $R=0.2,0.3, \ldots, 1.1$.
The mass and $p_T$ $R$-cores tend to carry almost identical information, and in the
end we use only $p_T$ $R$-cores for the final discriminant, since they work a little better.

Some distributions for mass and $p_T$ $R$-cores are shown in Figure~\ref{fig:ptRcores}. For large $R\gtrsim0.5$, we see that
the $W$-jets have their $p_T$ $R$-cores peaked much more sharply around 1 than the QCD-jet background. The longer
tail of the QCD-jets is characteristic of radiation being more diffuse away from the center of the jet, as expected from
the color-flow picture. As
$R$ is taken smaller, a larger fraction of events in the $W$-jet case have the two hard subjets separated by $\Delta R_\sub>R$. In this case, the $p_T$ of the hardest subjet measures the energy fraction of the splitting, similar to the
$z$-variable used in~\cite{Thaler:2008ju}.
Note that for this $p_T^\jet \sim 500$ GeV sample, the characteristic subjet separation is $ \Delta R_{\text{sub}} \sim 2 m_W/p_T \sim 0.32$.
The two-peak shape emerging around $R=0.3\sim \Delta R_\text{sub}$ is the result of splitting events
in which two hardest energy deposits are within $R$ or not. When they are within $R$, the $p_T$ of the subjet is close
to the $p_T$ of the fat jet.
The $R$-cores are useful in that they interpolate between a measure
of the color-flow induced radiation pattern, at larger $R$, and the hard splitting scales, at smaller $R$.

   \begin{figure}
    \includegraphics[width=0.5\textwidth]{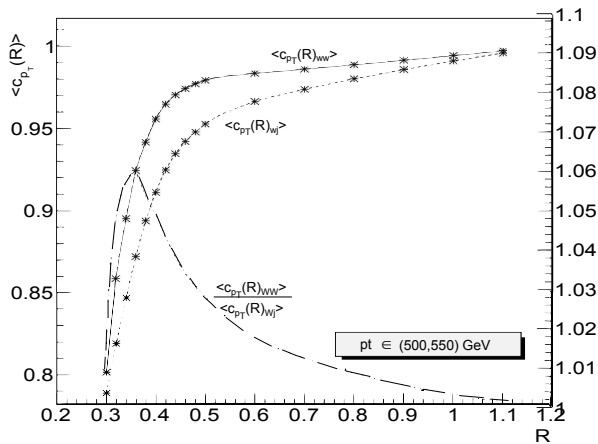}
   \caption{The average values of $p_T$ $R$-cores for $W$-jets and QCD-jets, gauged by the left axis, and the ratio of the two curves, gauged by the right axis.}
   \label{fig:Rmeans}
   \end{figure}
Another way to look at the $R$-cores is through their average values. Figure~\ref{fig:Rmeans} shows the average values of the $p_T$ $R$-cores
as a function of $R$ for the $W$-jet and the QCD-jet samples. For the same $R$, the $W$-jets tend to have a larger fraction of their $p_T$ in a single
subjet. Also shown is the ratio of these mean values, which peaks around $R=0.3\sim \Delta R_{\text{sub}}$. This transition point is another
way to estimate which $R$-core we expect to be most useful.

   \begin{figure}
    \includegraphics[width=0.4\textwidth]{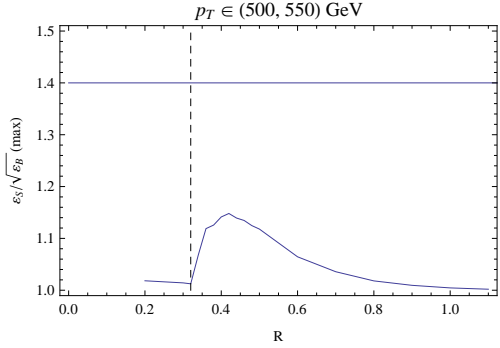}
   \caption{Maximal SIC as a function of $R$ when $c_{p_T}(R)$ is individually used, $p_T^\jet\in(500,550)\gev$. The solid horizontal line indicates the SIC when a set of 10 $p_T$ $R$-cores ($R$=0.2 to 1.1) are combined using BDTs; the dashed vertical line indicates the estimation of $\Delta R_{\sub}$ as $\sim2m_W/p_T$.}
   \label{fig:singleR}
   \end{figure}

    \begin{figure}
       \includegraphics[width=0.7\textwidth]{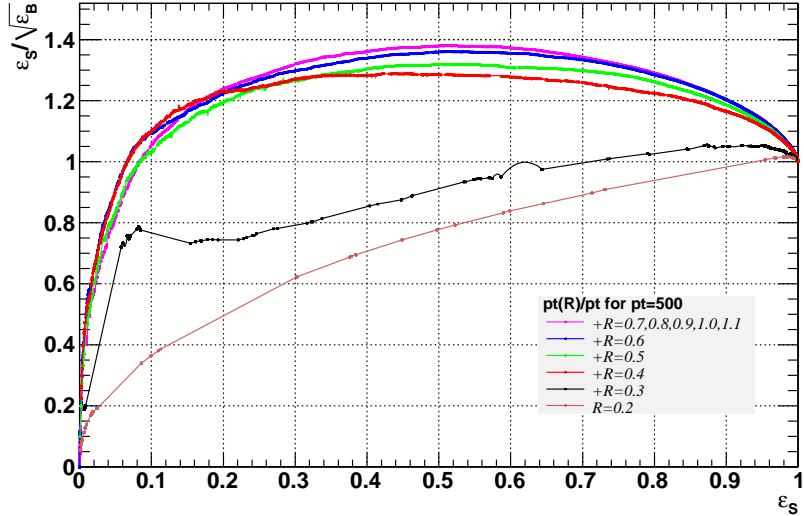}
       \caption{Gradual significance gain when adding $c_{p_T}(R)$ one by one, in the order of $R_i=0.2, 0.3, . . . , 1.1$, for $p_T^\jet\in(500,550)\gev$.}
       \label{Rgainall}
      \end{figure}

To see the usefulness of $R$-cores as discriminants, we show the maximal significance improvement characteristic as a
function of $R$ for the $p_T$ $R$-cores in Figure~\ref{fig:singleR}. We see that the best single $p_T$ $R$-core has
$R\sim 0.4$. This is close to the characteristic subjet separation, $\Delta R_{\text{sub}} \sim 0.32$. However, when multiple
$R$-cores are combined (with Boosted Decision Trees, see the next section), the significance improvement can be much larger,
as indicated by the horizontal line in the figure. Rather than a 15\% improvement in significance, which is the best
we can get from one variable, we find a 40\% improvement when the variables are combined.
The marginal improvement from adding 10 cores from $R=0.2$ to $R=1.1$ is shown in Figure~\ref{Rgainall}.

It would be nice if a single variable could substitute for the combination of $R$-cores. Clearly, any of the individual $R$-cores
will not do, as can been seen from Figure~\ref{fig:singleR}. The $R$-cores are combining to measure the full radiation profile
of the jet. Instead of looking at $R$-cores, one could try to look at individual jet shapes.
A reasonable candidate is {\bf girth}
 which is defined in \cite{Black:2010dq, Gallicchio:2010sw} as $g=\sum\frac{p_T^i|r_i|}{p_T^{\jet}}$. Girth can be understood as $p_T$ weighted average
 distance from the jet center, and is closely related to jet broadening. However, we find the gain from using girth is not
 comparable to that from the set of 10 $R$-cores.

 Finally, we show in Figure~\ref{fig:pTR-pT} the maximal significance improvement characteristic from the combined 10 $p_T$ $R$-cores
in different $p_T$ windows. The efficiency improves dramatically with higher $p_T$. This is expected because
the color-connected partons from $W$ decay are more collimated at high $p_T$, while the background color connections to the
beam remain roughly the same.
       \begin{figure}
       \includegraphics[width=0.4\textwidth]{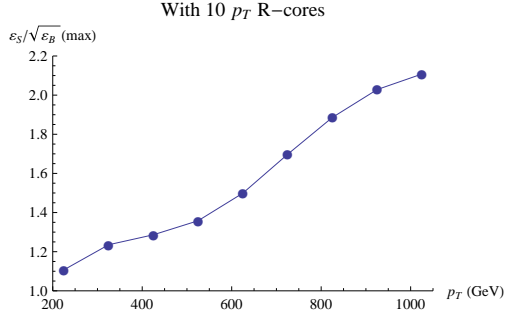}
       \caption{Maximal SICs for the whole set of $c_{p_T}(R)$ using BDTs as a function of $p_T$.}
       \label{fig:pTR-pT}
      \end{figure}

\subsection{Sensitivity to grooming procedures}
 As reviewed in Section~\ref{sec:grooming}, there are three recently developed general-purpose jet grooming procedures: filtering, trimming, pruning. Differing in details, these are all found to be efficient in removing soft QCD radiation from a fat initial jet. Because of the differences in details among various grooming procedures, the combination of them may give additional gain in significance compared to using one of them alone. This possibility was pointed out in~\cite{Soper:2010xk}, where a likelihood analysis was performed
on the original jet mass distribution for jets passing mass window cuts for two different grooming methods. It was also shown
in~\cite{Black:2010dq} that combining the mass from mildly and aggressively trimmed jets could improve upon the significance from a single
set of trimming parameters.

Here we use another way to combine information from different grooming procedures based on the sensitivity
to grooming.
 It is expected that for the same fat jet mass and $p_T$, radiation in QCD-jets has larger tendency to be
groomed away than radiation around a $W$-jet. The ratio of the jet mass or $p_T$ to its original value is
 therefore expected to be a good measure of this difference. We define dimensionless variables {\bf grooming sensitivities}
 \begin{equation}
    {\text{sens}}^m_\text{filt} \equiv \frac{m_\text{filt}}{m},\qquad
    {\text{sens}}^m_\text{trim} \equiv \frac{m_\text{trim}}{m},\qquad
    {\text{sens}}^m_\text{prun} \equiv \frac{m_\text{prun}}{m},
 \end{equation}
and similarly for $p_T$ grooming sensitivities.
To be clear, the sample that we test these on have already passed the filtered mass window cut $m_\text{filt} \in (60,100)$ GeV.
To calculate these sensitivities, we use the original jets, before filtering, but
which pass the filtered mass cuts. As expected, these ratios peak towards smaller value for QCD-jets than for $W$-jets (Figure~\ref{fig:groomedmass}).
      \begin{figure}
      \begin{tabular}{ccc}
      \includegraphics[width=0.3\textwidth]{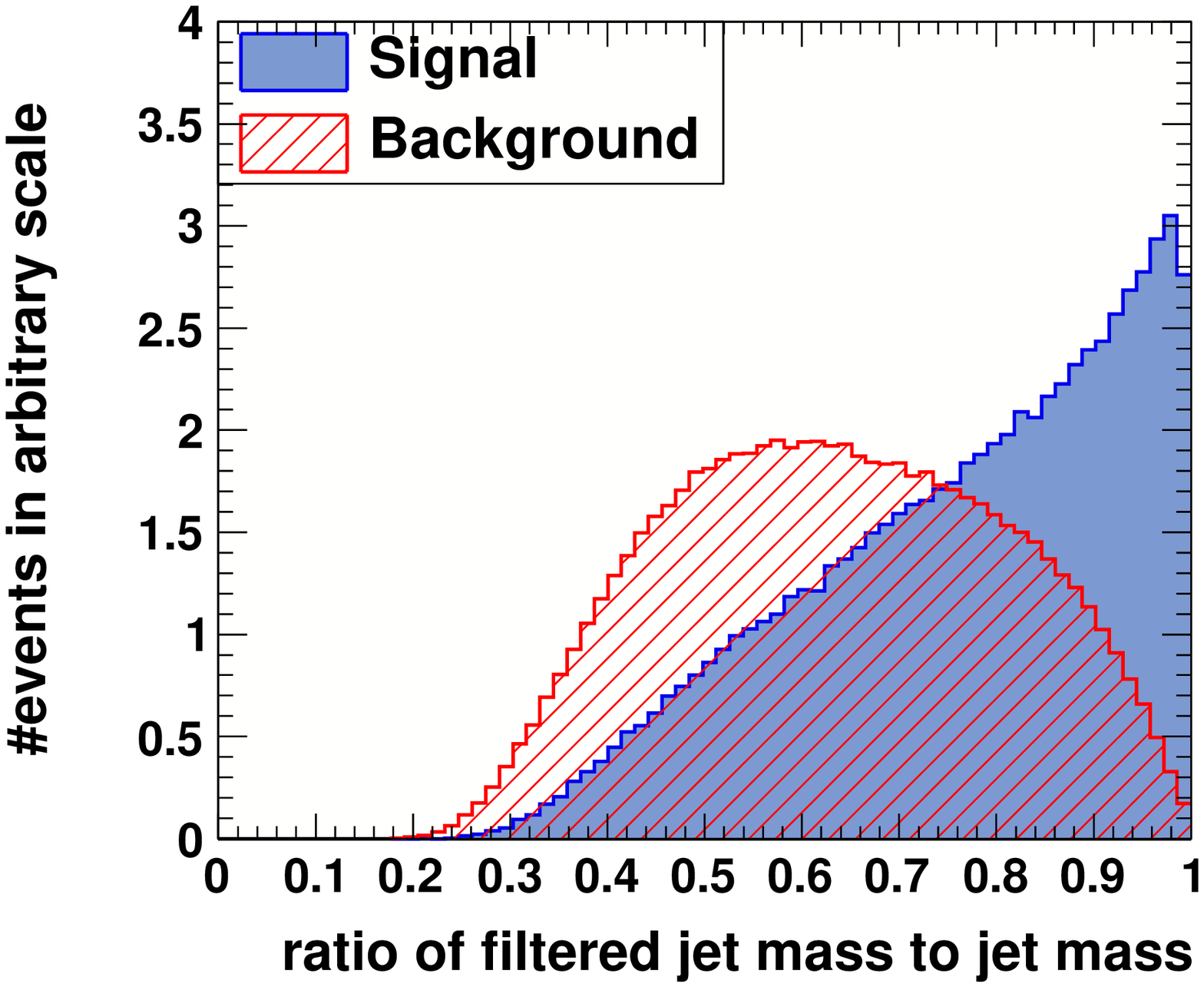}& \includegraphics[width=0.3\textwidth]{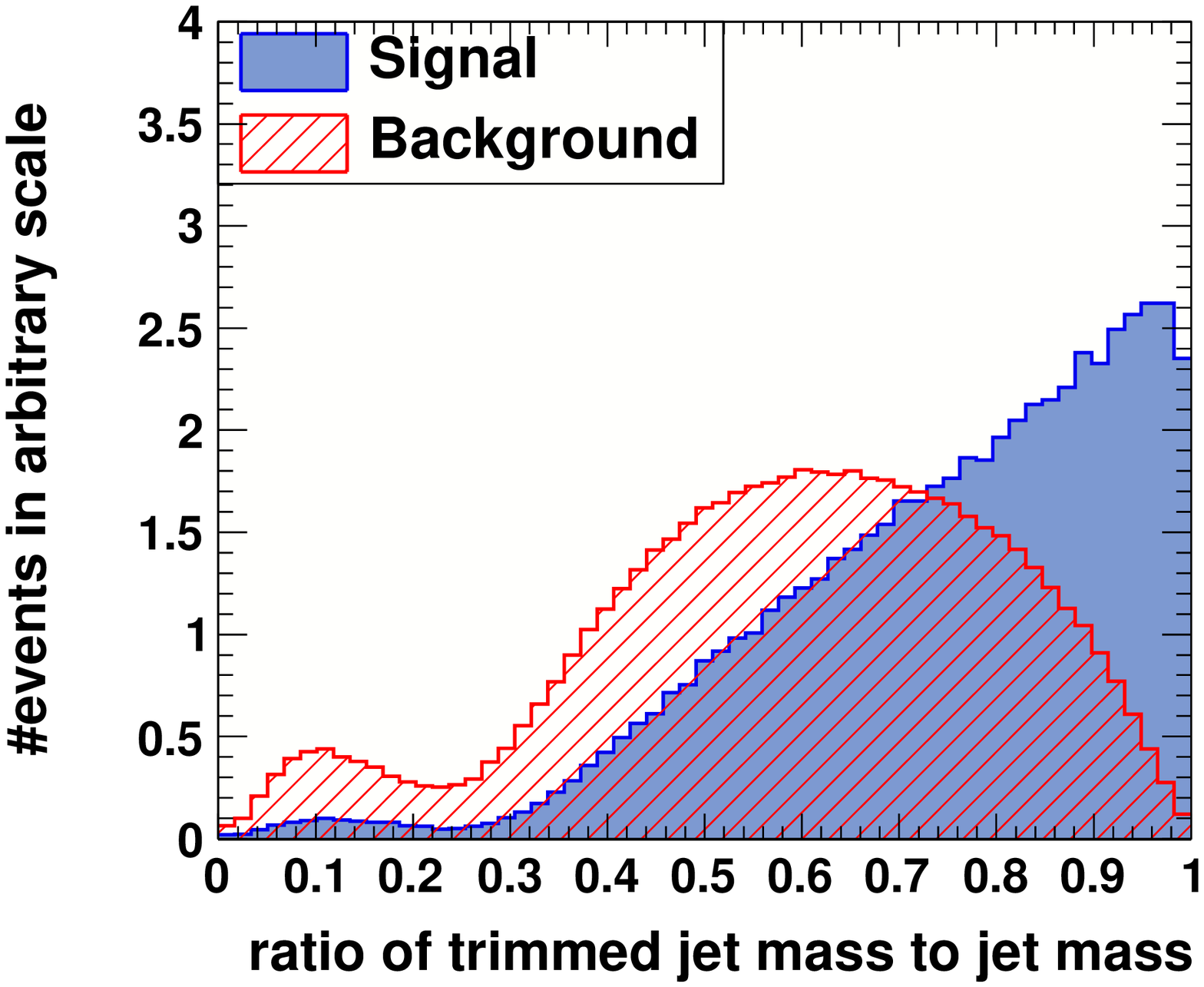}
      &\includegraphics[width=0.3\textwidth]{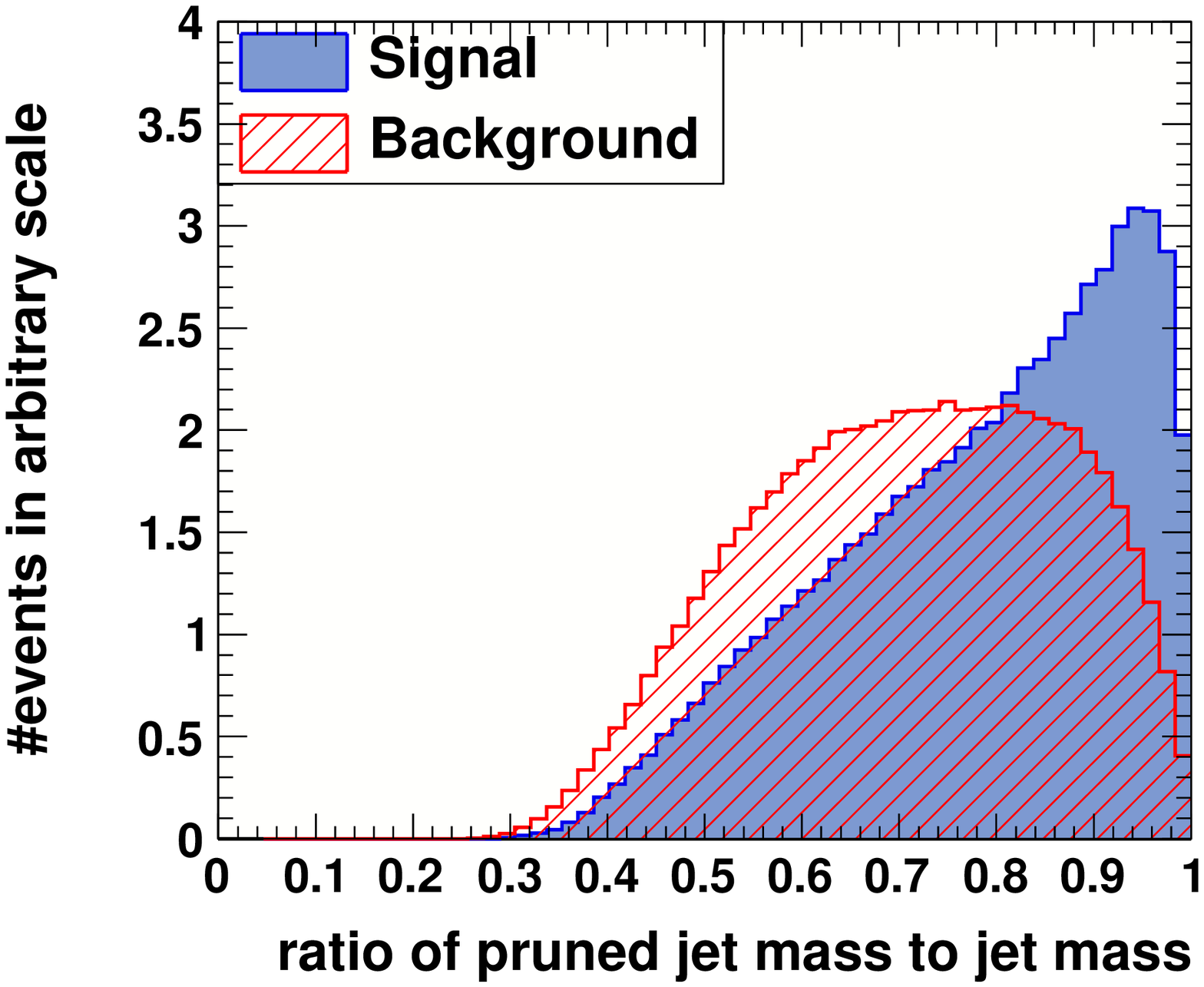}
      \end{tabular}
       \caption{Distributions of grooming sensitivities, ${\text{sens}}^m_\text{filt}$,  ${\text{sens}}^m_\text{trim}$, and ${\text{sens}}^m_\text{prun}$
for signal ($W$-jets) and background (QCD-jets) for $p_T^\jet\in(500,550)\gev$. All events satisfy $m_\text{filt} \in (60,100)$ GeV.}
      \label{fig:groomedmass}
      \end{figure}

\subsection{Planar flow}
 There have been attempts to discriminate jets from heavy particle decays against QCD-jets by using observables as functions of energy flow of the physical jet \cite{Almeida:2008yp,{Almeida:2010pa}}. One variable of such type that we found useful for our purpose is planar flow, $P_f$, which characterizes the geometric distribution of energy deposition from a jet. {\bf Planar flow} is defined as follows. For a given jet we first construct a matrix $I^{kl}_w=\frac{1}{m_\jet}\sum_i w_i\frac{p_{i,k}}{w_i}\frac{p_{i,l}}{w_i}$ where $m_\jet$ is the jet mass, $w_i$ is the energy of particle $i$ in the jet, $p_{i,k}$ is the $k^{th}$ component of its transverse momentum relative to the jet's momentum axis. $P_f$ is then defined based on $I_w$ as $P_f=\frac{4det(I_w)}{tr(I_w)^2}=\frac{4\lambda_1\lambda_2}{(\lambda_1+\lambda_2)^2}$ where $\lambda_{1,2}$ are eigenvalues of $I_w$.
For linear distributions, $P_f\to 0$, while for isotropic distributions, $P_f \to 1$.

       \begin{figure}
      \begin{tabular}{cc}
      \includegraphics[width=0.35\textwidth]{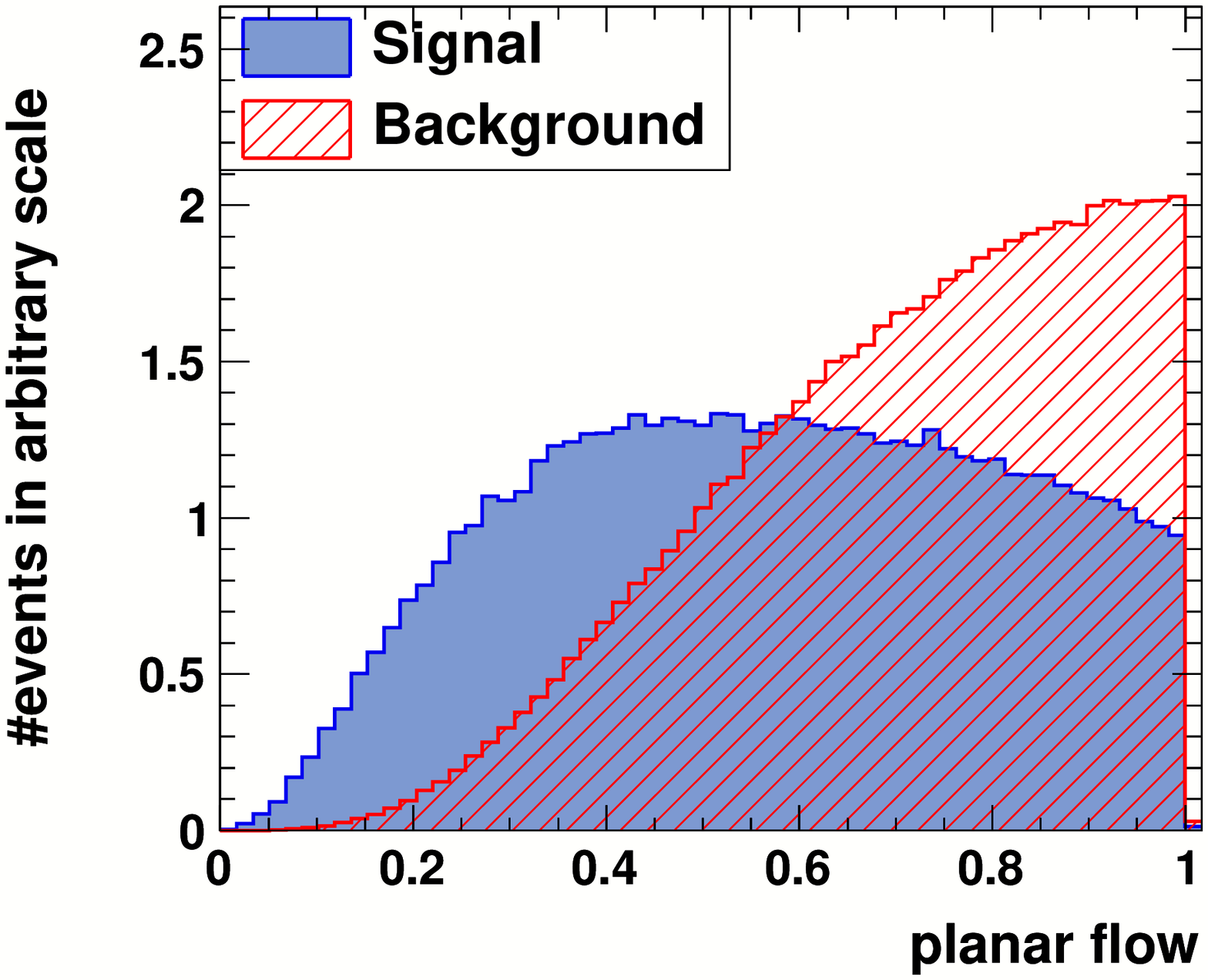}&\includegraphics[width=0.35\textwidth]{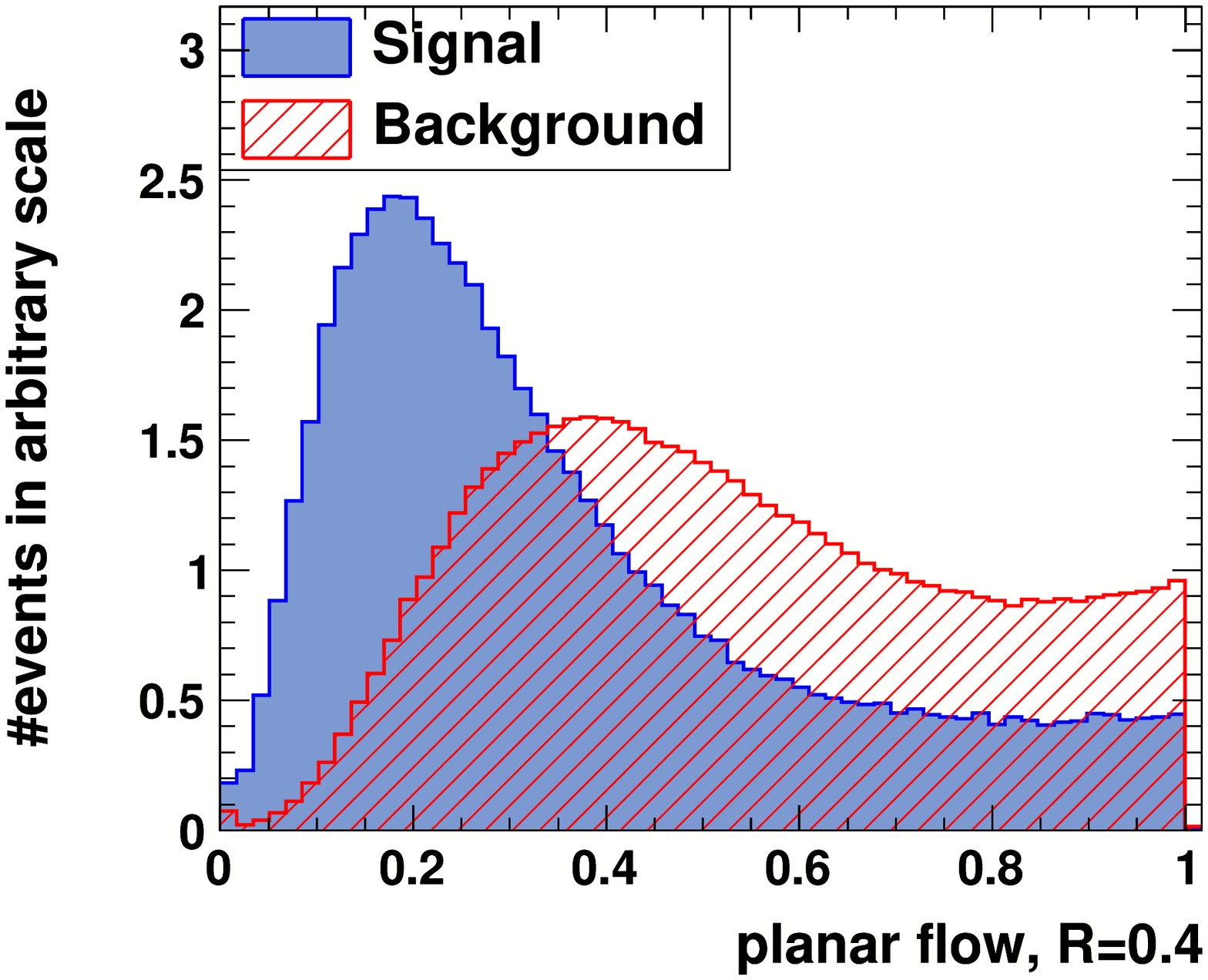}\\
      (a)&(b)
      \end{tabular}
      \caption{Signal vs. background planar flow ($P_f$) distributions for $p_T^\jet\in(500,550)\gev$: (a) $P_f$ for the fat jet ($R=1.2$); (b) $P_f$ for the leading subjet reclustered with $R=0.4$.}\label{planarflow}
      \end{figure}

 Planar flow has been suggested for top-tagging, since a boosted top jet should be more isotropic due to three hard prongs coming
from its on-shell decay. In contrast, a QCD-jet is more linear as it typically has two leading hard prongs.
Resonances decaying to two partons are more similar to QCD-jets in terms of $P_f$,
but as pointed out in \cite{Almeida:2010pa} with Higgs as an example: although both have two prongs and $P_f$ peaks towards $1$,
the prongs from the heavy particle decay are sharper and $P_f$ peaks at lower values than QCD. The planar flow distributions for $W$-jets
and their QCD-jet background are shown in Figure~\ref{planarflow}. We see that planar flow promises to still be a useful discriminant.
Planar flow becomes even more useful at higher $p_T$.

We find it useful to consider not just the planar flow of the original fat jet, $P_f$, but also the planar flow of the
the highest $p_T$ subjet resulting from reclustering with $R=0.4$, $P_f(0.4)$.
$R=0.4$ is more useful for high $p_T$ samples, while $R=1.2$ is more useful for low $p_T$ samples, which is related to the $p_T$-dependence of proper jet cone sizes.

\subsection{Features of Subjets}
After reclustering with smaller $R$ during filtering, we get a set of subjets from the original fat jet. Variables related to these subjets can further distinguish substructure of $W$-jets from that of QCD-jets.
It is known that the two subjets from the decay of a massive particle are more symmetric in $p_T$ than those from QCD.
In fact, the $y_{\text{cut}}$ parameter in the filtering algorithm is based on this consideration.
We call the subjet with the highest $p_T$ subjet 1 and the one
with the second highest $p_T$ subjet 2.

Two variables that we find useful are the ratios of the $p_T$'s of the two leading subjets to the original jet $p_T$:
 $p_T^{\text{sub1}}/p_T$ and $p_T^{\text{sub2}}/p_T$. These variables are more useful than $p_T^{\text{sub1}}/p_T^{\text{sub2}}$ alone.
  Another useful variable is the geometric distance in the $\eta$-$\phi$ plane between the two leading subjets
$\Delta R_\text{sub}$. For signal jets it peaks around smaller values than QCD-jets. Finally, the total number of subjets ($p_T>10\gev$) after the filtering process,
$\nsub$, can help. $\nsub$ concentrates around smaller values for $W$-jets than for QCD-jets. This is because compared with $W$-jets, QCD radiation is more diffusely distributed. For illustration plots, see Figure.\ref{subjet}.
\begin{figure}
\begin{tabular}{ccc}
\includegraphics[width=0.3\textwidth]{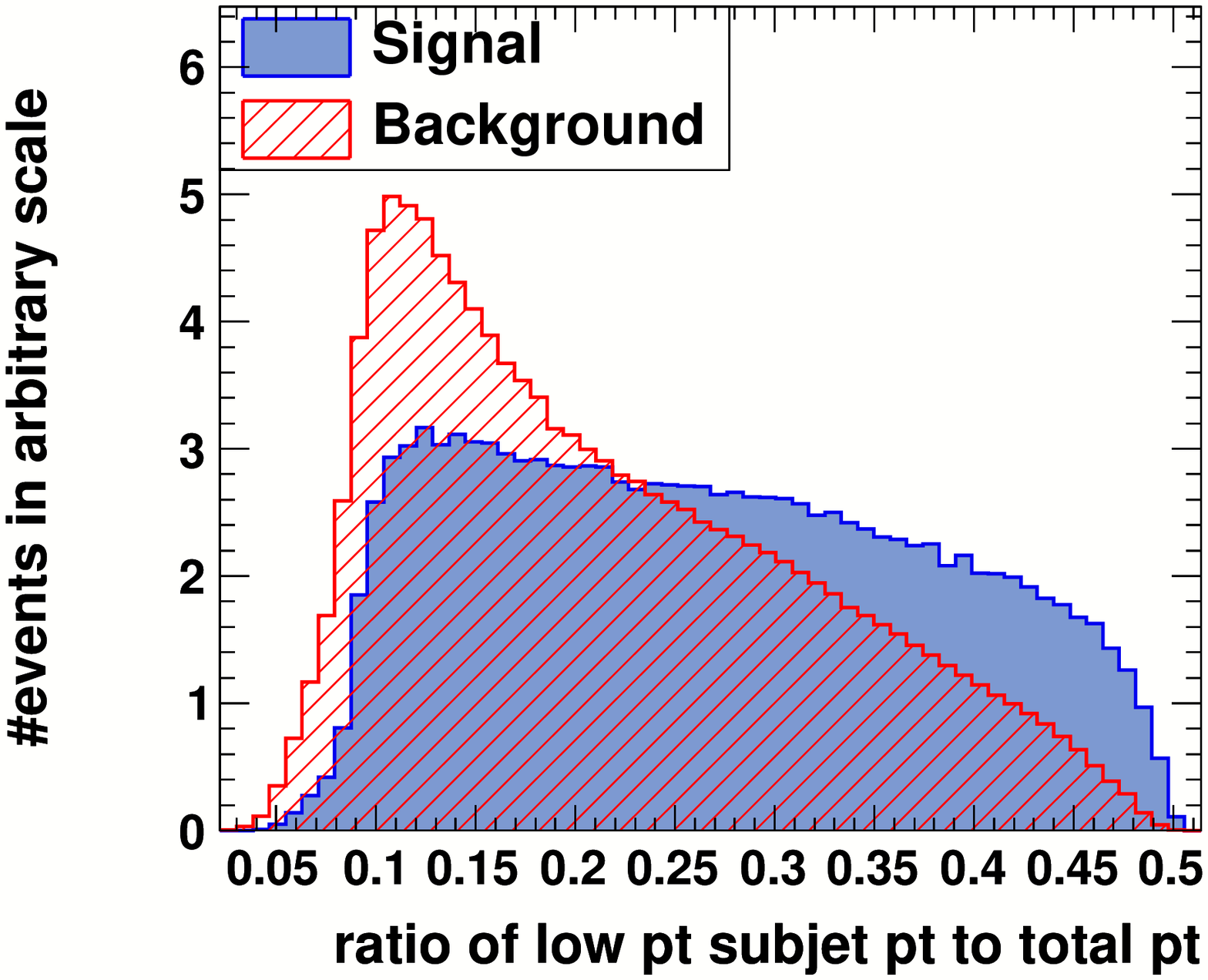} &
\includegraphics[width=0.3\textwidth]{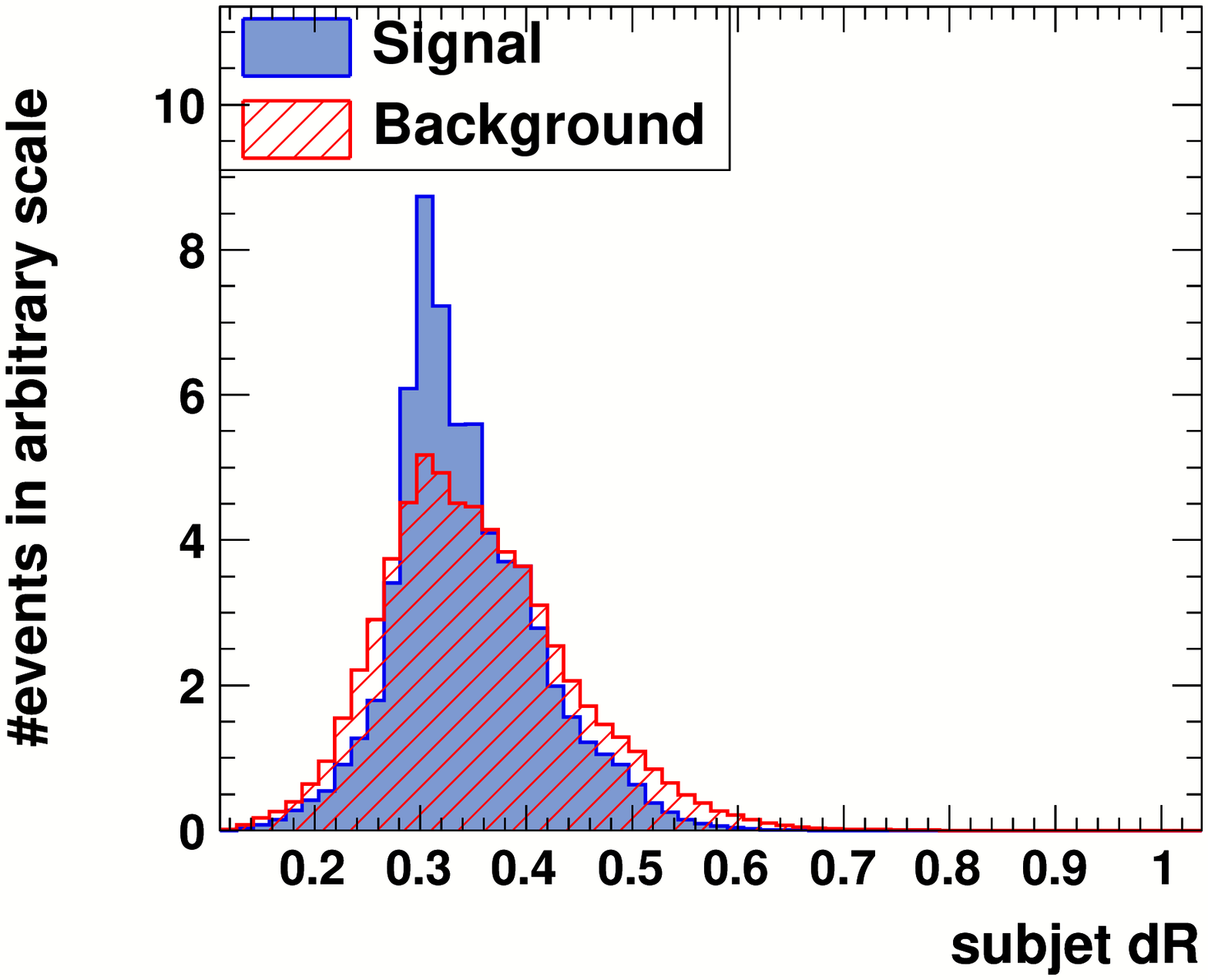} &
\includegraphics[width=0.3\textwidth]{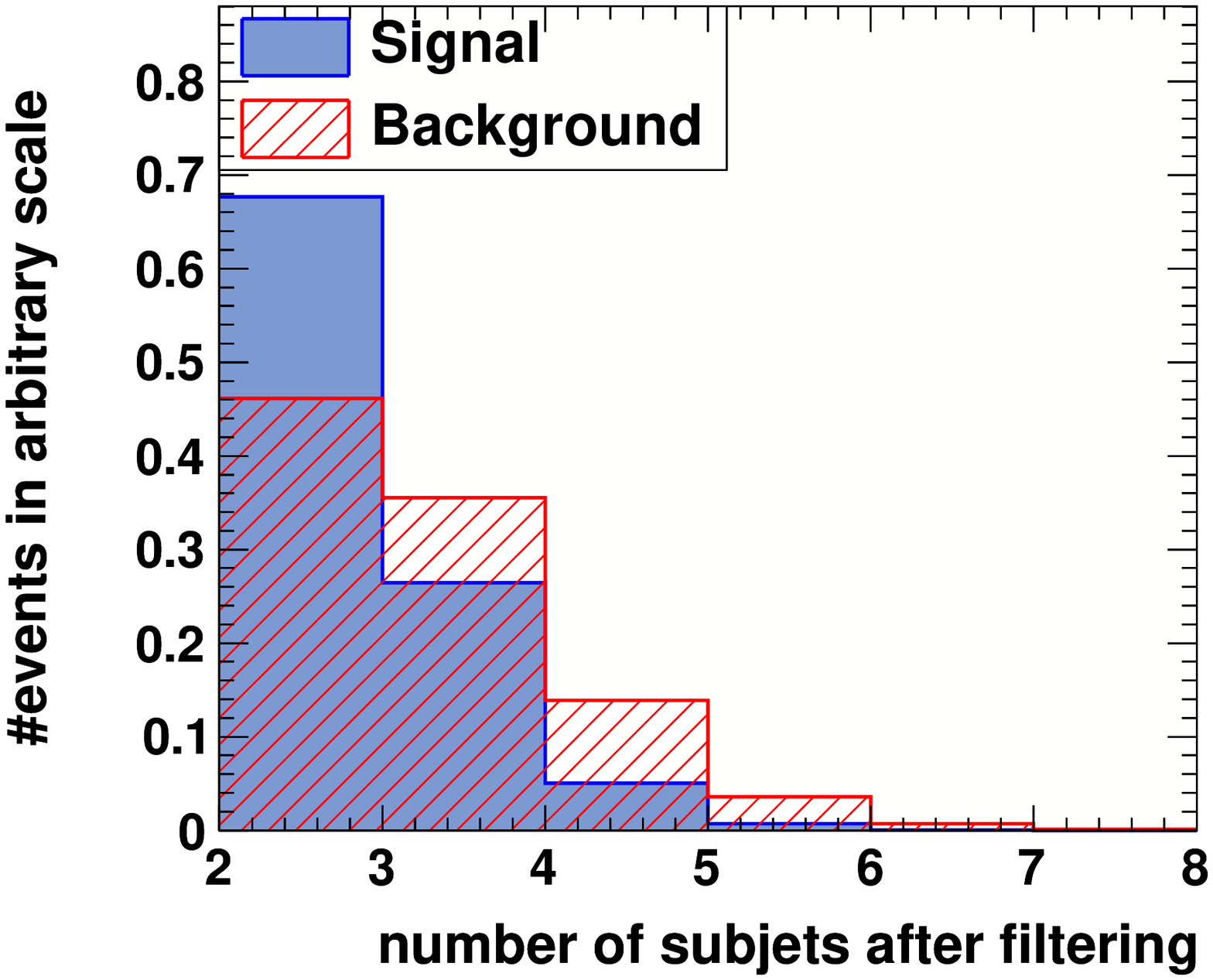}
\end{tabular}
\caption{Signal and background distributions of  $p_T^{\text{sub2}}/p_T$, $\Delta R_{\text{sub}}$ and $\nsub$
for $p_T^\jet\in(500,550)\gev$ samples in the filtered mass window.}
\label{subjet}
\end{figure}

\section{Multivariate analysis for optimal $W$-jet tagging}
\label{sec:MVA}
 So far we have seen how certain variables may help improve significance when individually used. A proper combination of different variables could optimize the discrimination power as it incorporates more details of radiation pattern. As before, we consider the SM $WW$ (semi-leptonic) and $Wj$ (leptonic $W$ decays) data samples which have been processed with filtering and then passed a $m_{\filt} \in (60,100)\gev$ mass window cut. After the mass window cut, the original unfiltered fat jets are used for subsequent analysis.

Simple rectangular cuts cannot make optimal use of multiple variables since they overlook the multidimensional
correlations. Instead we use more sophisticated multivariate techniques,
as implemented in TMVA (Toolkit for Multivariate Data Analysis with Root) \cite{TMVA},
to maximize the efficiencies. In particular, we use the Boosted Decision Trees (BDT) method which appears fast and reliable,
and particularly well suited for high energy theory analyses. Details of this method as used in particle physics can be found, for example, in~\cite{Roe:2004na}.
As we will see, using our variables and BDTs is significantly better than filtering alone,
with an additional factor of $2-3$ improvement in SIC.
One can then apply the cuts giving the maximal SIC to data samples from different processes (we will show two examples later: $Z'$ discovery and $Wj$ as signal vs. $jj$). Such applications also test the robustness of multivariate methods.

For various jet $p_T$'s we begin with $\sim10^5$ signal events and $\sim10^6$ background events after the filtered mass cut as input samples.
We first rank the individual variables based on the SIC when they are individually used.
Then among those at the top we try to find a combination of variables for
which the improvement in $S/\sqrt{B}$ almost saturates (adding even more variables on top has little effect).
Some variables, like the pull angles, girth, or mass $R$-cores tend not to help on top of other top variables,
so they are not used for the final list. A nice feature of the BDT method is adding useless variables
does not particularly downgrade the training speed or final efficiencies.
A set of 25 variables (all these variables have been defined in Section \ref{sec:variables}) that saturate the efficiencies is
\begin{equation}
m_\jet, \;
 c_{p_T}(0.2 - 0.11),  \;
{\text{sens}}^{m,p_T}_{\text{filt,trim,prun}}, \;
P_f, \;
P_f(0.4), \;
 \frac{p_T^{\text{sub1,sub2}}}{p_T},\;
\frac{m^{\text{sub1,sub2}}}{m},\;
 \Delta R_\sub,\; \nsub.
\end{equation}
We use 10 $p_T$ $R$-cores, from $R=0.2$ to $R=1.1$ by $0.1$ and 6 grooming sensitivities.

Figure~\ref{fig:mva-significance} shows the SIC curves ($\varepsilon_{S}/\sqrt{\varepsilon_{B}}$ functions of $\varepsilon_{S}$)
 for these variables, as each one (or set) are added. The curves are
cumulative. The big jumps in the lower curves come from adding 10 $R$-cores and then the two filtering sensitivities as groups.
Naturally, the discrimination efficiency of the variables is $p_T$ dependent, so plots
for $p_T\in(200,250), (500,550)$ and $(1000,1050)$ GeV are shown separately.
Figure~\ref{mva-pt} shows the maximal SIC using these 25 variables as a function of $p_T$. We see the improvement gets more appreciable towards higher $p_T$.
\begin{figure}
\begin{tabular}{c}
\includegraphics[width=0.6\textwidth]{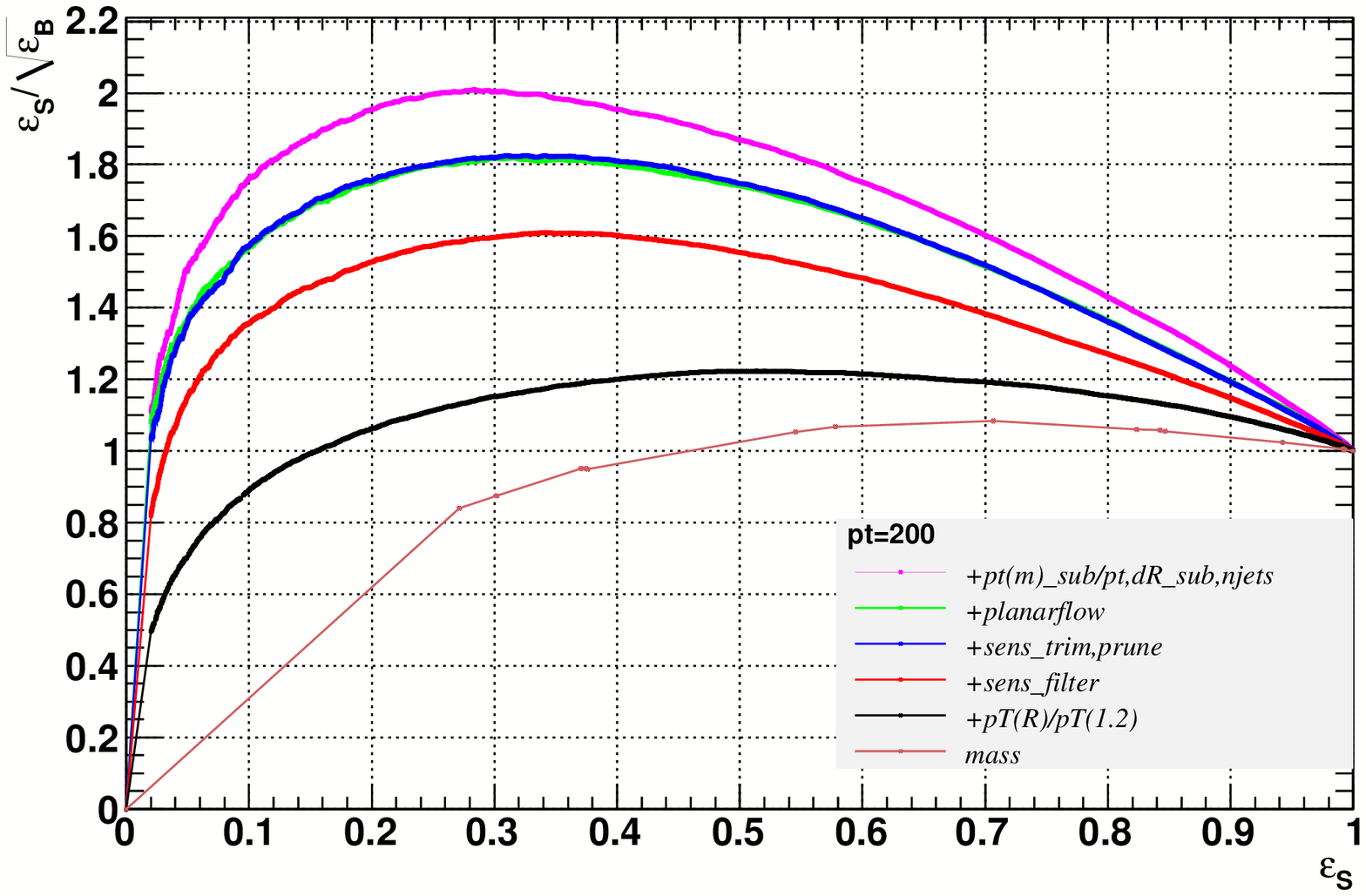} \\
\vspace{-0.1cm}
\includegraphics[width=0.6\textwidth]{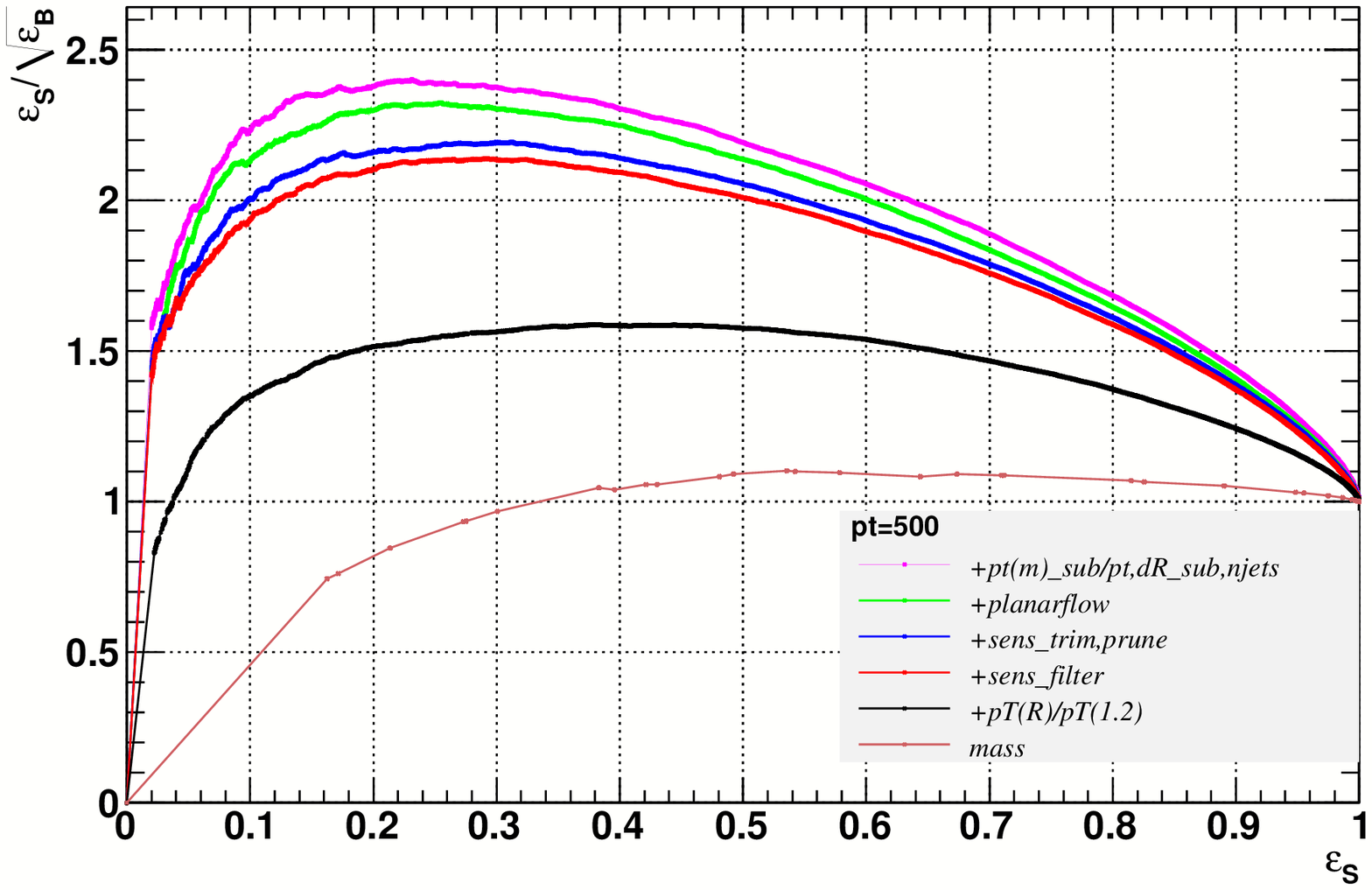} \\
\includegraphics[width=0.6\textwidth]{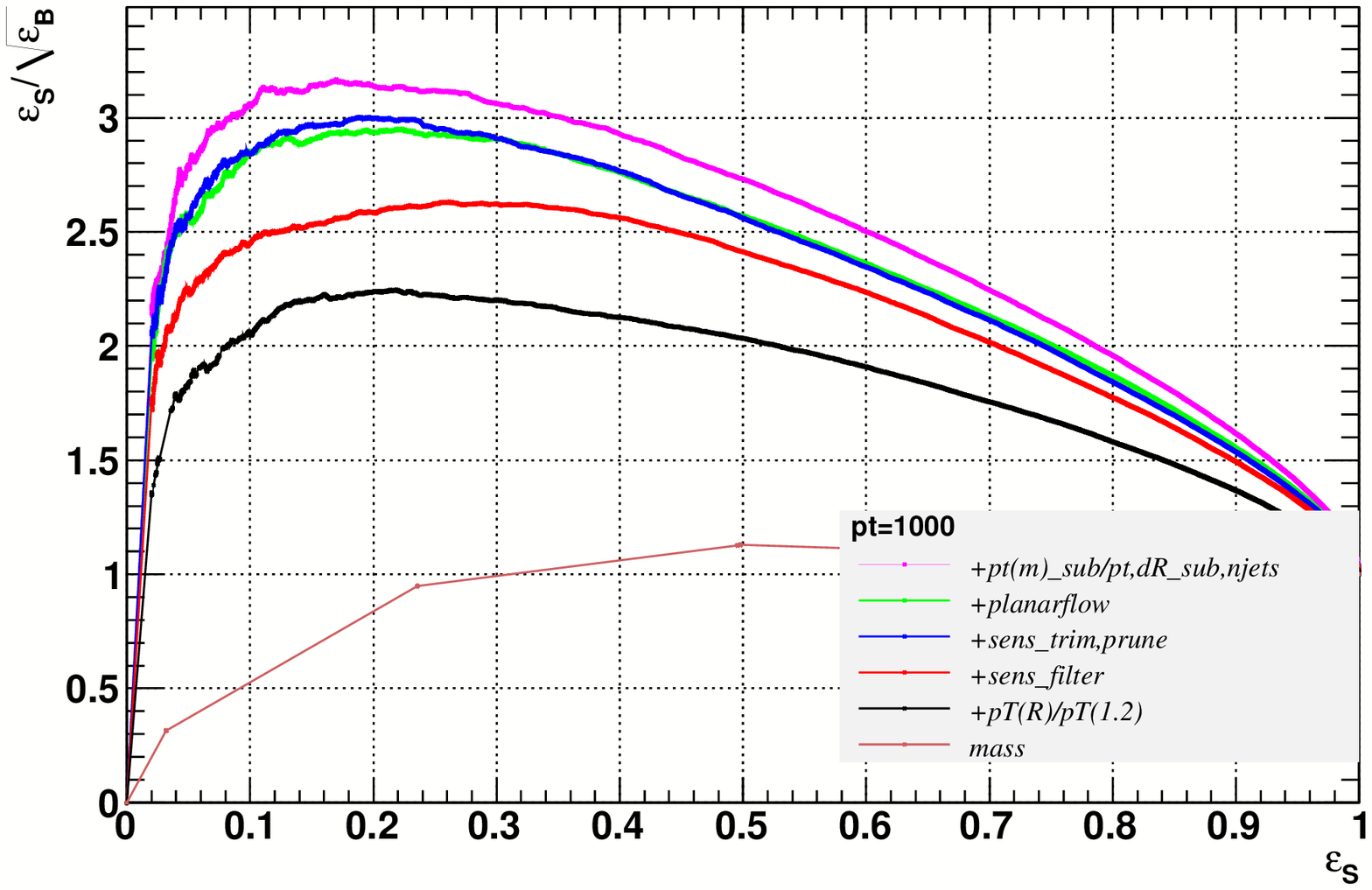}  \\
\end{tabular}
\caption{Significance gain from the multivariate analysis for $p_T^\jet=200, 500, 1000\gev$.}\label{fig:mva-significance}
\end{figure}
\begin{figure}
\includegraphics[width=0.6\textwidth]{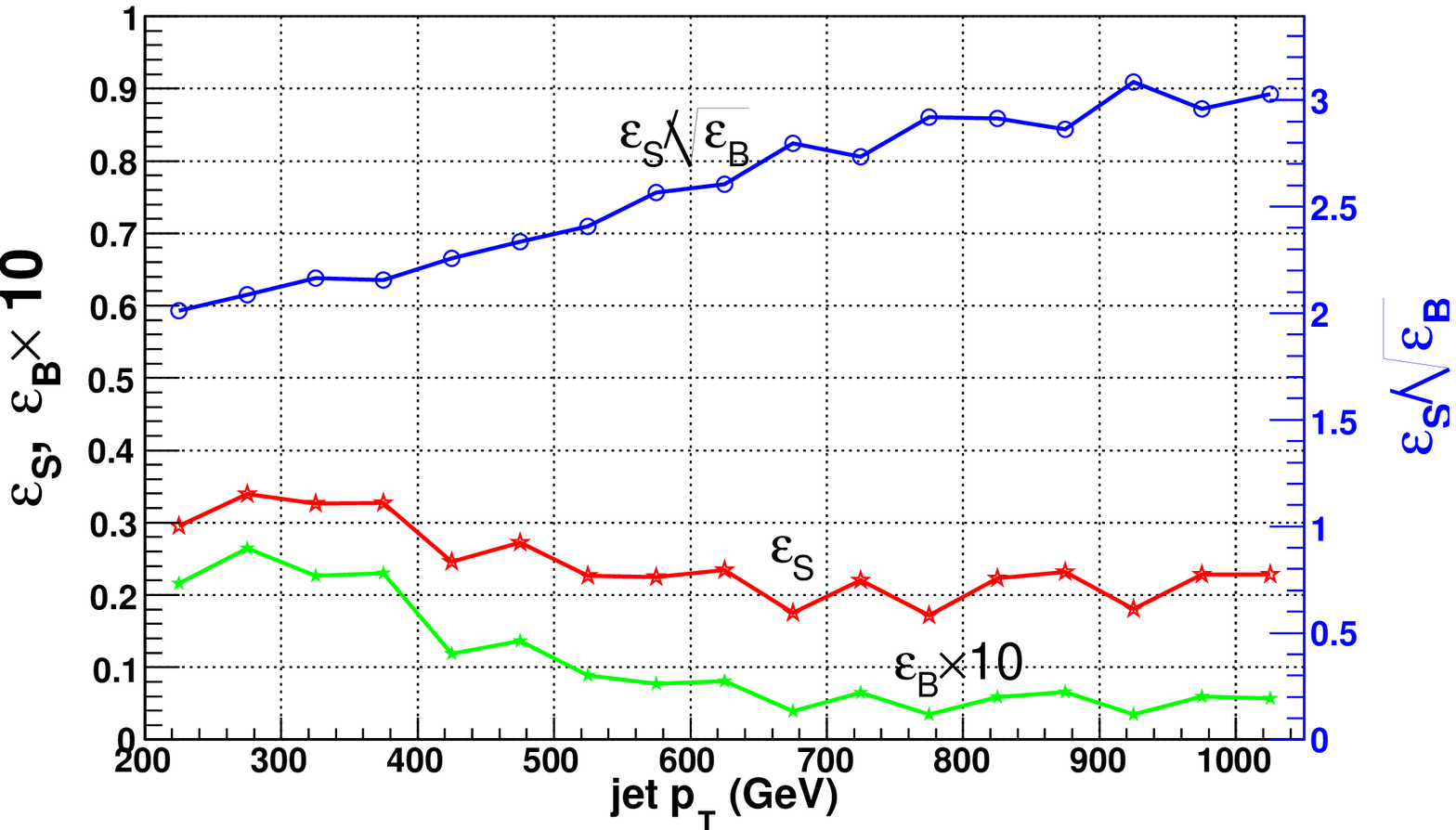}\\
\caption{The maximal SIC with MVA using all 25 principle variables as a function of jet $p_T$, and the corresponding signal, background efficiencies. The background efficiencies are multiplied by 10.}\label{mva-pt}
\end{figure}

In practice if one prefers to use fewer variables and be less ambitious about significance gain, one can do almost as well with
a subset of these variables. For example, if we take the 7 variables
\begin{equation}
 m(0.5),\;
 m(0.4),\;
m_{\text{filt}},\;
m^{\text{sub1}},\;
m^{\text{sub2}},\;
\frac{p_T^{\text{sub2}}}{p_T^{\text{sub1}}},\;
P_f(0.4), \label{eq7}
\end{equation}
we can achieve a significance gain of $\sim 1.9$ over the filtered sample, as compared to $\sim 2.4$ using the full 25 variables.
This particular subset of variables is partially motivated by having smaller sensitivity to the underlying event, as
will be discussed in Section~\ref{sec:mc} below.

\section{$W$-polarization dependence}
\label{sec:wpol}
As is well known, the distribution of $W$ decay products depends on the polarization of the $W$. This has an effect on the $W$-jet substructure and can therefore be exploited both to improve efficiency if the polarization of the sample
is known, or even to measure the $W$-polarization if the statistics are high enough. Similar ideas were used
for top-tagging in~\cite{Krohn:2009wm}.

Let us define $\theta$ as the angle between an up-type Fermion (including $u$ and $c$ quarks and neutrinos)
and the $W^+$ moving direction in the rest frame of $W^+$. Then the probability density of finding the Fermion is given by
\begin{equation}
P(\cos\theta)=\left\{\begin{array}{cc}\frac{3}{8}(1\mp\cos\theta)^2 & \  \ \mbox{for }h_{W^+}=\pm;\\\frac{3}{4}(1-\cos^2\theta) & \  \ \mbox{for }h_{W^+}=0.\end{array}\right.\label{eq:w_helicity}
\end{equation}
where $h_{W^+}$ is the helicity of the $W^+$ boson. For a down-type anti-fermion, $(1\mp\cos\theta)$ flips to $(1\pm\cos\theta)$ in the first line of Eq.~(\ref{eq:w_helicity}). The formula holds for $W^-$ too if we replace up-type with down-type.

These distributions imply that for transverse $W$'s the probability density is maximum at $\cos\theta\sim\pm1$, which means one of the decay products tends to go along the $W$ momentum and the other one against it. When the $W$ is boosted, this results in an unbalanced configuration for the two decay products' momenta, namely, one smaller than the other one. On the other hand, for longitudinal $W$'s, the probability density is maximum at $\cos\theta\sim0$, where the decay products' momenta are perpendicular to the $W$ momentum in the $W$ rest frame, and more balanced when boosted. Since a QCD splitting tends to produce unbalanced momentum configuration, transverse $W$'s behave more like QCD-jets than longitudinal $W$'s, and we expect better identification for longitudinal ones.
For the SM $W$-pair production, the $W$'s are dominantly transverse: about 92\% for $p_T^W > 200\gev$. Therefore, the results reported in the previous sections can be viewed to good approximation as for transverse $W$'s. There are also cases where the $W$'s are dominantly longitudinal, for example, $W$'s from a heavy SM-like Higgs decay or high energy $WW$ scattering.

\begin{figure}
\begin{center}
 \includegraphics[width=0.6\textwidth]{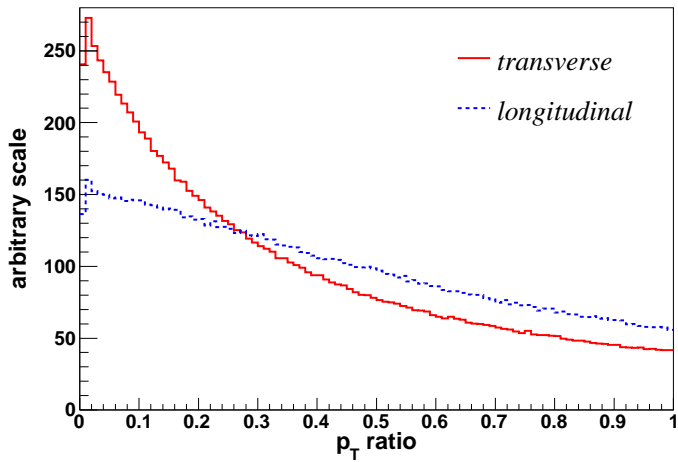}
\caption{\label{fig:ptratio_500} Ratio of the $p_T$ of the lower $p_T$ parton to $p_T$ of the higher $p_T$ parton
from a $W$ decay, for different $W$ polarizations. }
\end{center}
\end{figure}

To study the longitudinal case, we start by generating $WW$ pairs using Madgraph but this time we decay the $W$'s manually
according to $P(\cos\theta)\propto(1-\cos^2\theta)$. Note that in this way the spin correlation between the two $W$'s in the same event is not included, but it does not affect our results since the leptonic $W$ is excluded from jet clustering. In Figure~\ref{fig:ptratio_500}, we display the $p_T$ ratio between the two partons from a $W$ decay for $p_T^W\in(500, 550)\gev$. As expected, the momenta are more balanced for longitudinal $W$'s than transverse ones.
The events are then processed with Pythia 8 and we repeat the procedure described in Section \ref{sec:grooming} through \ref{sec:MVA}.

The filtering parameters which maximize the SIC for the longitudinal sample are shown in Figure~\ref{fig:filter_parameters_long}. The fact that the two subjets are more balanced allows us to use tighter cuts to cut more background events for the same signal efficiency, resulting in higher SIC than the transverse case. The multivariate analysis provides further a larger significance gain than for the transverse case, as can be seen in Figure~\ref{fig:compare_long_tran}. All together, after filtering and our MVA $W$-jet tagging, the maximal SIC is $\sim 7.0$ for longitudinal $W$'s, significantly larger than that of transverse $W$'s, $\sim 5.3$.
\begin{figure}
\begin{center}
\begin{tabular}{cc}
\includegraphics[width=0.5\textwidth]{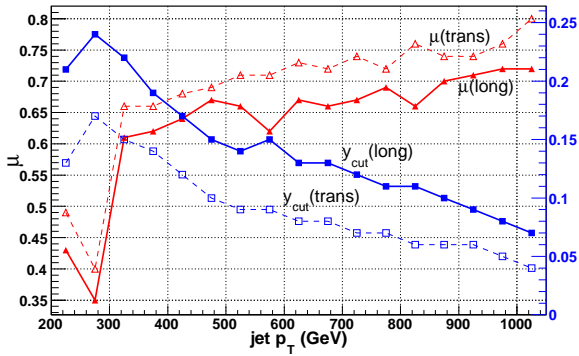}
&\includegraphics[width=0.5\textwidth]{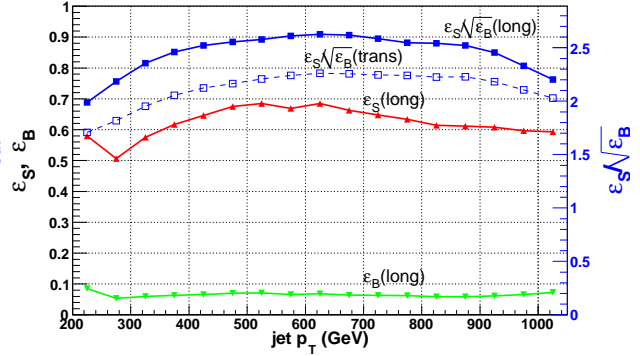}
\\(a) Optimized filtering parameters &
(b) Optimized efficiencies and SICs.
\end{tabular}
\caption{Tuning of filtering parameters for longitudinally polarized $W$-jets versus QCD jets.
For comparison, the results for transverse $W$'s from Figure~\ref{fig:filter_parameters} are reproduced here. \label{fig:filter_parameters_long}}
\end{center}
\end{figure}

\begin{figure}
\begin{center}
 \includegraphics[width=0.5\textwidth]{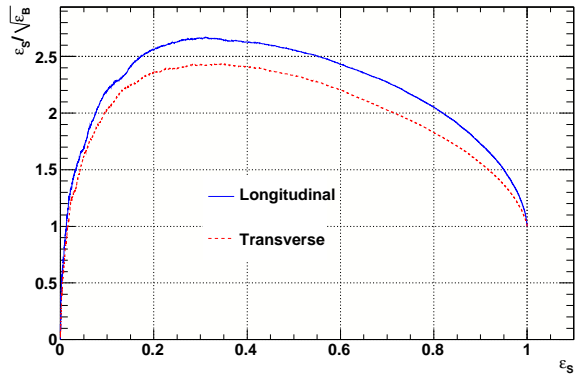}
\caption{\label{fig:compare_long_tran}
The SIC using BDTs as a function of signal efficiency for transverse and longitudinally
polarized $W$'s. This is for $p_T^\jet\in(500, 550)\gev$ and these gains are on top of the factors of $\sim 2$ or $\sim 2.5$ for
the two samples from filtering, as shown in Figure~\ref{fig:filter_parameters_long} (b).
}
\end{center}
\end{figure}

The polarization effect of the $W$ boson poses a question:
 what parameters/cuts should we use when looking for boosted $W$'s?
 This depends on our goal: if we are looking for $W$ bosons inclusively, we should be conservative and use relatively loose cuts obtained from transverse $W$'s; if we are interested in a particular process dominated by longitudinal $W$'s, we should use tighter cuts optimized for longitudinal ones.

\section{Differences in Monte Carlo tools}
\label{sec:mc}
In our analysis, we have extensively utilized the differences in radiation patterns between $W$-jets and QCD-jets.
These patterns have not been measured at high $p_T$ and we have been relying on Pythia 8 simulations.
It is important to cross-check using different Monte Carlo tools, which is the subject of this section.
It is also possible to compare the same event generator, with different tunes.
Up to now, all results have been obtained with the default tune of Pythia 8.142. We tried also the tune ``3C'',
which is a tune to the Tevatron and early LHC data for initial state radiation, multiple interaction and beam remnants.
There were no discernible differences between these tunes for our variables. So we restrict the discussion in this section
to a comparison of Pythia 8 and Herwig++. We perform the comparison by testing the cuts/parameters/BDTs trained on
Pythia 8 event samples on samples generated with Herwig++ {\sc v2.4.2} \cite{herwig++}.

As before we look at $WW$ and $W$+jet in the SM. With each Monte Carlo, we use the same jet algorithm (Cambridge/Aachen with $R=1.2$) to find the high $p_T$ jets. We consider only jets with $p_T\in(500, 550)$ GeV. We apply the filtering/pruning/trimming procedure using the parameters given in Table \ref{tab:jet_pars} in Appendix \ref{app:pars}. As before, only events passing the filtered mass window cut, $m_{\text{filt}} \in (60,\, 100)\gev$, are retained. For Herwig++ data samples, the efficiencies after the mass window cut for the signal and background jets are respectively 64.4\% and  8.68\%, yielding a significance gain of 2.18. The corresponding efficiencies for Pythia 8 are 65.8\% and 8.88\%, yielding a very similar significance gain of 2.21. So, as far as the filtering/mass-drop step is concerned, there is hardly any difference.

\begin{figure}
\begin{center}
\begin{tabular}{cc}
\includegraphics[width=0.5\textwidth]{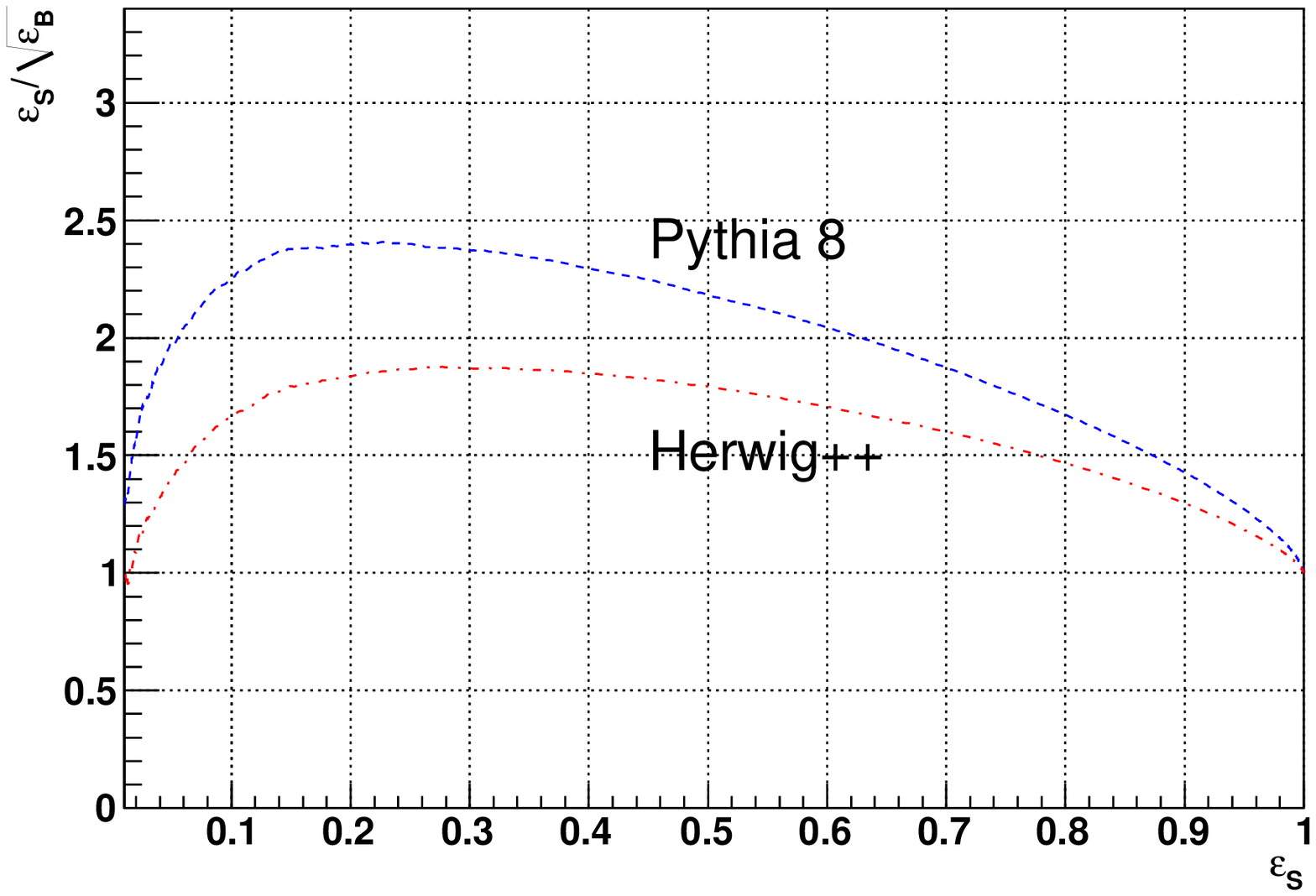}
& \includegraphics[width=0.5\textwidth]{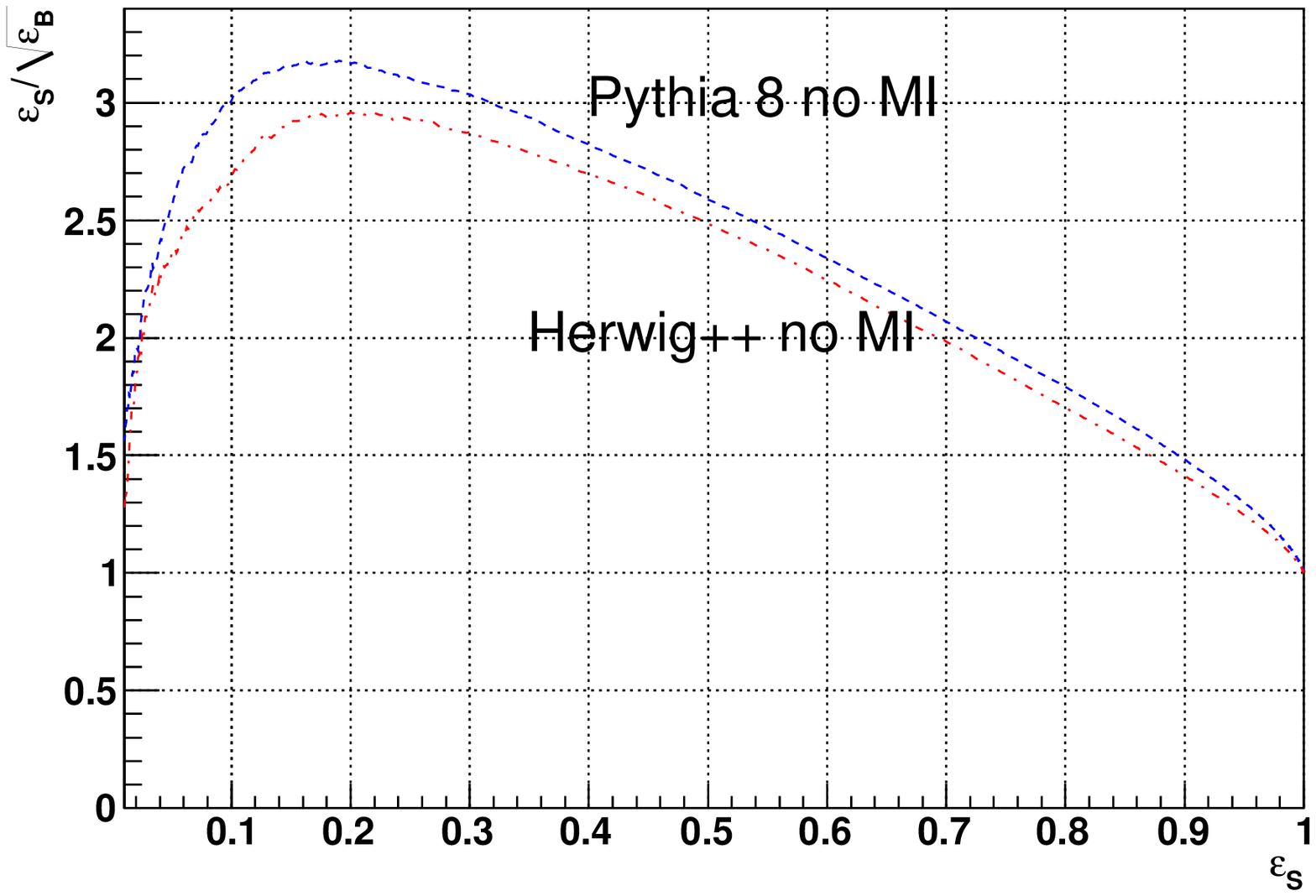}
\\(a) with underlying event& (b) no underlying event
\end{tabular}
\caption{\label{fig:compare_mc} Signficance improvements resulting from a boosted decision tree trained on Pythia 8, and tested
on Pythia 8 or Herwig++, for $p_T^{\mathrm{\jet}}\in(500, 550)\gev$.}
\end{center}
\end{figure}
We then obtain the values of the variables defined in Section \ref{sec:MVA} and evaluate the BDT response using weight files trained on Pythia 8 event samples. In Figure~\ref{fig:compare_mc} (a), we show the significance gain as a function of the signal efficiency, for jets with $p_T\in(500, 550)\gev$. From Figure~\ref{fig:compare_mc} (a), we see that the Pythia 8 results differ significantly from Herwig++. The most likely origin of the difference is in the modeling of the underlying event (UE), which can have an important effect on jet substructure.
To test this, we show in Figure~\ref{fig:compare_mc} (b) the result with UE turned off for both Pythia 8 and Herwig++\footnote{What are turned off are multiple interactions by using the switch ``PartonLevel:MI = off'' for Pythia 8 and ``set /Herwig/Shower/ShowerHandler:MPIHandler NULL'' for Herwig++.}.
For this figure, we retrained the BDT from the Pythia 8 sample and then tested it on both Pythia 8 and Herwig++.
The BDT responses without the underlying event are much less sensitive to the Monte Carlo.

\looseness -1 We can understand better the difference between the Monte Carlos
 by examining the contributions to our variables from the underlying event. Let us start with jet masses.
We have found that Herwig++ in general produces more radiation through the underlying event than Pythia 8, which can be seen from Figure~\ref{fig:mass_compare}.
 In Figure~\ref{fig:mass_compare} (a), we show the $W$-jet mass distributions in the signal sample after filtering.
For $p_T^\jet\in (500, 550)\gev$, the distance between the two subjets is only about $0.3\sim0.4$. Therefore, the filtered mass receives small contributions from initial state radiation and the underlying event, and as expected,
the two Monte Carlos give almost identical distributions. On the other hand, we see from Figure~\ref{fig:mass_compare} (b) that the original jet mass ($R=1.2$) from Herwig++ is larger than from Pythia 8. By using $R=1.2$ for jet clustering, we include ISR and UE contributions in a large region, which makes the difference of the two Monte Carlos manifest. For comparison, the jet mass without UE is given in Figure~\ref{fig:mass_compare} (c), showing opposite behavior in the mass tail, namely, the Herwig++ jet mass is lightly smaller.
This clearly shows that Herwig++ produces more radiation through UE. Consequently, for Herwig++, $W$-jets
look more like QCD-jets (compare Figure~\ref{fig:masses}), which explains the smaller significance improvement using Herwig++.
Similar behavior can be seen in other variables. For example, in Figure~\ref{fig:pf_pythia_herwig},  we compare the planar flow for two different $R$'s, $R=0.4$ and $R=1.2$ for signal jets with $p_T\in (500, 550)\gev$. We see very small differences between Pythia 8 and Herwig++ for $R=0.4$ but significant differences for $R=1.2$. Again, for the $R=1.2$ case more UE is included which explains the dramatic difference.

\begin{figure}
\begin{center}
\begin{tabular}{ccc}
\includegraphics[width=0.33\textwidth]{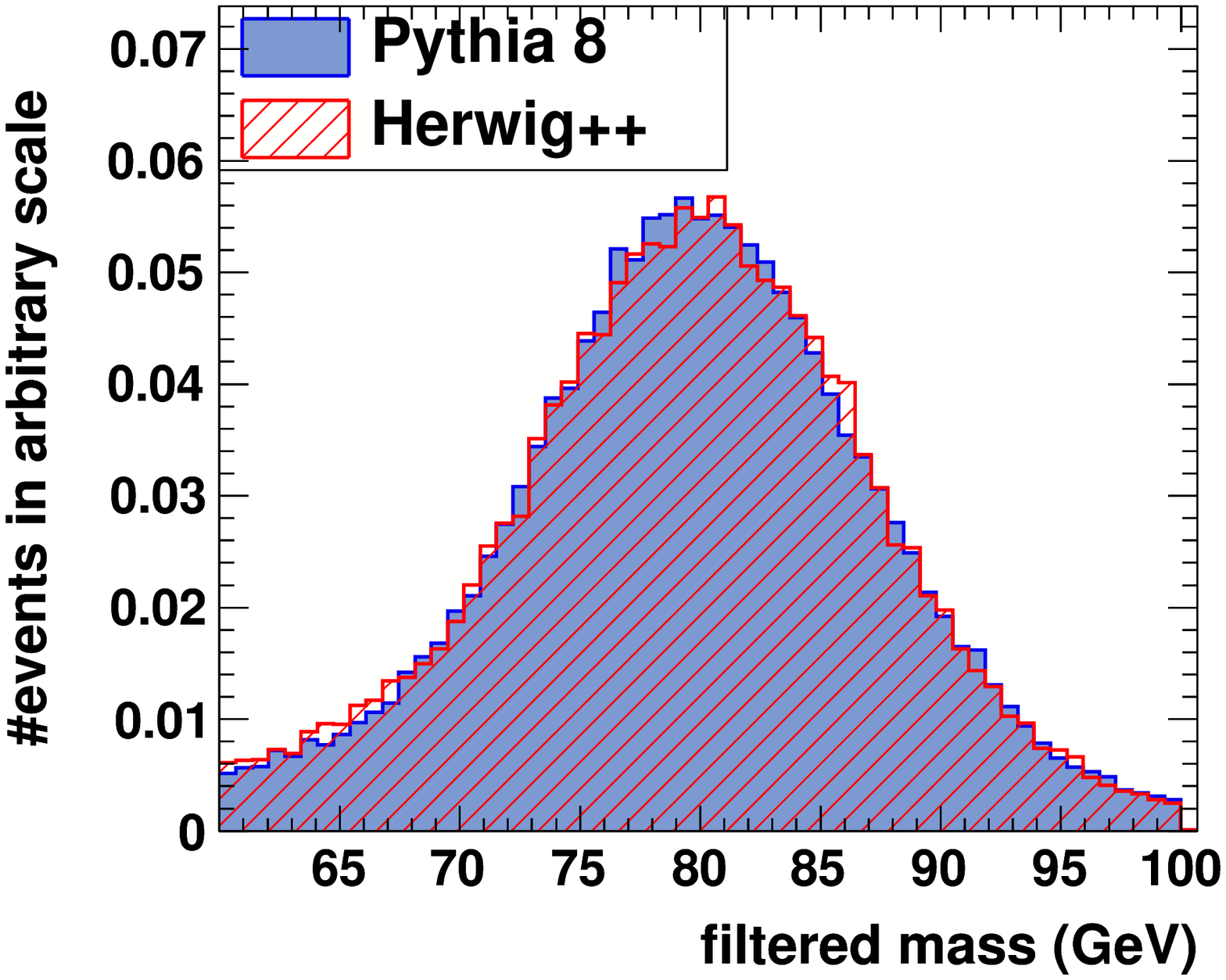}
&\includegraphics[width=0.33\textwidth]{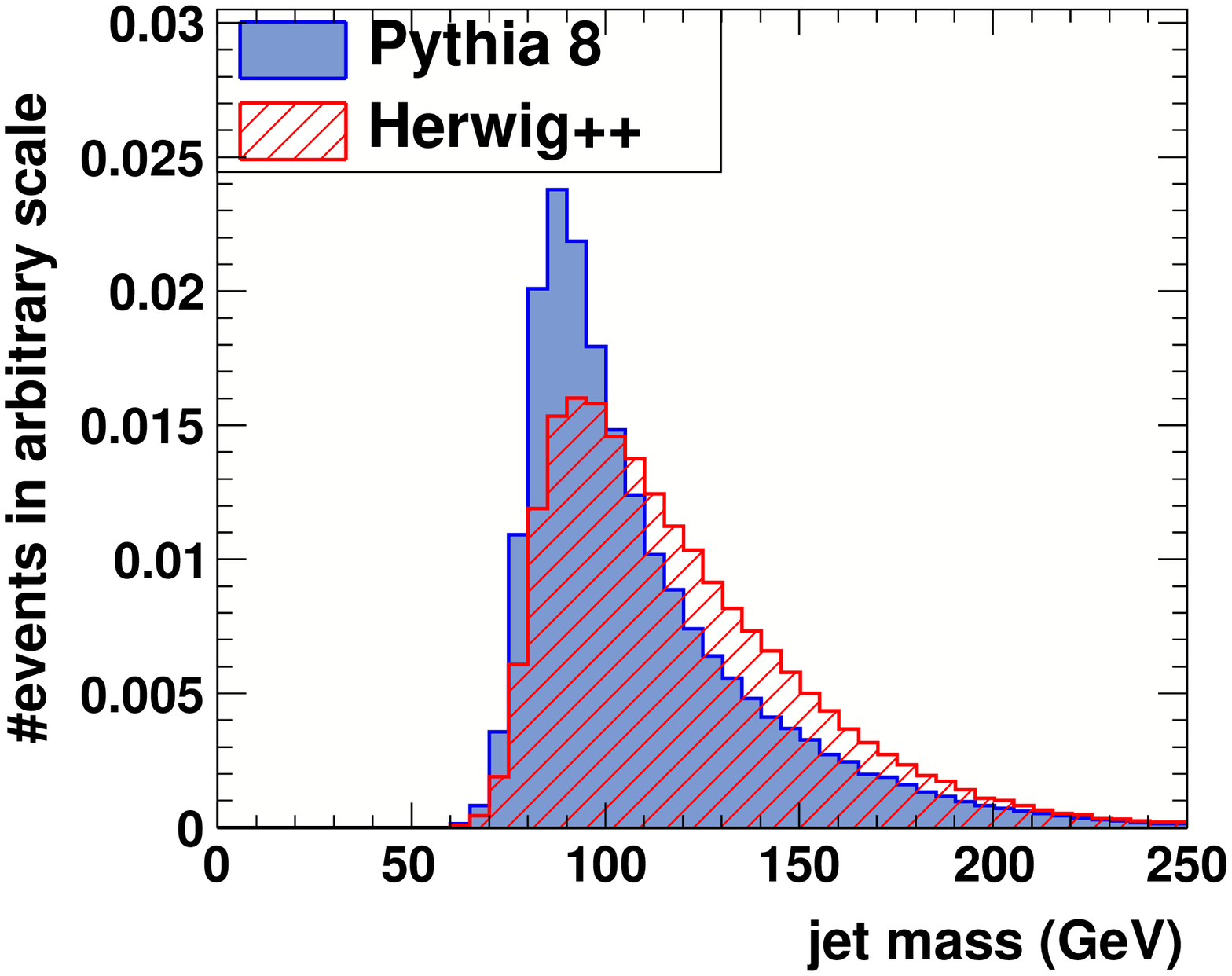}
&\includegraphics[width=0.33\textwidth]{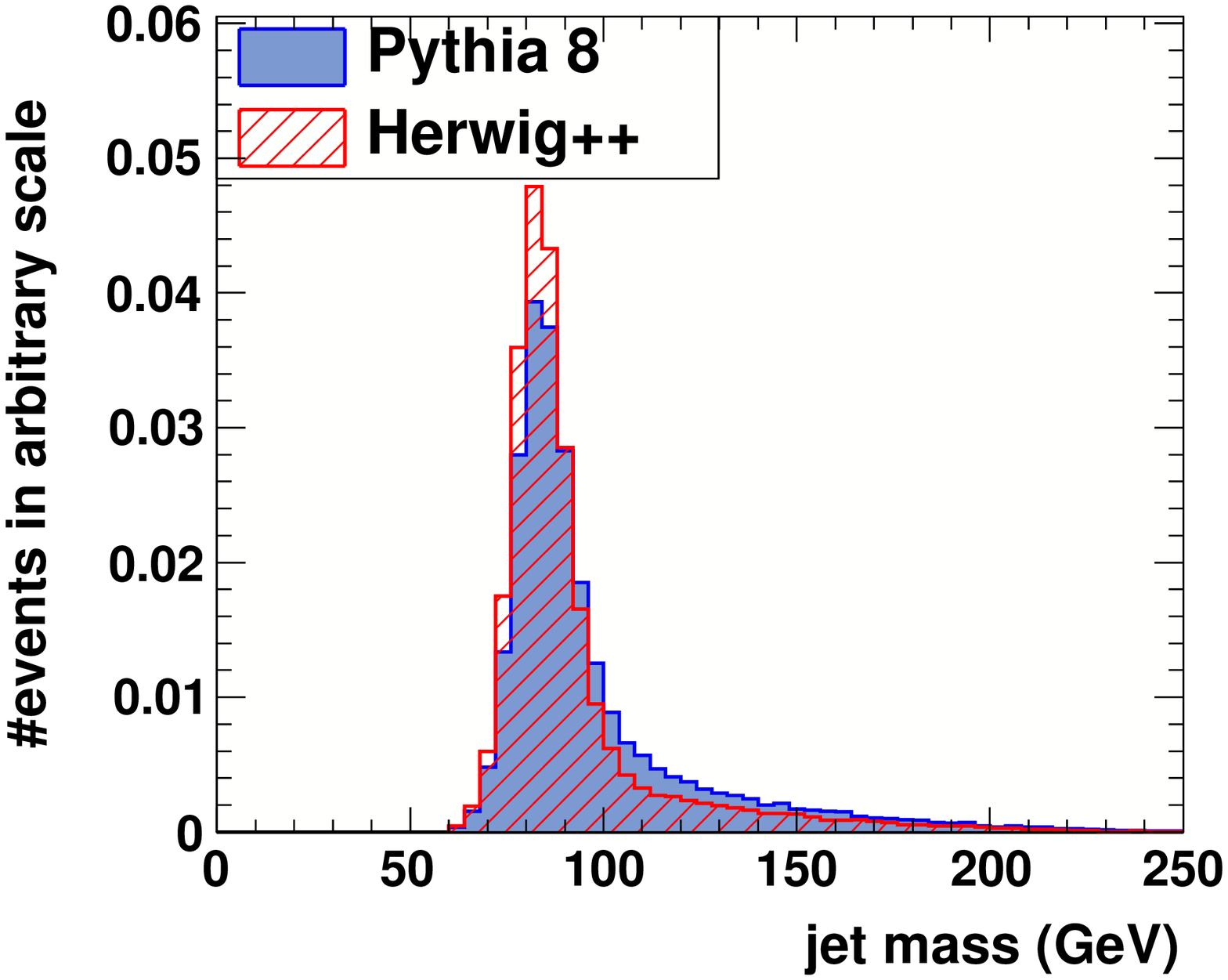}\\

(a) Filtered mass&(b)$R=1.2$ mass&(c) $R=1.2$ mass, no UE.
\end{tabular}
\caption{\label{fig:mass_compare}Simulation dependence of jet masses, for $W$-jets only. }
\end{center}
\end{figure}

\begin{figure}
\begin{center}
\begin{tabular}{cc}
\includegraphics[width=0.4\textwidth]{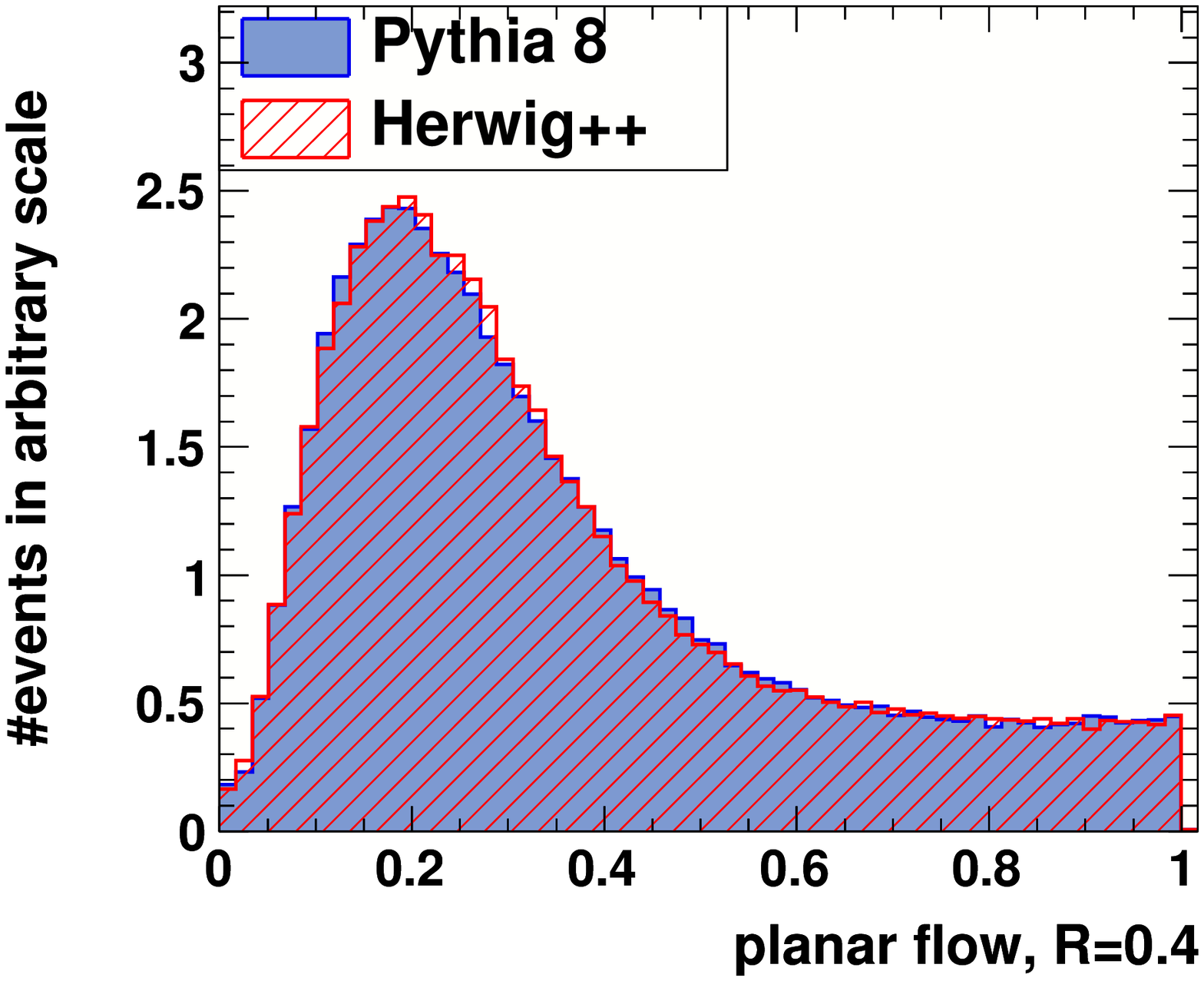}
&\includegraphics[width=0.4\textwidth]{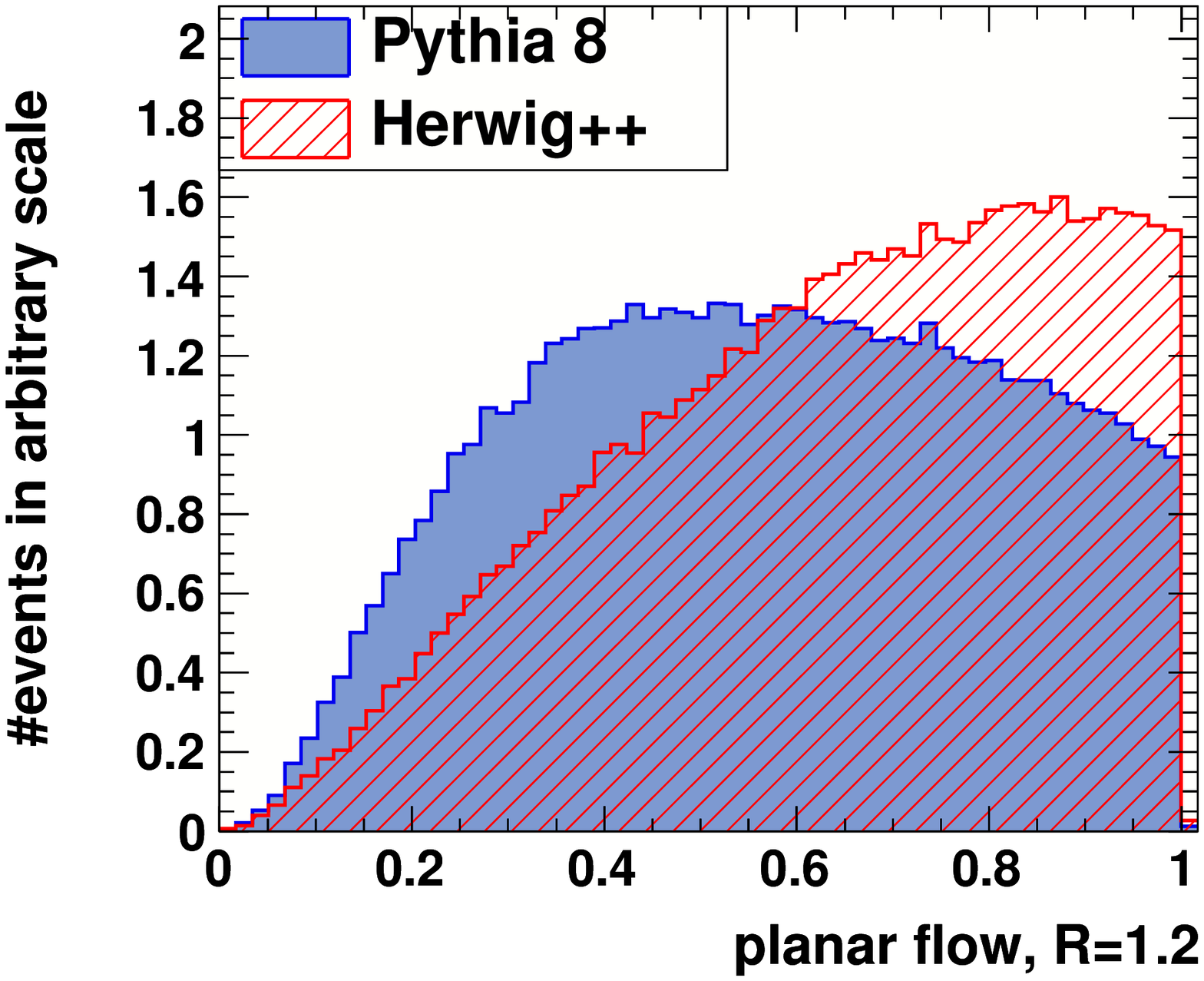}
\\
(a)&(b)
\end{tabular}

\caption{\label{fig:pf_pythia_herwig} Planar flow for $p_T^\jet\in(500, 550)\gev$, $W$-jets only: (a) $R=0.4$; (b) $R=1.2$.}
\end{center}
\end{figure}

\looseness -1 Another way to understand the effect is through the grooming sensitivities. In Figure~\ref{fig:compare_sens_trim}, we draw separately the trimming sensitivity, ${\rm sens}_{\rm trim}^m$ for $W$-jets and QCD-jets. We also draw distributions with UE turned off, and distributions with both UE and ISR turned off. In the latter case, the only contribution to the radiation is through final state radiation; we see that ${\rm sens}_{\rm trim}^m$ is much more concentrated around 1 for $W$-jets than QCD-jets, which means much less radiation is trimmed away for $W$-jets. After adding the ISR, the difference is still dramatic. When all contributions are included, the difference between $W$-jets and QCD-jets becomes smaller. This explains why one can obtain better discrimination power by turning off UE, as shown in Figure~\ref{fig:compare_mc}. Moreover, Figure~\ref{fig:compare_sens_trim} clearly shows that more radiation is trimmed away for Herwig++ than for Pythia 8, in both the signal and background distributions. The difference is more significant in the signal distributions and again, the Herwig++ result is more similar to the background.
\begin{figure}
\begin{center}
\begin{tabular}{cc}
\includegraphics[width=0.45\textwidth]{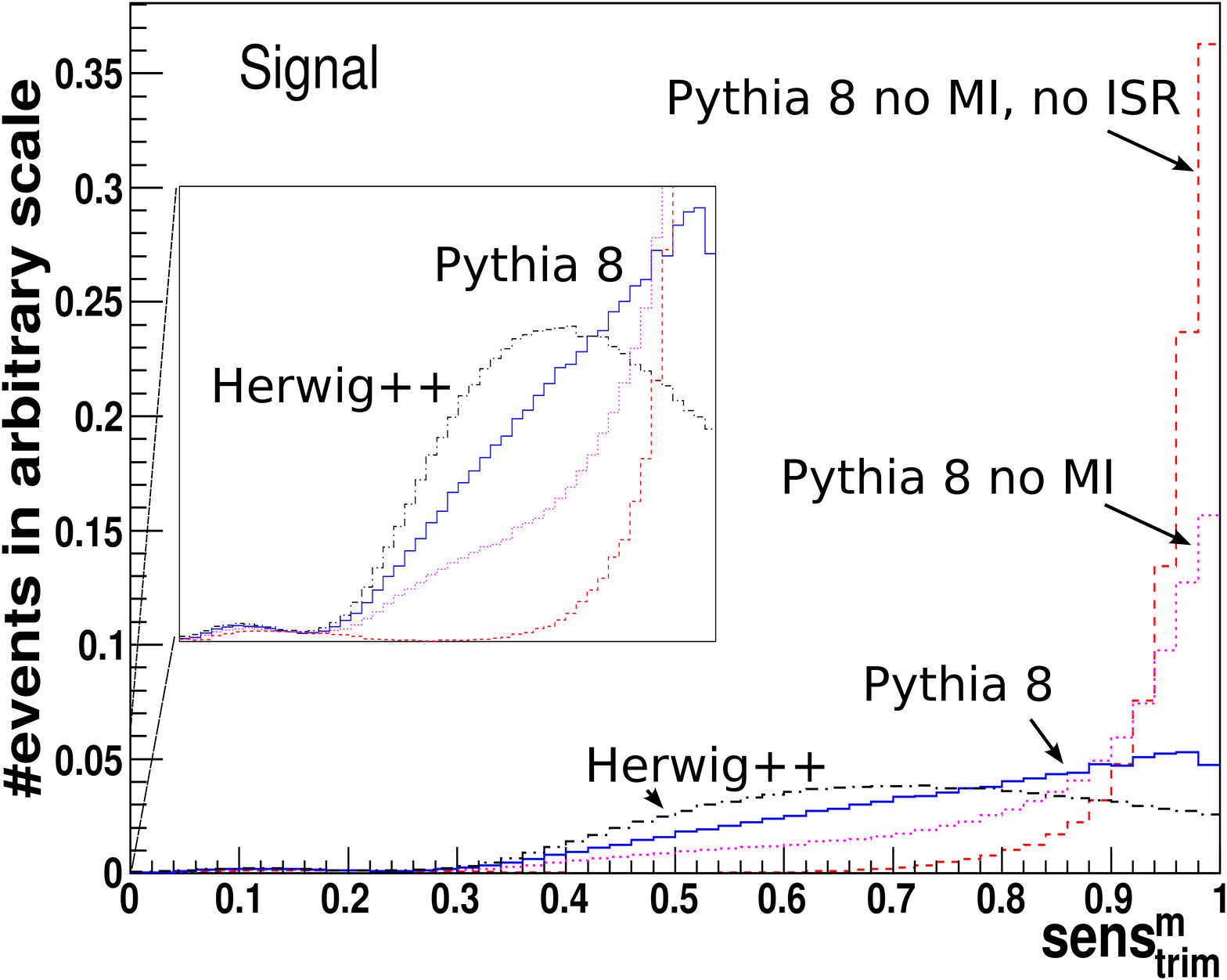}
&\includegraphics[width=0.45\textwidth]{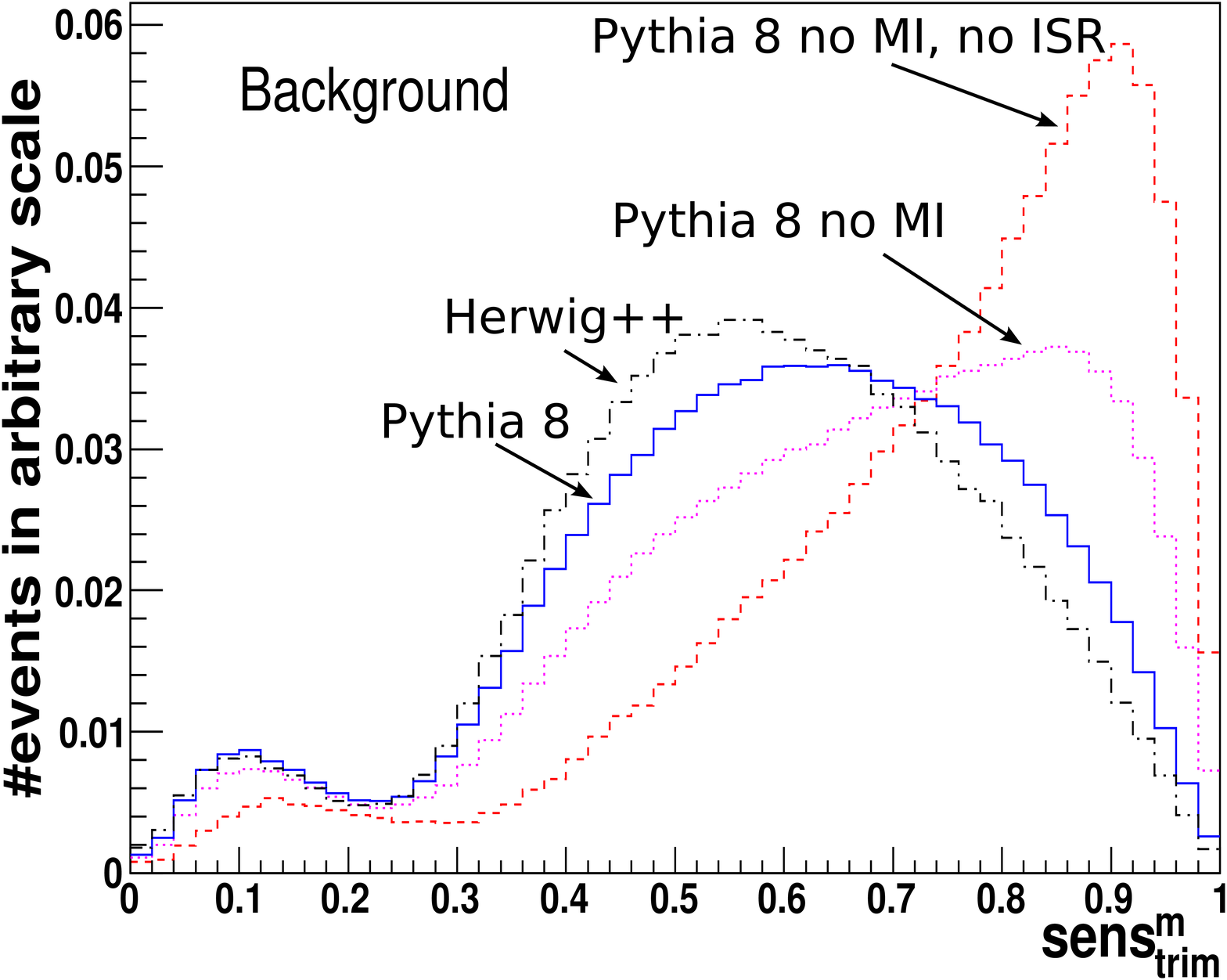}\\
(a)&(b)
\end{tabular}
\caption{\label{fig:compare_sens_trim} ${\rm sens}_{\rm trim}^m$ distributions for (a) signal; (b) background.}
\end{center}
\end{figure}

We have seen that the variables which have the larger difference involve larger, or unfiltered jets, and are
therefore more sensitive to the UE.
This motivates us to consider only variables defined within a small region around the candidate
$W$-jet direction. Such a set of variables was listed in Eq.~\eqref{eq7}.
 In Figure~\ref{fig:slim_set}, we show the significance improvement using this set.
The differences between the two Monte Carlos is clearly smaller than in Figure~\ref{fig:compare_mc},
but still visible.

\begin{figure}
\begin{center}
\begin{tabular}{cc}
\includegraphics[width=0.5\textwidth]{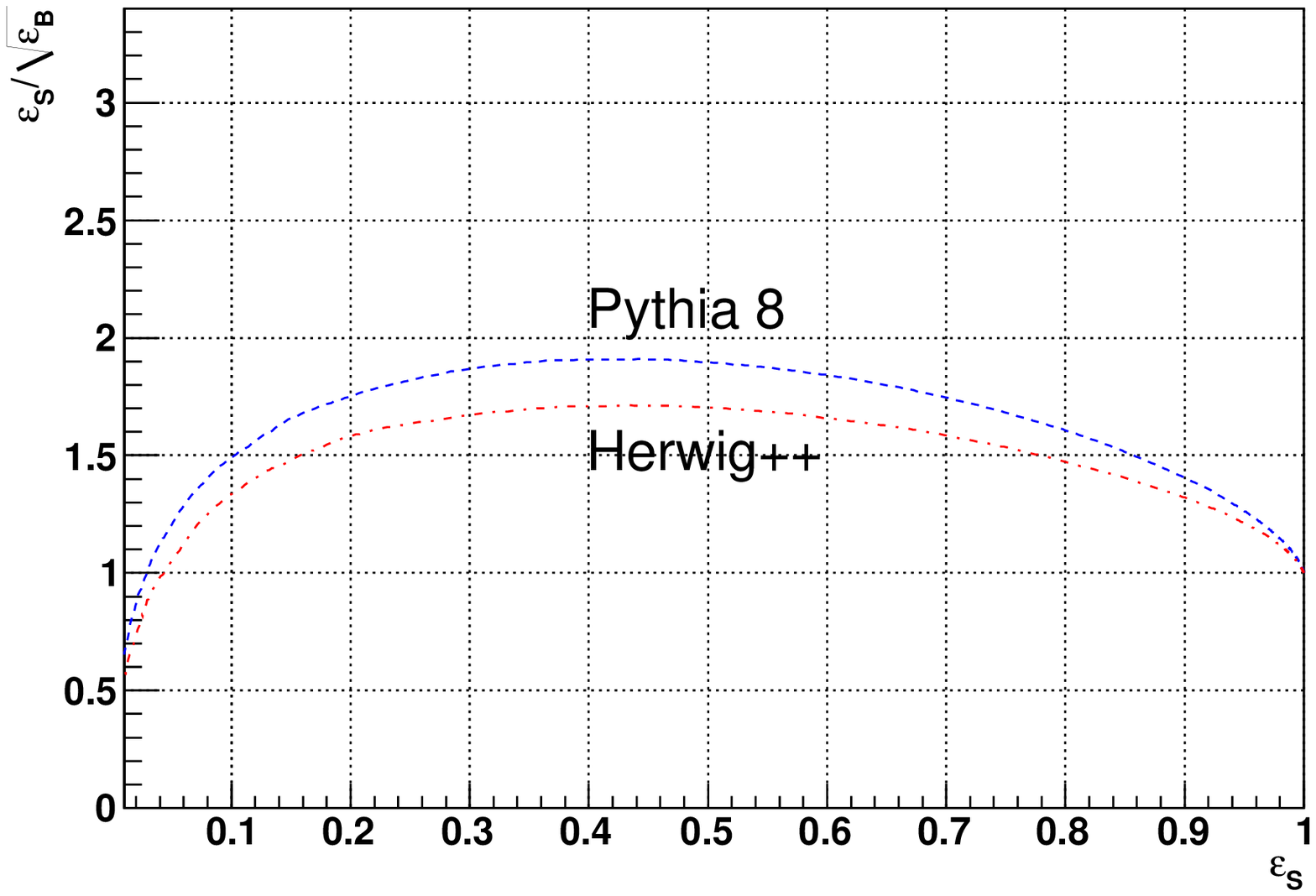}
&
\includegraphics[width=0.5\textwidth]{eff_significance_compare}\\
(a) 7 variables& (b) 25 variables
\end{tabular}
\caption{\label{fig:slim_set} SIC curves obtained using a smaller set of variables meant to reduce
dependence on modeling of the underlying event. The same BDTs trained on Pythia 8 are tested on Pythia 8 and Herwig++. For comparison,
Figure~\ref{fig:compare_mc} (a) is reproduced here in the second panel.
}
\end{center}
\end{figure}

\section{Applications}
\label{sec:applications}
In this section, we apply the method presented in the previous sections to other processes
involving boosted $W$-bosons. We demonstrate the robustness of our method as a general purpose $W$-tagger, and show the improvements compared to more conventional methods.

\subsection{$Z'\rightarrow W^+W^-\rightarrow l^\pm+j+\slashchar{E}_T$}
A well-motivated application of our W-jet tagging method is the search for new vector resonance $Z'$ via $pp\rightarrow Z'\rightarrow W^+W^-\rightarrow l\nu qq$. In addition to the general possibility that a new $Z'$ can have a significant coupling to $W^+W^-$, this channel is particularly important in models where electroweak symmetry breaking is related to strong dynamics. In technicolor or 5D Higgsless \cite{Csaki:2003zu} models, exchanging a tower of $Z'$ resonances is essential for restoring unitarity for high energy $W_LW_L$ scattering as a substitute of a light Higgs. For a $Z'$ with couplings similar to those of the $Z$,
direct searches and  electroweak precision constraints have pushed its allowed mass to be above $\sim 1$ TeV \cite{ewzprime}.
$W$ bosons produced from such heavy $Z'$ are expected to be highly boosted, therefore provide a natural arena to test our method.

In more conventional methods, the hadronic $W$ from a $Z'$ decay is either treated as two separate jets or one fat jet.
For example, the authors of Ref.~\cite{Alves:2009aa} demand two jets reconstructing the $W$ mass and separated by $\Delta R_{jj}>0.4$. This method eliminates a large fraction of the signal when $M_{Z'}\gtrsim$1 TeV due to the merging of the $W$ decay products to one jet. In the study of TeV scale Kaluza-Klein $Z'$ in Randall-Sundrum (RS) models in Ref.~\cite{Agashe:2007ki}, the authors use a simple jet mass cut around $M_W$ with jet size $R=0.4$. We will see that the latter gives us similar results as filtering, while using our $W$-jet tagging method, we obtain significantly better results in both $S/\sqrt{B}$ and $S/B$.

For concreteness we consider a $Z'$ which couples to the SM fermions and gauge bosons with the same Lorentz structure as the SM $Z$ boson, yet with rescaled strength. We choose the couplings $g_{Z'f\bar{f}}=0.2g_{Zf\bar{f}},~~ g_{Z'WW}=\frac{M_Z}{\sqrt{3}M_{Z'}}g_{ZWW}$, as in typical RS models
~\cite{Alves:2009aa}. We consider $Z'$ with a mass $M_{Z'}=1.5$ TeV and a width $\Gamma_{Z'}\approx125\gev$. We consider the 14 TeV run of the LHC, where the effective cross section for $Z'\rightarrow W^+W^-$ in the semileptonic channel is 26.4 fb. Note that for such a high mass $Z'$, 97.5\% events have a $\Delta R<0.4$ for the two quarks from the $W$ decay (parton level), making it very difficult to identify two separate jets. Therefore, we focus on the methods when the $W$'s are identified as single jets. The signal events therefore contain $l^\pm+1j+\slashchar{E}_T$. The major SM backgrounds are $W+1j$, $WW$ and $t\bar t$. All signal and background events are generated with Madgraph 4 at parton level. As before, the events are processed with Pythia 8 and jets are found with the C/A algorithm using $R=1.2$. The following kinematic cuts are then applied:
\beq
|\eta_l|<2.5, ~~~~~|\eta_j|<3,~~~~~p^l_T>100\gev, ~~~~~ p^j_T>500\gev, ~~~~~\slashchar{E}_T>100\gev,\label{starting-cut}
\eeq
where the $p_T$ cuts apply on the leading jet and lepton, which are assumed to be the $W$-jet and the lepton from the leptonic $W$ decay.

To efficiently reduce QCD backgrounds, especially the $t\bar{t}$ background we veto additional central jets with
\beq
|\eta_j|<3~~~~~~~~ {\rm and}~~~~~~~~~~~p^j_T>100\gev.\label{central-veto}
\eeq
We then apply our $W$-jet tagging procedure on the leading jet in events passing the above cuts to identify the hadronic $W$'s. In particular, we use the same parameters and BDT weight files obtained before from training the SM $WW$/$Wj$ samples.

The naive way of applying the BDT weight files is to impose the optimal BDT cuts for maximizing the SIC of $W$-jets {\it vs} QCD-jets, since $W$+jet is the dominant background in our $Z'$ search. However, our method is so efficient for reducing the QCD-jets such that after doing so, the $W$+jet background is comparable to the $WW$ and $t\bar t$ backgrounds which contain $W$-jets as well. Therefore, the optimal BDT cuts when all backgrounds are included are different from before. In order to obtain the best significance for $Z'$ search, we use the same BDT weight files while scan the BDT cuts for each $p_T$ bin to maximize $S/\sqrt{\sum{B}}$ where the sum is over the $Wj, WW, t\bar{t}$ SM backgrounds weighted by their cross-sections. The result presented below is then the optimal one from such scan.

The presence of only one neutrino in the final state allows the reconstruction of its momentum by requiring transverse momentum conservation and applying the $W$ mass constraint. In doing so, we obtain two solutions of the neutrino $p_z$, which, combined with the hadronic $W$ momentum, give rise to two reconstructed $WW$ masses. We take the minimum of the two reconstructed masses $M^{\text{min}}_{WW,rec}$. The resulting $M^{\text{min}}_{WW,rec}$ distributions are shown in Figure~\ref{fig:z'}, where an integrated luminosity of 2 ${\rm fb}^{-1}$ is assumed. We then apply a cut on the $Z'$ mass, $M^{\text{min}}_{WW,{\text{rec}}}\in (1300, 1700)$ GeV. The number of events within this window at various steps are given in Table \ref{tab:z'}, together with $S/\sqrt{B}$ and $S/B$.  For comparison, we have also included the results using conventional jet mass method, obtained by reclustering the events with $R=0.4$ and apply the kinematic cuts as well as a jet mass cut $(60, 100)\gev$ for candidate $W$-jets.

 \begin{figure}
\begin{center}
        \includegraphics[width=1\textwidth]{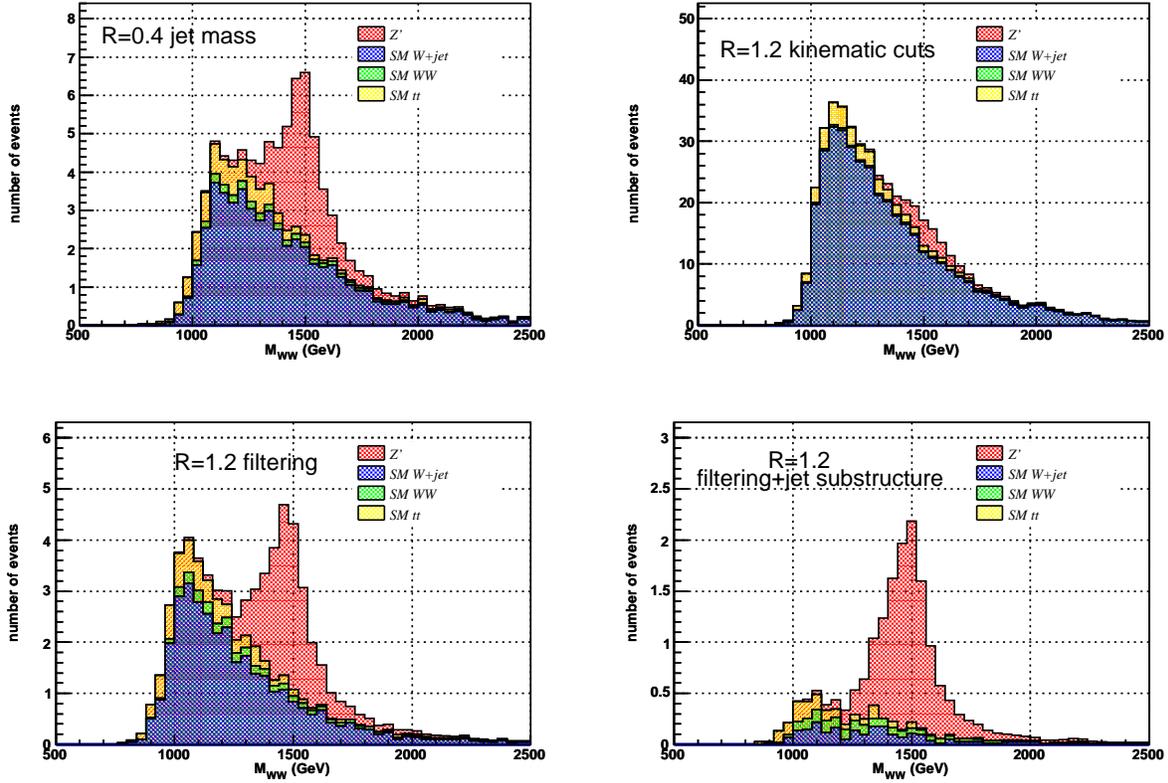}
\end{center}
\caption{Invariant mass distributions for signal ($Z'\rightarrow W^+W^-\rightarrow l^\pm+j+\slashchar{E}_T$) {\it vs} backgrounds. The upper left pane is for conventional jet mass method $(R=0.4)$.}
\label{fig:z'}
\end{figure}

\begin{table}
\caption{Number of events, $S/\sqrt{B}$ and $S/B$ at $2~\rm{fb}^{-1}$ for signals with $M_Z'=1.5~{\rm TeV}$ and major SM backgrounds. A $(1300, 1700)\gev$ mass window cut is imposed on the reconstructed $Z'$ mass. Numbers in parenthesis are for the case when only $Wj$ is taken as the background. \protect\footnote{Note that for small numbers of events, Poisson statistics should be used to extract the exact significance. Assuming an integer number of events closest to the expectation value of $S+\sum B$ are observed, we have the significances: 2.0, 4.3, 5.3 and 3.9.}}
  \begin{tabular}{|c|c|c|c|c||c|c|}
    \hline
    &  signal & $Wj$     & $t\bar{t}$ & $WW$ & $S/\sqrt{B}$ & $S/B$\\
    \hline
    Kinematic cuts    & $23$                & $148$    & $12$        & $2.1$ & 1.8(1.9) & 0.14 (0.15)\\
    \hline
    Filter  & $18$                & $10$      & $1.4$      &  $1.2$ & 5.0 (5.6) & 1.4 (1.7)\\
    \hline
    MVA     & $11$                 & $0.91$    &  $0.35$     & $0.68$ & $7.6 (11)$ & $5.5 (11)$  \\
    \hline
    $R=0.4$ mass cut &  $22$          & $22$     & $2.4$     & $1.4$ & $4.3 (4.6)$ & $0.85 (1.0)$ \\
    \hline
    \end{tabular}
  \label{tab:z'}
  \end{table}

From Table \ref{tab:z'}, we see the traditional jet mass method gives similar $S/\sqrt{B}$ as filtering, while using our $W$-jet tagging method, we obtain significantly better results in both $S/\sqrt{B}$ and $S/B$. Note that the signal efficiency after filtering is larger than those given in Table \ref{tab:jet_pars} because the $W$'s from $Z'$ decays are dominantly longitudinal.

\subsection{Dijet versus $W+$jet}
Our last test and application of the method is to consider the possibility of identifying boosted $W$-bosons in dijet events at the early LHC. We consider the 7 TeV run with 1 $\mbox{fb}^{-1}$ integrated luminosity\footnote{A similar study using the filtering method alone has been performed in \cite{Wj-Boost2010}.}. We will not include systematic uncertainties such as from QCD dijet cross-section calculation, since the main purpose here is to test the robustness of our method. In this process, there is no way to distinguish hadronic $W$ and $Z$ bosons except for the mass difference. If one would like to identify both $W$'s and $Z$'s, it is better to rerun the optimization procedure including both $W$'s and $Z$'s. For example, we should probably use a wider filtered mass window and also include both $W$'s and $Z$'s when training the BDT. As a direct test of our method, we apply exactly the same cuts/weight files obtained above and treat $Z$+jet as a background.

We consider jets with $p_T>400$ GeV. The jet mass distributions for $W$+jet, QCD dijet and $Z$+jet event samples (generated with Pythia 8) are shown in Figure~\ref{fig:dijet}. The corresponding numbers of jets, $S/\sqrt{B}$ and $S/B$ are shown in Table \ref{tab:dijet}. Note that in the $W$+jet sample, only half of the high $p_T$ jets come from a $W$ decay. If only the $W$'s are counted as signal, $S/\sqrt{B}$ and $S/B$ in the first row of Table \ref{tab:dijet} should be cut in half to 1.1 and 0.0016 respectively. Then we see filtering increases the significance by a factor of $\sim 2$, which is increased further by a factor of 2.2 after MVA. This is in line with the results given in Section \ref{sec:MVA}, although the processes and center of mass energy are different.

 \begin{figure}
\begin{center}
 \includegraphics[width=0.8\textwidth]{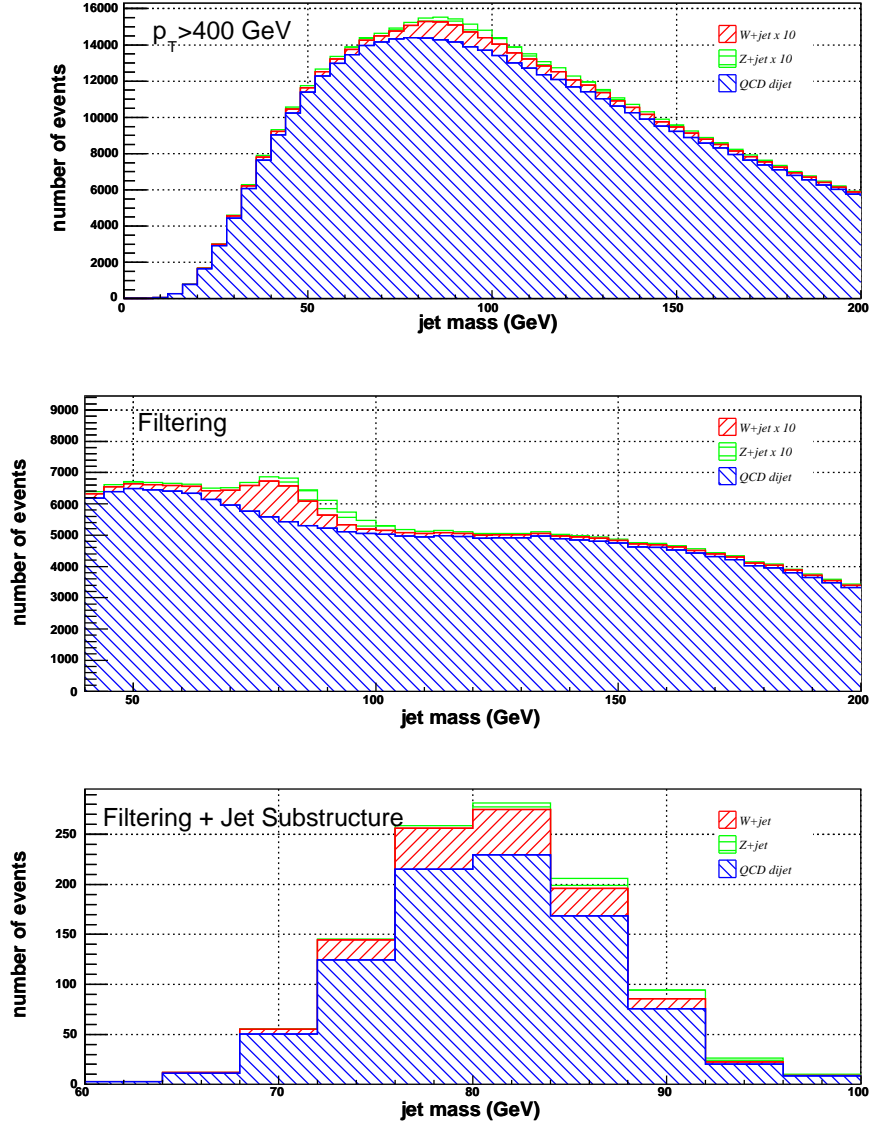}
\caption{\label{fig:dijet} Application of $W$-jet tagging to hadronic-$W$+jet search. Top:
jets with $p_T>400\gev$; middle: after filtering+mass-drop; bottom: after multivariate analysis. The $W$+jet and $Z$+jet contributions are multiplied by 10 in the top two panels to make them visible.}
\end{center}
\end{figure}
\begin{table}[htb]
\begin{center}
\begin{tabular}{|c|c|c|c||c|c|}
\hline
&$W$+jet & QCD dijet & $Z$+jet & $S/\sqrt{B}$ & $S/B$\\
\hline
 $p_T>400\gev$ & 1570 & 490k& 753& 2.2&0.0032\\
\hline
filtering&594 &67k &250 &2.3 & 0.0088\\
\hline
MVA &153 &906 & 34 &5.1 &0.17\\
\hline
\end{tabular}
\end{center}
\caption{\label{tab:dijet} Number of jets in different dijet samples for 7 TeV LHC with 1 $\mbox{fb}^{-1}$ integrated luminosity.}
\end{table}

\section{Conclusion and discussions}
\label{sec:conclusion}

In this article, we have investigated the differences between QCD-jets and highly boosted hadronically decaying color singlet particles. We have shown that excellent distinguishing power can be achieved by utilizing a multivariate method: for jets with $p_T>200\gev$, we obtain a factor of $\sim5$ improvement in the statistical significance. We have considered $W$ bosons as an example, and the same method can be used on highly boosted $Z$ bosons or Higgs bosons as well.

\looseness -1 There are two major differences between a $W$-jet and a QCD-jet. First, the two subjets initiated by the two quarks from a $W$ decay tend to carry momenta of similar size with their angular distance determined by the $W$ mass and momentum. If the $W$ boson is not too boosted ($p_T \lesssim 1200\gev$), two clean subjets can be identified using usual jet algorithms but with smaller radius. On the other hand, due to collinear and soft divergences, a QCD splitting tends to produce either two partons too close to be identified as two separate subjets, or two separate partons with hierarchical momenta. Therefore, we can distinguish a $W$-jet from a QCD-jet by requiring two subjets with balanced momenta. This is the idea behind the jet grooming algorithms proposed for identifying boosted decaying particles. However, as we mentioned in the introduction, jet grooming alone cannot give us the optimal discriminating power because information regarding radiation patterns is discarded.

Indeed, the second difference between $W$-jets and QCD-jets lies in the different patterns of final state radiation, which have not been explored sufficiently in the literature. For example, the radiation of a boosted color singlet particle such as a $W$ is mostly concentrated within a small region around its momentum. In this article, we have identified a set of efficient jet substructure variables and combined them in a multivariate analysis. We have found much better discriminating power than using jet grooming alone: a factor of $2\sim3$ improvement in the statistical significance is achieved on top of the filtering results.

We have used the SM $WW\rightarrow l\nu q q$ and $Wj\rightarrow l\nu j$ processes to optimize the discrimination power. It turns out that the variables we use characterize generic properties of high $p_T$ jets, independent of the specific process. We have illustrated this by considering two interesting applications. The first one is a $Z'$ search at the LHC with center of mass energy of 14 TeV, with the $Z'$ decaying to a $W$ pair and the $W$'s decaying semileptonically. The second one is searching for hadronic-$W$+jet events in dijet events at the 7 TeV LHC. In both processes, we have identified the boosted $W$'s using the same multivariate $W$-jet tagging algorithm trained to distinguish the SM $WW$ events from the SM $Wj$ events. We have found significant improvement over existing methods, consistent with the SM $WW$/$Wj$ results.

We have obtained our results using Pythia 8 simulations. As another test, we have applied exactly the same cuts obtained from Pythia 8 on data samples simulated with Herwig++. We have found a 25\% difference in the maximal significance, with Herwig++ giving the smaller value. As we have verified, most of the difference comes from the different treatment of the underlying event in the two Monte Carlo tools, which should be resolved once both Monte Carlos are tuned to the LHC measurements. We have also shown by using a subset of the variables that are less sensitive to the underlying event, we obtain more robust results which are almost as good as using the whole set.

Finally, we point out that the code for $W$-jet tagging is publicly available at http://jets.physics.harvard.edu/wtag. This code
contains the trained boosted decisions trees and can be used immediately in applications. Users can also conveniently use the provided routines to examine the jet substructure variables and/or train their own event samples.

\acknowledgments
We thank Jason Gallicchio for comments on the manuscript. The computations in this paper were run on the Odyssey cluster supported by the FAS Sciences Division Research Computing Group at Harvard University. YC is supported by NSF grant PHY-0855591 and the Harvard Center for Fundamental Laws of Nature. ZH is supported in part by NSF grant PHY-0804450. MS is supported in part by the Department of Energy under Grant DE-SC003916.

\appendix

\section{Filtering/trimming/pruning}
\label{app:pars}

All of the three jet grooming algorithms start from a jet found with some recombination algorithm such as $k_t$, anti-$k_t$ and Cambridge/Aachen (C/A) algorithms. It turns out filtering with mass drop gives us slightly better significance than pruning and trimming. For filtering, the C/A jet algorithm works the best, so we will fix the jet algorithm to C/A, except for trimming (see below). Starting from a jet with relatively large size $R$, the jet grooming algorithms act on the fat jet as follows
\begin{enumerate}
\item Filtering with mass drop \cite{filtering}: For a given jet found with recombination parameter $R$, we first look for a significant ``mass drop'' by the following procedure:
\begin{enumerate}
\item Undo the last step of jet clustering for jet $j$. The two resulting subjets $j_1$, $j_2$ are ordered such that $m_{j_1}>m_{j_2}$.
\item Stop the algorithm if a significant mass drop is found and the splitting is not too asymmetric, {\it i.e.}, if the following conditions are met:
\begin{equation}
m_{j_1} < \mu m_j \  \ \mbox{ and }\  \ y\equiv\frac{\min(p_{Tj_1}^2, p_{Tj_2}^2)}{m_j^2}\Delta R_{j_1.j_2}^2>y_\cut,
\end{equation}
where $\mu$ and $y_\cut$ are free parameters.
\item Otherwise redefine subjet $j_1$ as $j$ and repeat.
\end{enumerate}
When a mass drop is found, we use $R_{\filt}= \min(0.3, R_{j_1, j_2}/2)$ to recluster particles contained in $j_1$ and $j_2$.  The three hardest subjets are retained and combined as the new ``filtered'' jet. It is possible to do the reclustering procedure without the mass drop algorithm. Nevertheless, in our analysis mass drop is always included, and implicitly assumed whenever we refer to filtering.

\item Pruning \cite{pruning}: For a given jet, we recluster it with C/A algorithm, but when trying to merge subjets $i,j\rightarrow p$, the following condition is checked:
\begin{equation}
z\equiv\frac{\min(p_{Ti}, p_{Tj})}{p_{Tp}} < z_\cut \mbox{ and } \Delta R_{ij} > D_\cut,
\end{equation}
where $z_\cut$ and $D_\cut$ are free parameters. If the condition is met, do not merge the two subjets and the one with smaller $p_T$ is discarded. Continue until all particles are clustered or discarded.
In the code provided in Ref.~\cite{pruningcode}, $D_{\cut}$ is determined from another parameter, $R_\cut^{\rm{factor}}$, by $D_\cut = 2R_{\cut}^{\rm{factor}} m_p/p_{Tp}$.

\item Trimming \cite{trimming}:
For a given jet, we recluster it using $k_t$ algorithm with radius $R_\sub$ to identify the subjets. We then discard subjets $i$ with
\begin{equation}
p_{T,i} < f_{\cut}\, p_{T, \jet},
\end{equation}
where $p_{T,\jet}$ is the $p_T$ of the original jet. We see the difference between filtering and trimming is that we keep fixed number of subjets in filtering, while in trimming whether we keep a subjet is determined by the subjet's $p_T$.

\end{enumerate}

All three grooming algorithms involve two parameters in addition to the initial jet radius $R$. In our analysis, we fix $R=1.2$ and scan the other parameters to maximize $\varepsilon_S/\sqrt{\varepsilon_B}$ in the mass window $(60, 100)\gev$. As examples, the significance gain for pruning and trimming are shown in Figure~\ref{fig:prune_trim_contour} for jet $p_T\in(500, 550)\gev$. The optimal parameters for all $p_T$ bins we consider are given in Table \ref{tab:jet_pars}.
\begin{figure}
\begin{center}
\begin{tabular}{cc}
\includegraphics[width=0.5\textwidth]{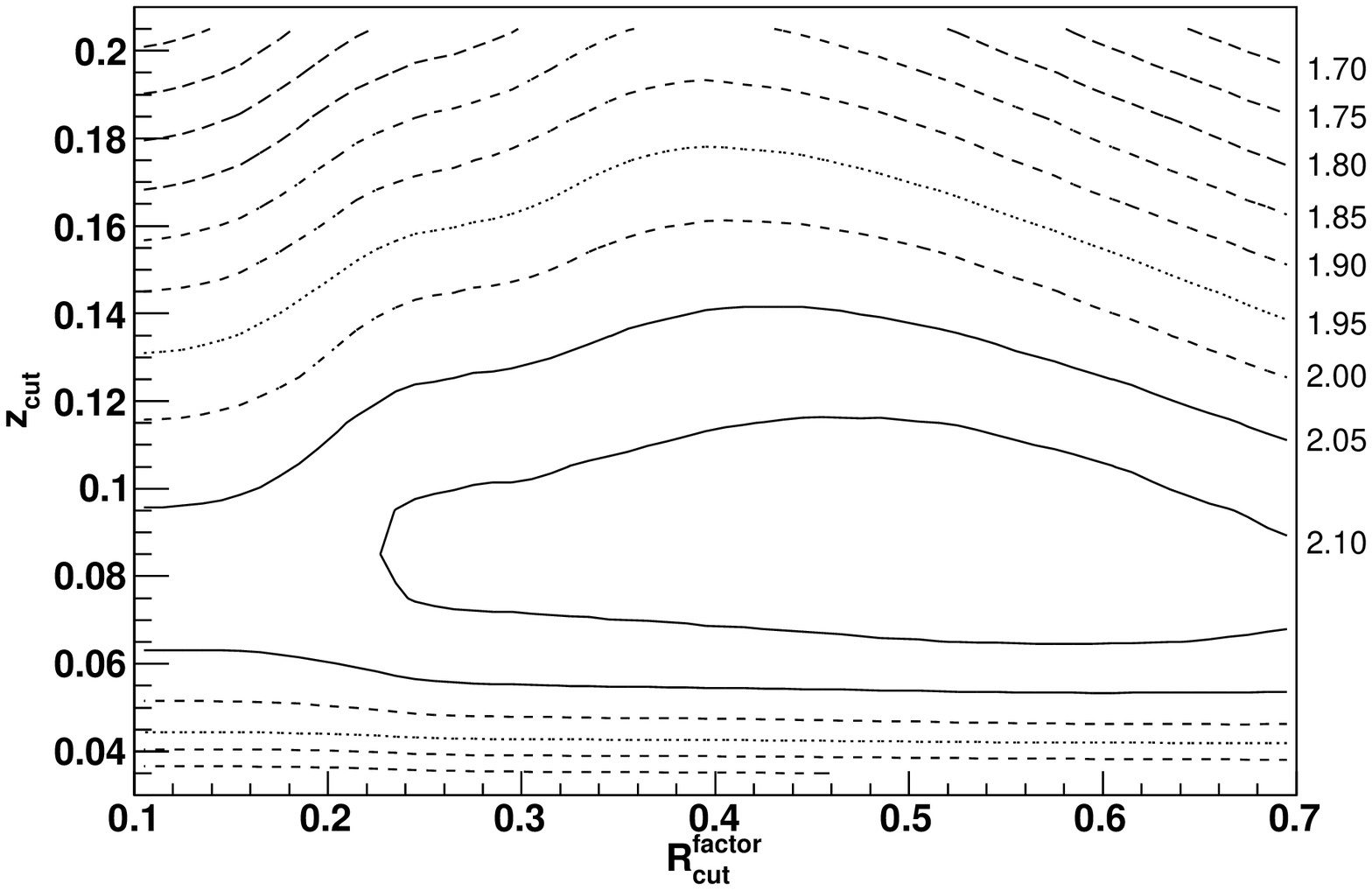}
&\includegraphics[width=0.5\textwidth]{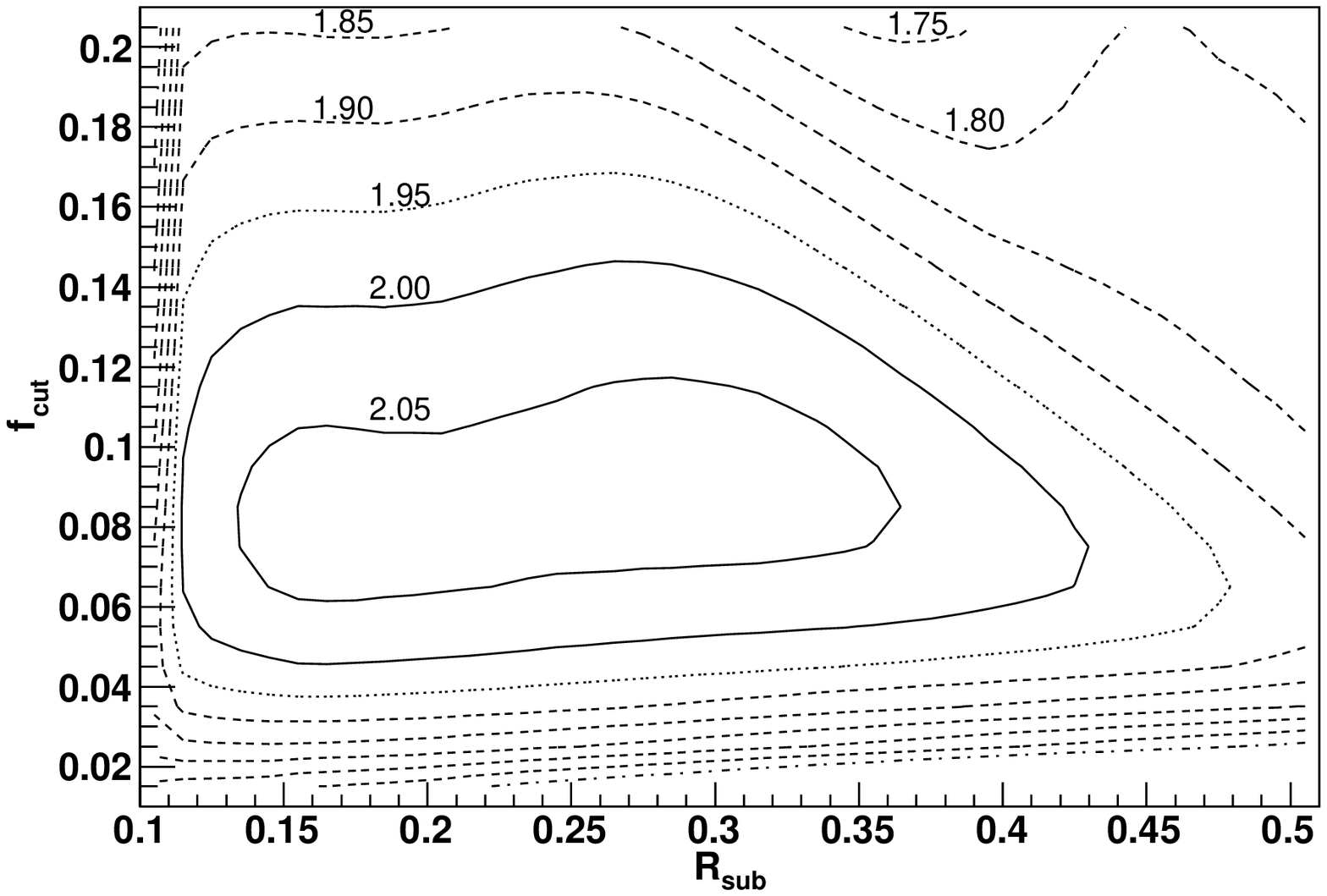}
\\(a) &(b)
\end{tabular}
\caption{For jet $p_T\in(500, 550)\gev$: (a) significance gain as a function of the pruning parameters; (b) significance gain as a function of the trimming parameters. \label{fig:prune_trim_contour}}
\end{center}
\end{figure}

\begin{table}[htb]
\begin{center}
{\footnotesize
\begin{tabular}{|c|c||c|c|c|c|c|c|c|c|c|c|c|c|c|c|c|c|c|}
\hline
\multicolumn{2}{|c||}{$p_T$ (GeV)} & 200 & 250 & 300 & 350 & 400 & 450 & 500 & 550 & 600 & 650 & 700 & 750 &800 & 850 & 900 &950 & 1000\\
\hline
\multirow{5}{*}{filt} & $\mu$ & .49 & .4 & .66 & .66 & .68 & .69 & .71 & .71 & .73 & .72 & .74 & .72 & .76 & .74 & .74 & .76 & .8\\\cline{2-19}
 & $y_\cut$  & .13 & .17 & .15 & .14 & .12 & .1 & .09 & .09 & .08 & .08 & .07 & .07 & .06 & .06 & .06 & .05 & .04\\\cline{2-19}
 &$\varepsilon_S$ & .61 & .52 & .57 & .58 & .61 & .64 & .66 & .65 & .66 & .64 & .64 & .61 & .61 & .6 & .58 & .58 & .59 \\\cline{2-19}&$\varepsilon_B$ & .13 & .082 & .084 & .079 & .083 & .088 & .089 & .085 & .086 & .081 & .08 & .075 & .076 & .073 & .072 & .077 & .084 \\\cline{2-19} &sig & 1.7 & 1.8 & 2 & 2.1 & 2.1 & 2.2 & 2.2 & 2.2 & 2.3 & 2.3 & 2.2 & 2.2 & 2.2 & 2.2 & 2.2 & 2.1 & 2\\\hline
\multirow{5}{*}{trim} & $R_\sub$ & .17 & .22 & .22 & .21 & .17 & .17 & .16 & .15 & .15 & .15 & .15 & .16 & .16 & .15 & .16 & .15 & .17\\\cline{2-19}
 & $f_{\cut}$  & .08 & .1 & .11 & .1 & .09 & .08 & .08 & .07 & .07 & .06 & .05 & .05 & .05 & .05 & .04 & .04 & .03\\\cline{2-19}
 &$\varepsilon_S$ & .58 & .61 & .6 & .62 & .64 & .67 & .67 & .69 & .7 & .72 & .74 & .74 & .74 & .74 & .74 & .71 & .7 \\\cline{2-19}&$\varepsilon_B$ & .1 & .11 & .1 & .1 & .1 & .11 & .11 & .11 & .11 & .12 & .12 & .13 & .13 & .13 & .14 & .14 & .15 \\\cline{2-19} &sig & 1.8 & 1.8 & 1.9 & 1.9 & 2 & 2 & 2.1 & 2.1 & 2.1 & 2.1 & 2.1 & 2.1 & 2.1 & 2.1 & 2 & 1.9 & 1.8\\\hline
\multirow{5}{*}{prun} & $R_{\cut}^{factor}$ & .48 & .54 & .56 & .53 & .55 & .52 & .52 & .52 & .49 & .32 & .33 & .35 & .37 & .39 & .29 & .17 & .16\\\cline{2-19}
 & $z_{\cut}$  & .17 & .15 & .13 & .12 & .1 & .09 & .08 & .07 & .07 & .06 & .06 & .05 & .05 & .04 & .04 & .04 & .03\\\cline{2-19}
 &$\varepsilon_S$ & .55 & .57 & .6 & .62 & .66 & .68 & .69 & .71 & .72 & .73 & .72 & .73 & .72 & .73 & .72 & .7 & .67 \\\cline{2-19}&$\varepsilon_B$ & .1 & .098 & .099 & .097 & .1 & .1 & .11 & .11 & .11 & .11 & .11 & .11 & .11 & .12 & .12 & .12 & .12 \\\cline{2-19} &sig & 1.7 & 1.8 & 1.9 & 2 & 2.1 & 2.1 & 2.1 & 2.1 & 2.2 & 2.2 & 2.2 & 2.2 & 2.2 & 2.1 & 2.1 & 2 & 1.9\\\hline
\end{tabular}

}
\end{center}
\caption{Jet grooming parameters maximizing the significance. \label{tab:jet_pars}}
\end{table}


\begin{thebibliography}{99}
\bibitem{Chanowitz:1985hj}
  M.~S.~Chanowitz and M.~K.~Gaillard,
  Nucl.\ Phys.\  B {\bf 261}, 379 (1985).

\bibitem{Butterworth:2002tt}
  J.~M.~Butterworth, B.~E.~Cox and J.~R.~Forshaw,
  Phys.\ Rev.\  D {\bf 65}, 096014 (2002)
  [arXiv:hep-ph/0201098].

\bibitem{Almeida:2008yp}
  L.~G.~Almeida, S.~J.~Lee, G.~Perez, G.~F.~Sterman, I.~Sung and J.~Virzi,
  Phys.\ Rev.\  D {\bf 79}, 074017 (2009)
  [arXiv:0807.0234 [hep-ph]].

\bibitem{pruning}
  S.~D.~Ellis, C.~K.~Vermilion and J.~R.~Walsh,
  Phys.\ Rev.\  D {\bf 80}, 051501 (2009)
  [arXiv:0903.5081 [hep-ph]];
S.~D.~Ellis, C.~K.~Vermilion and J.~R.~Walsh,
  Phys.\ Rev.\  D {\bf 81}, 094023 (2010)
  [arXiv:0912.0033 [hep-ph]].


\bibitem{Hackstein:2010wk}
  C.~Hackstein and M.~Spannowsky,
  arXiv:1008.2202 [hep-ph].

\bibitem{Katz:2010mr}
  A.~Katz, M.~Son and B.~Tweedie,
  arXiv:1010.5253 [hep-ph].


\bibitem{Thaler:2010tr}
  J.~Thaler and K.~Van Tilburg,
  arXiv:1011.2268 [hep-ph].

 \bibitem{filtering}
  J.~M.~Butterworth, A.~R.~Davison, M.~Rubin and G.~P.~Salam,
  Phys.\ Rev.\ Lett.\  {\bf 100}, 242001 (2008)
  [arXiv:0802.2470 [hep-ph]].



\bibitem{Plehn:2009rk}
  T.~Plehn, G.~P.~Salam and M.~Spannowsky,
  Phys.\ Rev.\ Lett.\  {\bf 104}, 111801 (2010)
  [arXiv:0910.5472 [hep-ph]].


\bibitem{Kribs:2009yh}
  G.~D.~Kribs, A.~Martin, T.~S.~Roy and M.~Spannowsky,
  Phys.\ Rev.\  D {\bf 81}, 111501 (2010)
  [arXiv:0912.4731 [hep-ph]].

\bibitem{Soper:2010xk}
  D.~E.~Soper and M.~Spannowsky,
  JHEP {\bf 1008}, 029 (2010)
  [arXiv:1005.0417 [hep-ph]].

\bibitem{Chen:2010wk}
  C.~R.~Chen, M.~M.~Nojiri and W.~Sreethawong,
  arXiv:1006.1151 [hep-ph].

\bibitem{Falkowski:2010hi}
  A.~Falkowski, D.~Krohn, L.~T.~Wang, J.~Shelton and A.~Thalapillil,
  arXiv:1006.1650 [hep-ph].

\bibitem{Kribs:2010hp}
  G.~D.~Kribs, A.~Martin, T.~S.~Roy and M.~Spannowsky,
  arXiv:1006.1656 [hep-ph].

\bibitem{Almeida:2010pa}
  L.~G.~Almeida, S.~J.~Lee, G.~Perez, G.~Sterman and I.~Sung,
  arXiv:1006.2035 [hep-ph].

\bibitem{Katz:2010iq}
  A.~Katz, M.~Son and B.~Tweedie,
  arXiv:1011.4523 [hep-ph].

\bibitem{Kim:2010uj}
  J.~H.~Kim,
  arXiv:1011.1493 [hep-ph].

\bibitem{Thaler:2008ju}
  J.~Thaler and L.~T.~Wang,
  JHEP {\bf 0807}, 092 (2008)
  [arXiv:0806.0023 [hep-ph]].

\bibitem{Kaplan:2008ie}
  D.~E.~Kaplan, K.~Rehermann, M.~D.~Schwartz and B.~Tweedie,
  Phys.\ Rev.\ Lett.\  {\bf 101}, 142001 (2008)
  [arXiv:0806.0848 [hep-ph]].

\bibitem{Almeida:2008tp}
  L.~G.~Almeida, S.~J.~Lee, G.~Perez, I.~Sung and J.~Virzi,
  Phys.\ Rev.\  D {\bf 79}, 074012 (2009)
  [arXiv:0810.0934 [hep-ph]].

\bibitem{Krohn:2009wm}
  D.~Krohn, J.~Shelton and L.~T.~Wang,
  JHEP {\bf 1007}, 041 (2010)
  [arXiv:0909.3855 [hep-ph]].


\bibitem{Plehn:2010st}
  T.~Plehn, M.~Spannowsky, M.~Takeuchi and D.~Zerwas,
  arXiv:1006.2833 [hep-ph].

\bibitem{Bhattacherjee:2010za}
  B.~Bhattacherjee, M.~Guchait, S.~Raychaudhuri and K.~Sridhar,
  Phys.\ Rev.\  D {\bf 82}, 055006 (2010)
  [arXiv:1006.3213 [hep-ph]].

\bibitem{Rehermann:2010vq}
  K.~Rehermann and B.~Tweedie,
  arXiv:1007.2221 [hep-ph].


\bibitem{atlas}
ATL-PHYS-PUB-2009-088. ATL-COM-PHYS-2009-345.

\bibitem{cms}
  G.~Giurgiu  [for the CMS collaboration],
  arXiv:0909.4894 [hep-ex].


\bibitem{trimming}
  D.~Krohn, J.~Thaler and L.~T.~Wang,
  JHEP {\bf 1002}, 084 (2010)
  [arXiv:0912.1342 [hep-ph]].

\bibitem{Black:2010dq}
  K.~Black, J.~Gallicchio, J.~Huth, M.~Kagan, M.~D.~Schwartz and B.~Tweedie,
  arXiv:1010.3698 [hep-ph].


\bibitem{madgraph}
  J.~Alwall {\it et al.},
  JHEP {\bf 0709}, 028 (2007)
  [arXiv:0706.2334 [hep-ph]].


\bibitem{pythia8}
  T.~Sjostrand, S.~Mrenna and P.~Z.~Skands,
  Comput.\ Phys.\ Commun.\  {\bf 178}, 852 (2008)
  [arXiv:0710.3820 [hep-ph]].

\bibitem{Cacciari:2005hq}
  M.~Cacciari and G.~P.~Salam,
  Phys.\ Lett.\  B {\bf 641}, 57 (2006)
  [arXiv:hep-ph/0512210].


\bibitem{herwig++}
  M.~Bahr {\it et al.},
  Eur.\ Phys.\ J.\  C {\bf 58}, 639 (2008)
  [arXiv:0803.0883 [hep-ph]].


\bibitem{pruningcode}
C.~K.~Vermilion, FastPrune (2009), URL http://www.phys.washington.edu/groups/lhcti/pruning/.
  \bibitem{Gallicchio:2010sw}
  J.~Gallicchio and M.~D.~Schwartz,
  Phys.\ Rev.\ Lett.\  {\bf 105}, 022001 (2010)
  [arXiv:1001.5027 [hep-ph]].



\bibitem{TMVA}
http://tmva.sourceforge.net/.

\bibitem{Roe:2004na}
  B.~P.~Roe, H.~J.~Yang, J.~Zhu, Y.~Liu, I.~Stancu and G.~McGregor,
  Nucl.\ Instrum.\ Meth.\  A {\bf 543}, 577 (2005)
  [arXiv:physics/0408124].




\bibitem{Csaki:2003zu}
  C.~Csaki, C.~Grojean, L.~Pilo and J.~Terning,
  Phys.\ Rev.\ Lett.\  {\bf 92}, 101802 (2004)
  [arXiv:hep-ph/0308038].

\bibitem{ewzprime}
  C.~Amsler {\it et al.} [ Particle Data Group Collaboration ],
  Phys.\ Lett.\  {\bf B667}, 1 (2008);
  E.~Salvioni, G.~Villadoro, F.~Zwirner,
  JHEP {\bf 0911}, 068 (2009).
  [arXiv:0909.1320 [hep-ph]].

\bibitem{Alves:2009aa}
  A.~Alves, O.~J.~P.~Eboli, D.~Goncalves, M.~C.~Gonzalez-Garcia and J.~K.~Mizukoshi,
  Phys.\ Rev.\  D {\bf 80}, 073011 (2009)
  [arXiv:0907.2915 [hep-ph]].

\bibitem{Agashe:2007ki}
  K.~Agashe {\it et al.},
  Phys.\ Rev.\  D {\bf 76}, 115015 (2007)
  [arXiv:0709.0007 [hep-ph]].

\bibitem{Birkedal:2004au}
  A.~Birkedal, K.~Matchev and M.~Perelstein,
  Phys.\ Rev.\ Lett.\  {\bf 94}, 191803 (2005)
  [arXiv:hep-ph/0412278].
  
\bibitem{D0pull}
[D0 Collaboration], D0Note 6087-CONF,
``Search for the standard model
Higgs boson in the ZH$\rightarrow$bbvv channel in 6.4 $\rm{fb}^{-1}$ of ppbar collisions
at sqrt(s)=1.96 TeV'', Preliminary Results for Summer 2010 Conferences,
http://www-d0.fnal.gov/Run2Physics/WWW/results/prelim/HIGGS/H90/ , August 2010.

\bibitem{Wj-Boost2010}
A.~Davison and J.~Butterworth,``Boosted W/Z in Early Data'', presentation at BOOST 2010 conference,
http://indico.cern.ch/getFile.py/access?contribId=24$\&$sessionId$=$7$\&$resId$=0\&$materialId$=$slides$\&$confId$=74604$, June 2010.


\end{thebibliography}
\end{document}